\documentclass[reprint,amsmath,amssymb,aps,superscriptaddress,showpacs,floatfix,byrevtex,longbibliography]{revtex4-1}
\usepackage{amsmath}
\usepackage{amssymb}
\usepackage{amstext}
\usepackage{amsopn}
\usepackage{amsfonts}
\usepackage{amsxtra}
\usepackage[english]{babel}
\usepackage{graphicx}
\usepackage{float}
\usepackage{bm}
\usepackage{multirow}
\usepackage{dcolumn}
\usepackage{color}
\usepackage{hyperref}

\begin{document}

\preprint{Draft --- not for distribution}

%
%
\title{Optical properties of \emph{A}Fe$_\mathbf{2}$As$_\mathbf{2}$ (\emph{A}$=\,$Ca, Sr, and Ba) single crystals}
\author{Y. M. Dai}
\email{ymdai@lanl.gov}
\affiliation{Condensed Matter Physics and Materials Science Division,
  Brookhaven National Laboratory, Upton, New York 11973, USA}%
\author{Ana Akrap}
\affiliation{DQMP -- University of Geneva, CH-1211 Geneva 4, Switzerland}
\author{S. L. Bud'ko}
\author{P. C. Canfield}
\affiliation{Ames Laboratory, U.S. DOE, and Department of Physics and Astronomy,
  Iowa State University, Ames, Iowa 50011, USA}
\author{C. C. Homes}
\email{homes@bnl.gov}
\affiliation{Condensed Matter Physics and Materials Science Division,
  Brookhaven National Laboratory, Upton, New York 11973, USA}%
\date{\today}

%
%
\begin{abstract}
The detailed optical properties have been determined for the iron-based materials
$A$Fe$_2$As$_2$, where $A=\,$Ca, Sr, and Ba, for light polarized in the iron-arsenic
(\emph{a-b}) planes over a wide frequency range, above and below the magnetic and
structural transitions at $T_N = 172$, 195, and 138~K, respectively.  The real and
imaginary parts of the complex conductivity are fit simultaneously using two Drude
terms in combination with a series of oscillators.
Above $T_N$, the free-carrier response consists of a weak, narrow Drude term, and a
strong, broad Drude term, both of which show only a weak temperature dependence.
Below $T_N$ there is a slight decrease of the plasma frequency but a dramatic drop
in the scattering rate for the narrow Drude term, and for the broad Drude term
there is a significant decrease in the plasma frequency, while the decrease in the
scattering rate, albeit significant, is not as severe.  The small values observed for
the scattering rates for the narrow Drude term for $T\ll{T_N}$ may be related to the
Dirac cone-like dispersion of the electronic bands.
Below $T_N$ new features emerge in the optical conductivity that are associated with
the reconstruction Fermi surface and the gapping of bands at $\Delta_1 \simeq 45-80$~meV,
and $\Delta_2 \simeq 110-210$~meV.
The reduction in the spectral weight associated with the free carriers is captured by
the gap structure, specifically, the spectral weight from the narrow Drude term appears
to be transferred into the low-energy gap feature, while the missing weight from the broad
term shifts to the high-energy gap.
%
%
\end{abstract}
%
%
\maketitle

%
%
%
\section{Introduction}
The discovery of the iron-based superconductors has resulted in an intensive
investigation of this class of materials in the hope of discovering new
compounds with high superconducting critical temperatures ($T_c$'s)
\cite{johnston10,paglione10,canfield10,si16}.
The iron-arsenic materials are characterized by Fe-As sheets separated by
layers of different elements or chemical structures.  One material, BaFe$_2$As$_2$,
is particularly useful as superconductivity can be induced by the application
of pressure \cite{ishikawa09,alireza09,colombier09,yamazaki10}, as well as
electron \cite{sefat08,ni08a,chu09,canfield09}, hole \cite{rotter08b,torikachvili08,chenh09},
or isovalent doping \cite{jiang09,rullier10,thaler10,nakai10}, with $T_c$'s as high as 40~K
in the hole-doped material.
At room temperature, BaFe$_2$As$_2$ is a paramagnetic metal with a tetragonal
structure.  The resistivity in the planes decreases with temperature until it
drops anomalously  as the material undergoes a magnetic transition at $T_N \simeq 138$~K
to a spin-density-wave (SDW)-like antiferromagnetic ground state that is also accompanied
by a structural transition to an orthorhombic phase \cite{rotter08a,wang09,fisher11}.
The resistivity also displays a slight anisotropy close to and below $T_N$,
being slightly larger along the \emph{b} axis than along \emph{a} \cite{chu10};
however, this anisotropy decreases dramatically if the samples are annealed
\cite{ishida13}.
The related materials CaFe$_2$As$_2$ and SrFe$_2$As$_2$ have similar transport
properties due to magnetic and structural transitions that occur at
$T_N\simeq 172$ and 195~K, respectively \cite{tegel08,zhao08,goldman08,tanatar09,
tanatar10,blomberg11}.

The broad interest in this family of materials has resulted in a number of
optical studies \cite{hu08,akrap09,pfuner09,chen10,schafgans11,moon12,nakajima11,
dusza11,charnukha13,nakajima14,wang14}.  Early investigations treated
the free-carrier response using only a single band.  However, a minimal
description of the electronic structure of the iron-arsenic materials
consists of hole and electron pockets at the center and corners of the
Brillouin zone, respectively \cite{singh08,fink09}.  As a result, more recent
studies consider a two band approach (the so-called two-Drude model)
in which the electron and hole pockets are treated as separate electronic
subsystems \cite{wu10a}.  Above $T_N$, this model reveals the presence of
a narrow Drude response that has a strong temperature dependence in
combination with a broad Drude term that is essentially temperature
independent.
Below $T_N$, the optical conductivity undergoes dramatic changes due to the
reconstruction of the Fermi surface \cite{yi09,yang09b,richard10,shimojima10,jensen11,dhaka14,
gofryk14}; however, one of the most detailed optical studies on this family of materials
restricts the two-Drude analysis to $T\gtrsim T_N$ \cite{charnukha13}.

%
%
In this work we have determined the detailed temperature dependence of the
complex optical properties in the \emph{a-b} planes of single crystals of
BaFe$_2$As$_2$, SrFe$_2$As$_2$, and CaFe$_2$As$_2$, above and below $T_N$.
This has allowed us to track the evolution of the electronic properties and
the SDW gap-like features $(T < T_N)$ with the different alkali earth atoms.
%
%
The complex conductivity has been modeled using the two-Drude model.
Above $T_N$, the complex conductivity is described by a strong, broad Drude
response, and a narrow, less intense term, both of which exhibit only a weak
temperature dependence, in combination with a strong interband feature at
about 0.5~eV;  these results are in good agreement with other works.
Below $T_N$, there is a dramatic narrowing of the Drude responses and a
suppression of the low-frequency conductivity as spectral weight is
transferred to two new gap-like features that appear below about 200~meV.
To avoid the difficulties of false convergence typically associated with
the extra degrees of freedom due to these new gap features, we introduce
the constraint that the spectral weight due to the Drude terms above
$T_N$ must be captured by the Drude terms and the two oscillators used
to describe the gap features below $T_N$.  Using this approach, reliable
convergence is achieved and we are able to track the detailed temperature
dependence of the Drude and Lorentz parameters below $T_N$.

We observe the same response in all three materials; namely, just below $T_N$ there
is a slight reduction of the plasma frequency for the narrow term, but a dramatic
decrease of the scattering rate, while for the broad Drude term there is a
significant reduction of both the plasma frequency and the scattering rate.
The missing spectral weight in the narrow Drude term appears to be transferred
to the low-energy gap feature ($\Delta_1 \simeq 45-80$~meV), while the missing
weight from the broad Drude term appears to be transferred to the high-energy
gap ($\Delta_2 \simeq 110-210$~meV).  We also note that the increasing values
for the plasma frequencies and the gap values are scaling roughly with the
electron affinities of the alkali earth atoms.

%
%

%
%
\begin{figure*}[t]
%
%
\includegraphics[width=2.50in]{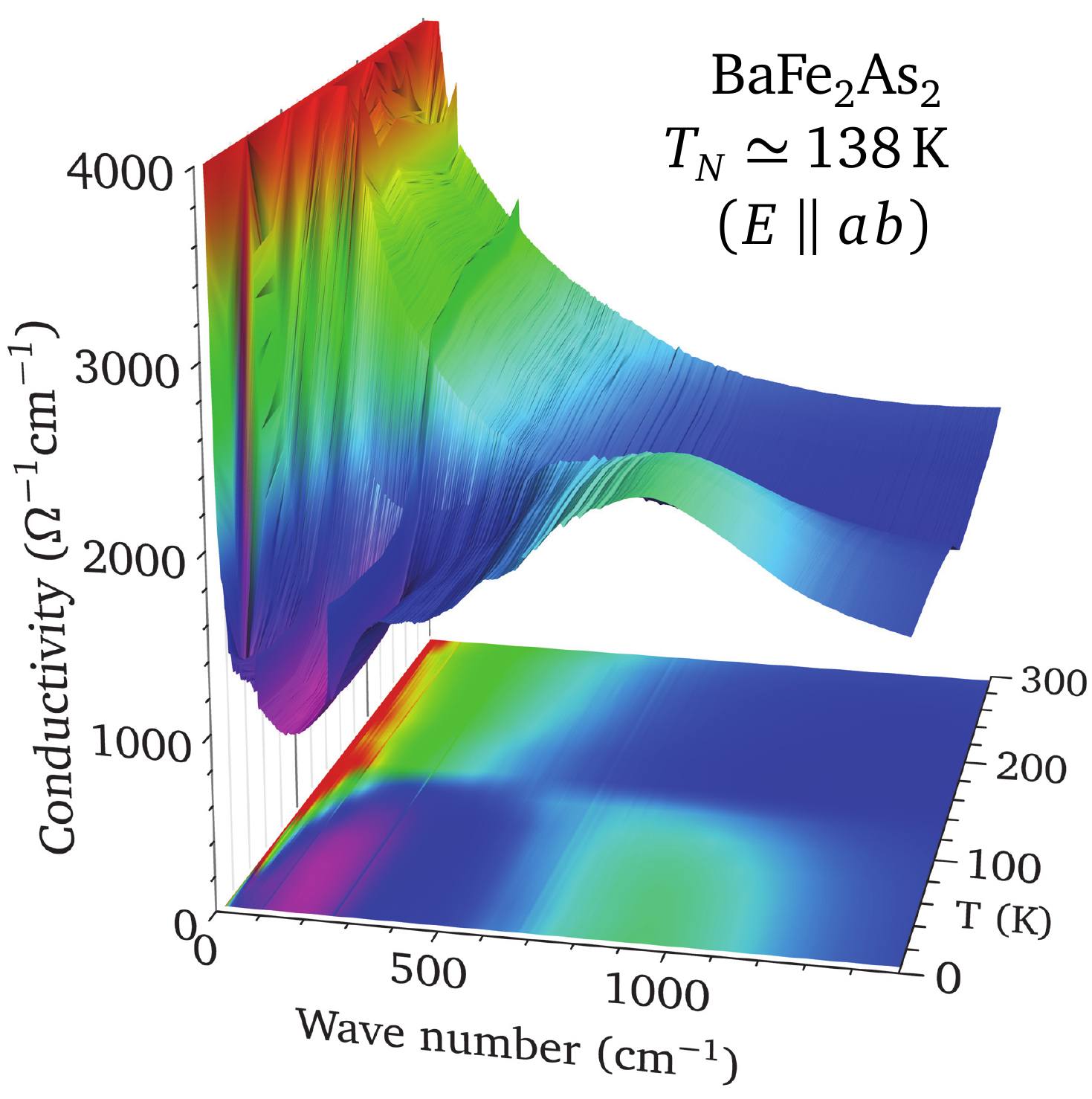} \hfil
\includegraphics[width=2.05in]{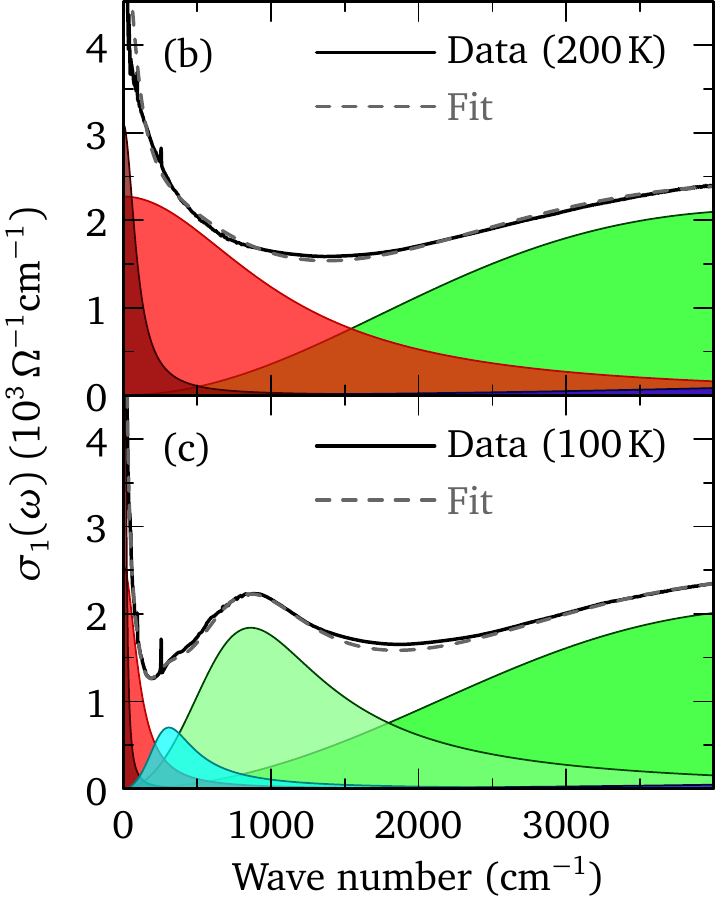} \hfil
\includegraphics[width=2.20in]{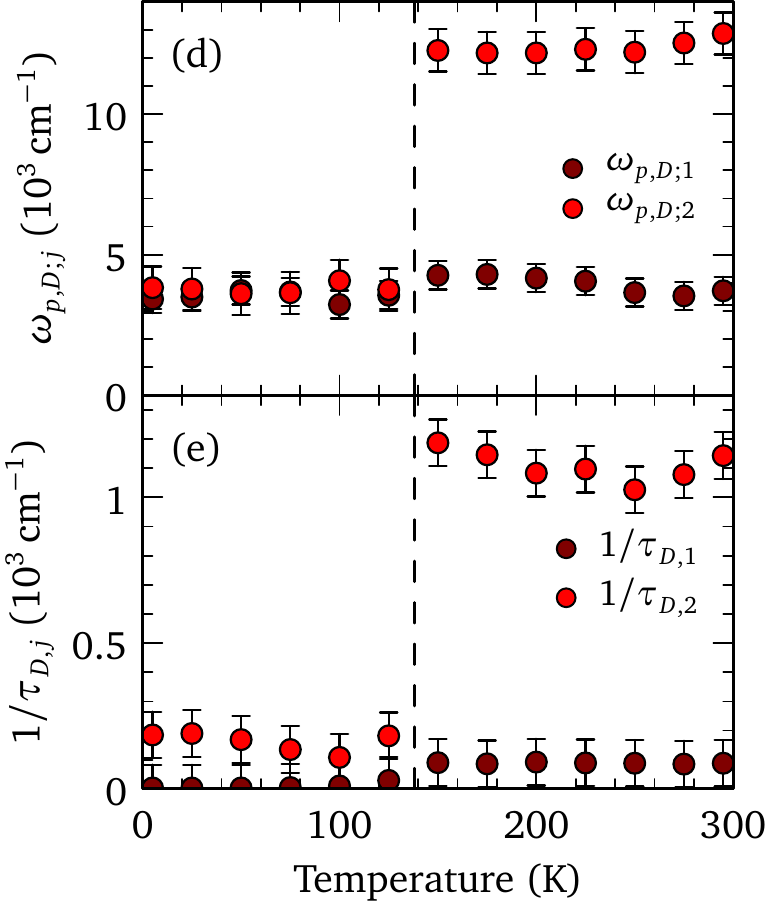}
\caption{(Color online) (a) The temperature dependence of the real part of the optical
conductivity for light polarized in the \emph{a-b} planes of BaFe$_2$As$_2$ above and
below $T_N=138$~K showing the dramatic redistribution of spectral weight below $T_N$.
The results of the fits to the complex conductivity are compared to the real part
of the conductivity at (b) 200~K, and (c) 100~K; the fit is decomposed into the
contributions from the narrow and broad Drude components, as well as several of
the Lorentz oscillators.
The temperature dependence of the (d) plasma frequencies and (e) scattering
rates for the broad and narrow Drude components above and below $T_N$.
}
\label{fig:ba122}
\end{figure*}
%

%
%
\section{Experiment}
Single crystals of $A$Fe$_2$As$_2$ ($A=\,$Ba, Sr, or Ca) were grown using conventional
high-temperature solution growth techniques either out of self flux ($A =\,$Ba) \cite{ni08a},
or out of Sn flux ($A =\,$ Sr, Ca) \cite{yan08,ni08b} and characterized by scattering and
bulk physical measurements.  These crystals have not been annealed.

The reflectance of mm-sized, as-grown crystal faces has been measured at a
near-normal angle of incidence for light polarized in the \emph{a-b} planes
over a wide frequency range from the far infrared ($\simeq 2$~meV) to the
ultraviolet ($\simeq 4 - 5$~eV) for a wide variety of temperatures above and
below $T_N$ using an \emph{in situ} evaporation technique \cite{homes93}.
The complex optical conductivity has been determined from a Kramers-Kronig
analysis of the reflectance \cite{dressel-book}, which requires extrapolations
for $\omega\rightarrow 0,\infty$.  At low frequency, the material is
always metallic, so the Hagen-Rubens form for the reflectance is employed,
$R(\omega)=1-a\sqrt{\omega}$, where $a$ is chosen to match the data at the
lowest-measured frequency point.  Above the highest-measured frequency, the
reflectance is typically assumed to be constant up to $8\times 10^4$~cm$^{-1}$,
above which a free-electron gas asymptotic reflectance extrapolation $R(\omega)
\propto 1/\omega^4$ is assumed \cite{wooten}.

%
%
\section{Results and Discussion}
For simplicity, the multiple hole and electron bands are gathered into single
electron and hole pockets that are treated as two separate electronic subsystems
using the so-called two-Drude model \cite{wu10a} with the complex dielectric
function $\tilde\epsilon=\epsilon_1+i\epsilon_2$,
\begin{equation}
  \tilde\epsilon(\omega) = \epsilon_\infty - \sum_{j=1}^2 {{\omega_{p,D;j}^2}\over{\omega^2+i\omega/\tau_{D,j}}}
    + \sum_k {{\Omega_k^2}\over{\omega_k^2 - \omega^2 - i\omega\gamma_k}},
  \label{eq:eps}
\end{equation}
where $\epsilon_\infty$ is the real part at high frequency.  In the first sum
$\omega_{p,D;j}^2 = 4\pi n_je^2/m^\ast_j$ and $1/\tau_{D,j}$ are the square of the
plasma frequency and scattering rate for the delocalized (Drude) carriers in the $j$th
band, respectively, and $n_j$ and $m^\ast_j$ are the carrier concentration and effective mass.
In the second summation, $\omega_k$, $\gamma_k$ and $\Omega_k$ are the position, width, and
strength of the $k$th vibration or bound excitation.  The complex conductivity is
$\tilde\sigma(\omega) = \sigma_1 +i\sigma_2 = -2\pi i \omega [\tilde\epsilon(\omega) -
\epsilon_\infty ]/Z_0$ (in units of $\Omega^{-1}$cm$^{-1}$); $Z_0\simeq 377$~$\Omega$ is
the impedance of free space.

%
%
\begin{figure}[tbh]
%
%
\includegraphics[width=2.8in]{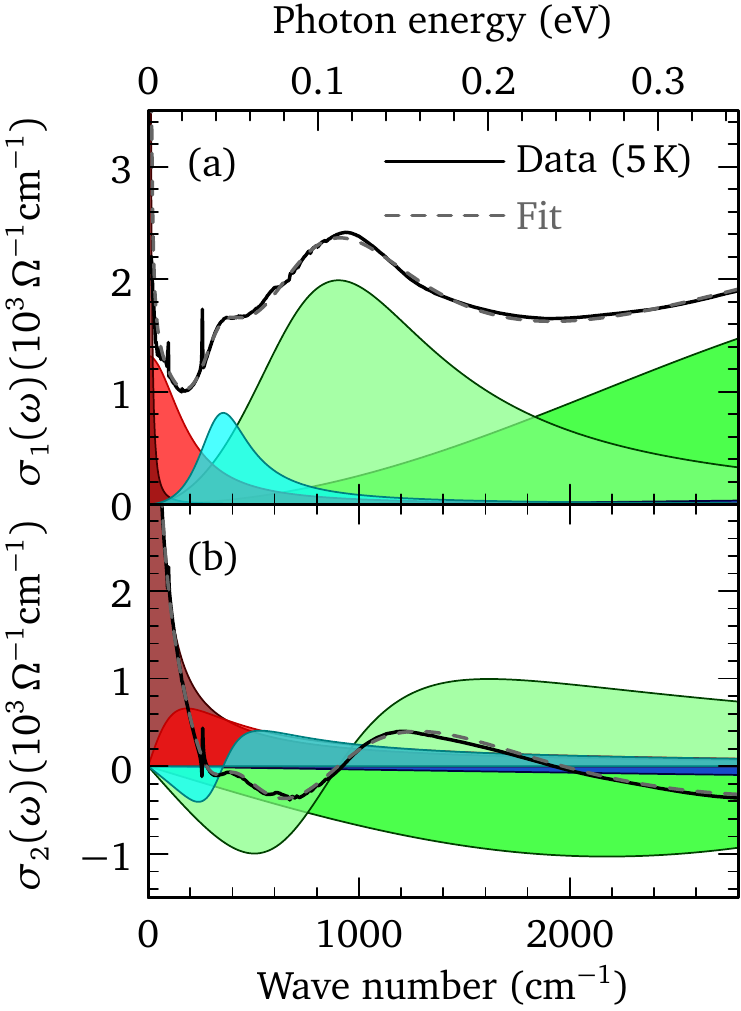}
\caption{(Color online) The comparison of the fit to the complex conductivity
of BaFe$_2$As$_2$ at 5~K for light polarized in the planes showing the contributions
form the Drude and Lorentz components.
(a) The real part of the optical conductivity; note that most of the spectral
weight from the narrow Drude component lies outside of the experimental data.
(b) The imaginary part of the optical conductivity; note that the response of
the narrow Drude component now extends to much higher frequencies, allowing
a precise determination of the scattering rate.
}
\label{fig:drude}
\end{figure}
%

%
%
\subsection{BaFe$_\mathbf{2}$As$_\mathbf{2}$}
The temperature dependence of the real part of the optical conductivity of BaFe$_2$As$_2$ for
light polarized in the \emph{a-b} planes is shown in Fig.~\ref{fig:ba122}(a) in the infrared
region.
Above $T_N$, the conductivity is metallic with a Drude-like free carrier component
that slowly gives way with increasing frequency to a series of interband transitions (the
reflectance and the optical conductivity are shown over a broader energy range in Figs.~S1
and S2 in the Supplemental Material).  Below $T_N$, a narrow free-carrier response is
observed at low frequency and there is a dramatic suppression of the conductivity in
the far-infrared region with a commensurate transfer of spectral weight to the peaks
that emerge at $\simeq 360$ and 900~cm$^{-1}$.  The spectral weight is defined as the
area under the conductivity curve over a given interval,
\begin{equation}
  S(\omega)=\int_0^{\omega} \sigma_1(\omega^\prime)d\omega^\prime .
\end{equation}
These results are in good agreement with other optical studies of this
material \cite{hu08,akrap09,pfuner09,chen10,schafgans11,moon12,nakajima11,dusza11,charnukha13,nakajima14}.

%
%
\begin{table*}[tbh]
\caption{The results of the non-linear least-squares fit of the two-Drude model with Lorentz oscillators to the
complex conductivity of BaFe$_2$As$_2$, SrFe$_2$As$_2$, and CaFe$_2$As$_2$ at all measured temperatures. The
terms $D_1$ and $D_2$ denote the two Drude contributions, while $L_1$ and $L_2$ are the two low-frequency
Lorentzian oscillators; the two new oscillators that appear below $T_N$ are denoted $L_{01}$ and $L_{02}$.
The oscillators above 0.5~eV display relatively little temperature dependence.  The estimated errors for the
location and width are typically 2\% or less, and 5\% or less for the plasma frequency (oscillator strength).
All units are in cm$^{-1}$ unless otherwise indicated.$^a$}
\begin{ruledtabular}
\begin{tabular}{c cc cc ccc ccc ccc ccc}
%
%
  & & & & & & & & & & & & & & & & \\
  \multicolumn{17}{c}{(a) BaFe$_2$As$_2$ ($T_N\simeq 138\,$K)} \\
  & \multicolumn{2}{c}{$D_1$} & \multicolumn{2}{c}{$D_2$} & \multicolumn{3}{c}{$L_{01}$}  & \multicolumn{3}{c}{$L_{02}$}
  & \multicolumn{3}{c}{$L_1$} & \multicolumn{3}{c}{$L_2^a$} \\
 T (K) & $1/\tau_{D,1}$ & $\omega_{p,D;1}$ & $1/\tau_{D,2}$ &  $\omega_{p,D;2}$ &
   $\omega_{01}$ & $\gamma_{01}$ & $\Omega_{01}$ & $\omega_{02}$ & $\gamma_{02}$ & $\Omega_{02}$
   & $\omega_1$ & $\gamma_1$ & $\Omega_1$ & $\omega_2$ & $\gamma_2$ & $\Omega_2$ \\
 \cline{1-5} \cline{6-11} \cline{12-17}
 295 &  88 & 3717 & 1143 & 12870 &     &     &      &     &      &       & 4222 & 8883 & 33260 &  13745 & 19380 & 21710 \\
 275 &  85 & 3535 & 1080 & 12530 &     &     &      &     &      &       & 4278 & 8920 & 32981 &  13863 & 19443 & 21678 \\
 250 &  88 & 3654 & 1026 & 12203 &     &     &      &     &      &       & 4325 & 8772 & 32843 &  14128 & 19691 & 21932 \\
 225 &  89 & 4058 & 1097 & 12306 &     &     &      &     &      &       & 4377 & 8455 & 32753 &  13518 & 19943 & 23146 \\
 200 &  92 & 4168 & 1083 & 12177 &     &     &      &     &      &       & 4438 & 8383 & 32663 &  13778 & 20311 & 23488 \\
 175 &  86 & 4299 & 1146 & 12176 &     &     &      &     &      &       & 4504 & 8410 & 32582 &  13943 & 20681 & 23660 \\
 150 &  90 & 4267 & 1187 & 12265 &     &     &      &     &      &       & 4571 & 8305 & 32483 &  13607 & 21023 & 22294 \\
\hline
 125 &  29 & 3558 &  182 &  3759 & 100 & 354 & 4010 & 752 & 1289 & 11527 & 4515 & 7415 & 30253 &  15906 & 21292 & 21705 \\
 100 &  10 & 3227 &  108 &  4074 & 308 & 382 & 4010 & 863 & 1150 & 11274 & 4572 & 7455 & 30394 &  15334 & 21729 & 21374 \\
  75 & 4.6 & 3674 &  135 &  3648 & 339 & 335 & 3941 & 882 & 1137 & 11521 & 4582 & 7417 & 30413 &  15319 & 22000 & 22168 \\
  50 & 3.2 & 3727 &  169 &  3614 & 350 & 334 & 3902 & 896 & 1140 & 11599 & 4589 & 7457 & 30429 &  15877 & 22293 & 22443 \\
  25 & 3.1 & 3495 &  190 &  3781 & 359 & 305 & 3778 & 904 & 1115 & 11500 & 4596 & 7522 & 30466 &  14199 & 22464 & 21225 \\
   5 & 3.2 & 3424 &  185 &  3828 & 356 & 294 & 3787 & 902 & 1112 & 11527 & 4593 & 7515 & 30476 &  14177 & 22570 & 21292 \\
%
%
  & & & & & & & & & & & & & & & & \\
  & & & & & & & & & & & & & & & & \\
  \multicolumn{17}{c}{(b) SrFe$_2$As$
_2$ ($T_N\simeq 195\,$K)} \\
  & \multicolumn{2}{c}{$D_1$} & \multicolumn{2}{c}{$D_2$} & \multicolumn{3}{c}{$L_{01}$}  & \multicolumn{3}{c}{$L_{02}$}
  & \multicolumn{3}{c}{$L_1$} & \multicolumn{3}{c}{$L_2$} \\
 T (K) & $1/\tau_{D,1}$ & $\omega_{p,D;1}$ & $1/\tau_{D,2}$ &  $\omega_{p,D;2}$ &
   $\omega_{01}$ & $\gamma_{01}$ & $\Omega_{01}$ & $\omega_{02}$ & $\gamma_{02}$ & $\Omega_{02}$
   & $\omega_1$ & $\gamma_1$ & $\Omega_1$ & $\omega_2$ & $\gamma_2$ & $\Omega_2$ \\
 \cline{1-5} \cline{6-11} \cline{12-17}
 295 & 475 & 5205 & 2331 & 17738 &     &     &      &      &      &       & 4760 & 7303 & 26955 &  9718 & 25017 & 36660 \\
 275 & 432 & 5312 & 2324 & 17633 &     &     &      &      &      &       & 4733 & 7277 & 26965 &  9687 & 25095 & 36636 \\
 250 & 381 & 5287 & 2317 & 17673 &     &     &      &      &      &       & 4813 & 7039 & 27052 & 10399 & 25402 & 36889 \\
 225 & 326 & 5254 & 2366 & 17675 &     &     &      &      &      &       & 4885 & 7017 & 27166 & 10553 & 25985 & 37078 \\
 200 & 294 & 5075 & 2354 & 17699 &     &     &      &      &      &       & 4932 & 6895 & 27442 & 11154 & 26393 & 37057 \\
 \hline
 175 &  48 & 2850 &  340 &  6964 & 482 & 371 & 2025 & 1261 & 1983 & 15382 & 5062 & 6879 & 29995 & 13022 & 24306 & 36057 \\
 150 &  41 & 3318 &  308 &  6101 & 480 & 500 & 3186 & 1322 & 1909 & 15317 & 5119 & 6900 & 30219 & 12977 & 25588 & 35411 \\
 125 &  38 & 3543 &  291 &  5586 & 482 & 495 & 3737 & 1376 & 1855 & 15383 & 5054 & 6349 & 27297 & 11197 & 23999 & 38812 \\
 100 &  31 & 3620 &  285 &  5252 & 473 & 483 & 4036 & 1405 & 1841 & 15540 & 5065 & 6348 & 27836 & 11731 & 23994 & 38232 \\
  75 &  26 & 3693 &  332 &  5249 & 482 & 476 & 4042 & 1429 & 1837 & 15577 & 5193 & 6351 & 28240 & 12176 & 24007 & 37803 \\
  50 &  20 & 3720 &  374 &  5274 & 479 & 470 & 4062 & 1445 & 1827 & 15573 & 5198 & 6360 & 28482 & 12424 & 24026 & 37620 \\
  25 &  14 & 3626 &  359 &  5179 & 473 & 465 & 4266 & 1448 & 1819 & 15657 & 5182 & 6372 & 28517 & 12487 & 24056 & 37586 \\
   5 &  13 & 3597 &  333 &  5022 & 469 & 469 & 4144 & 1446 & 1816 & 15645 & 5145 & 6385 & 28685 & 12597 & 24093 & 37510 \\
  & & & & & & & & & & & & & & & & \\
  & & & & & & & & & & & & & & & & \\
  \multicolumn{17}{c}{(c) CaFe$_2$As$_2$ ($T_N\simeq 172\,$K)} \\
  & \multicolumn{2}{c}{$D_1$} & \multicolumn{2}{c}{$D_2$} & \multicolumn{3}{c}{$L_{01}$}  & \multicolumn{3}{c}{$L_{02}$}
  & \multicolumn{3}{c}{$L_1$} & \multicolumn{3}{c}{$L_2$} \\
 T (K) & $1/\tau_{D,1}$ & $\omega_{p,D;1}$ & $1/\tau_{D,2}$ &  $\omega_{p,D;2}$ &
   $\omega_{01}$ & $\gamma_{01}$ & $\Omega_{01}$ & $\omega_{02}$ & $\gamma_{02}$ & $\Omega_{02}$
   & $\omega_1$ & $\gamma_1$ & $\Omega_1$ & $\omega_2$ & $\gamma_2$ & $\Omega_2$ \\
 \cline{1-5} \cline{6-11} \cline{12-17}
 295  & 732 & 8714 & 3219 & 20010 &     &      &      &      &      &       & 5605 &  9466 & 32127 &  12156 & 20080 & 36919 \\
 275  & 646 & 8485 & 3151 & 20040 &     &      &      &      &      &       & 5766 &  9470 & 32864 &  12560 & 20097 & 36119 \\
 250  & 598 & 8879 & 3263 & 20006 &     &      &      &      &      &       & 5828 &  9475 & 33126 &  12801 & 20121 & 35786 \\
 225  & 531 & 8893 & 3278 & 19982 &     &      &      &      &      &       & 5873 &  9470 & 33292 &  12862 & 20159 & 35578 \\
 200  & 491 & 9029 & 3122 & 19332 &     &      &      &      &      &       & 5843 &  9488 & 33925 &  12958 & 20216 & 35345 \\
 175  & 402 & 8524 & 2712 & 18636 &     &      &      &      &      &       & 5736 &  9504 & 34486 &  12912 & 20287 & 35335 \\
 \hline
 150  &  74 & 3752 &  332 &  6723 & 667 &  912 & 6023 & 1414 & 1770 & 14526 & 5111 &  7856 & 32118 &  11432 & 16402 & 36034 \\
 125  &  55 & 4084 &  284 &  5686 & 619 &  919 & 6112 & 1488 & 1765 & 14910 & 5288 &  7835 & 32694 &  11832 & 16330 & 35332 \\
 100  &  44 & 4353 &  304 &  5207 & 625 &  903 & 6129 & 1553 & 1757 & 15146 & 5396 &  7813 & 32914 &  12061 & 16264 & 35032 \\
  75  &  32 & 4428 &  312 &  4872 & 623 &  884 & 6275 & 1601 & 1745 & 15222 & 5452 &  7801 & 33060 &  12188 & 16221 & 34827 \\
  50  &  21 & 4530 &  489 &  5398 & 658 &  854 & 5748 & 1643 & 1733 & 15294 & 5526 &  7792 & 33200 &  12325 & 16200 & 34615 \\
  25  &  13 & 4414 &  497 &  5428 & 661 &  828 & 5703 & 1668 & 1722 & 15313 & 5532 &  7788 & 33223 &  12343 & 16186 & 34587 \\
   5  &  11 & 4395 &  518 &  5498 & 661 &  823 & 5651 & 1673 & 1719 & 15310 & 5535 &  7786 & 33253 &  12362 & 16175 & 34552 \\
\end{tabular}
\end{ruledtabular}
\footnotetext[1] {A convenient conversion is 1~eV = 8065.5~cm$^{-1}$.}
\label{tab:fits}
\end{table*}

The real and imaginary parts of the complex optical conductivity have been fit
simultaneously with the Drude-Lorentz model using a non-linear least-squares technique.
For $T>T_N$, the data was initially fit using a single Drude component and a series
of Lorentzian oscillators to reproduce the interband transitions; however, even with
extremely overdamped oscillators, the returned fits were of poor quality.  A low-frequency
Lorentz oscillator was introduced to improve the quality of the fit; however, the best
result was obtained when the frequency of this oscillator went to zero and a
second Drude component was recovered.  Given the multiband nature of this material,
this is a natural result.  While there are as many as five Fermi surfaces, we
have adopted a minimal description that only considers two sets of carriers; this
approach has the advantage of producing good fits, as well as keeping the total
number of parameters (degrees of freedom) relatively low, an approach that typically
results in fits with good convergence.

%
%
The fit to the real part of the optical conductivity using the two-Drude model above
$T_N$ at 200~K is shown in Fig.~\ref{fig:ba122}(b) where the individual Drude and
Lorentz contributions are shown.  The data is reproduced quite well by a narrow and a
broad Drude term, with $\omega_{p,D;1}\simeq 4200$~cm$^{-1}$ and $1/\tau_{D,1}\simeq
90$~cm$^{-1}$, and $\omega_{p,D;2} \simeq 12\,200$~cm$^{-1}$ and $1/\tau_{D,2} \simeq
1100$~cm$^{-1}$, respectively, as well as bound excitations at $\omega_1\simeq 4450$~cm$^{-1}$
and $\omega_2\simeq 13\,800$~cm$^{-1}$, which are attributed to interband transitions
\cite{valenzuela13,calderon14}; the parameters for the Drude components, as well
as the first two Lorentzian oscillators, are listed in Table~\ref{tab:fits}(a).  The
observation of narrow and broad terms in the two-Drude analysis is in agreement with other
optical studies \cite{nakajima11,dusza11,charnukha13,nakajima14}, and appears to be a general
result for most of the iron-based materials that incorporate iron-arsenic sheets.  If
the plasma frequencies for the free carriers are added in quadrature [Eq.~(\ref{eq:wp})],
then the derived plasma frequency $\omega_{p} \simeq 12\,900$~cm$^{-1}$ is in good
agreement with a previous study on this material that considered only a single Drude
component \cite{hu08}.

%
%
This approach has also been applied to the optical conductivity below $T_N$, where two
new oscillators at $\omega_{01}\simeq 360$~cm$^{-1}$ and $\omega_{02}\simeq 900$~cm$^{-1}$
have been included to reproduce the broad peaks that emerge in the optical conductivity at
low temperature.  However, the six extra degrees of freedom introduced by the two new
oscillators can lead to a false convergence with non-unique solutions.  By observing
that the loss of spectral weight of the free carriers below $T_N$ appears to be captured
by the two new oscillators, we can introduce the constraint that the redistribution
of the spectral weight among these fitted parameters must be a conserved quantity.  This
is equivalent to the statement that below $\simeq 5000$~cm$^{-1}$ the spectral weight
is roughly constant.  Thus, above and below $T_N$,
\begin{equation}
 \omega_p\simeq \left\{\!\! \begin{array}{ll}
  \sqrt{\omega_{p,D;1}^2+\omega_{p,D;2}^2} & ( T>T_N)  \\
  &  \\
  \sqrt{\omega_{p,D;1}^2+\omega_{p,D;2}^2+\Omega_{01}^2+\Omega_{02}^2} & (T<T_N)
   \end{array}
   \right.
\label{eq:wp}
\end{equation}
where $\Omega_{01}$ and $\Omega_{02}$ are the strengths of the two new oscillators;
this constraint leads unique solutions (this is explored in more detail in the Supplemental
Material).  The results of the fit to the data at 100~K using this approach are shown in
Fig.~\ref{fig:ba122}(c); the Drude components have narrowed and lost spectral weight,
which has shifted to the two new oscillators; the excitation $\omega_1$ has shifted
upwards slightly to $\simeq 4600$~cm$^{-1}$, but otherwise shows relatively little
temperature dependence below $T_N$ [Table~\ref{tab:fits}(a)].  These results are in
good agreement with the optical conductivity of a detwinned sample at 5~K \cite{nakajima11},
which show that the broad Drude component is only observed along the \emph{a} axis,
while the narrow Drude component observed in both the \emph{a} and \emph{b} directions.
The fact that the narrow Drude component is isotropic suggests that it is not related
to the magnetic order in this material.

%
%
\begin{figure*}[thb]
%
%
\includegraphics[width=2.50in]{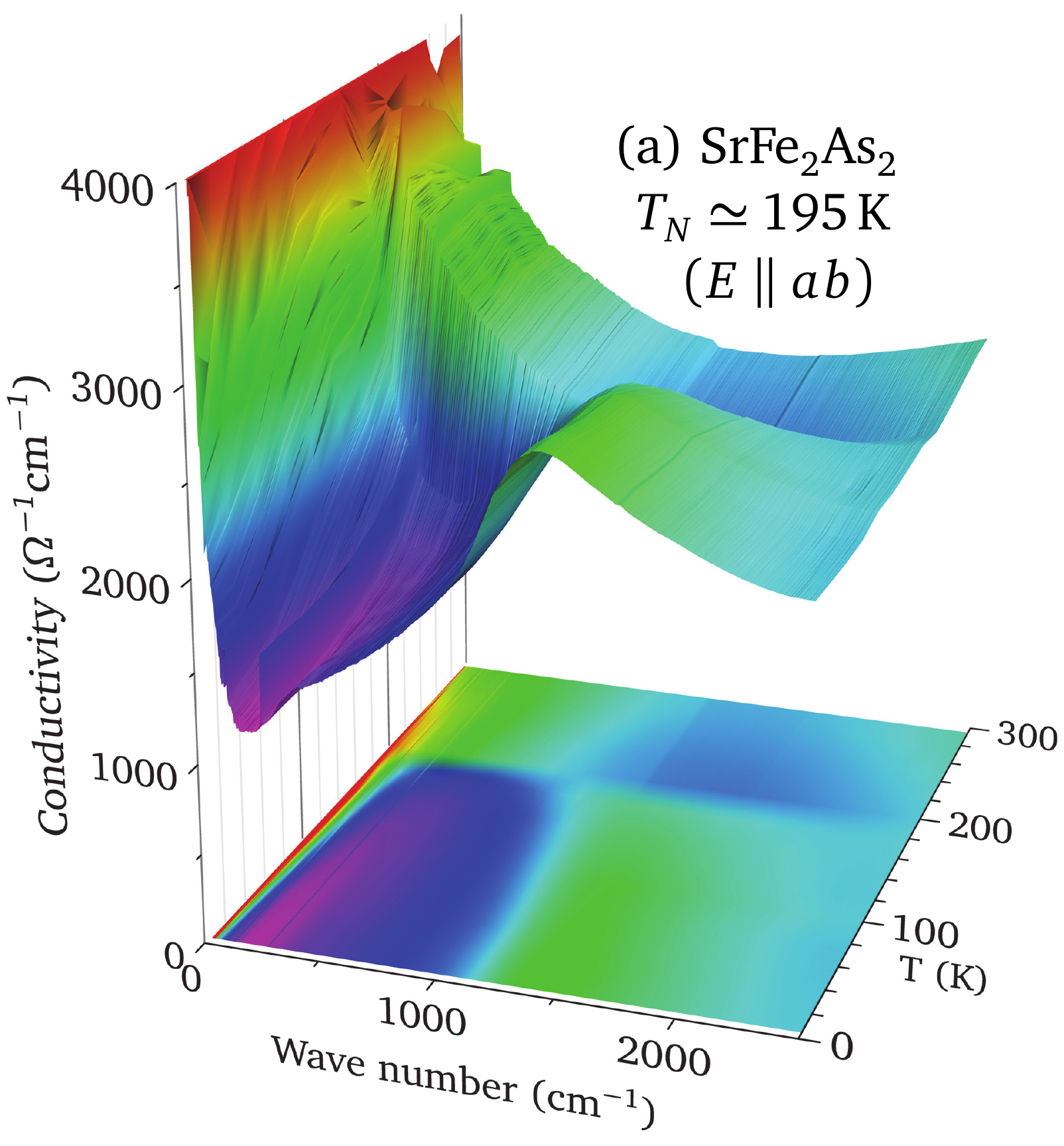} \hfil
\includegraphics[width=2.05in]{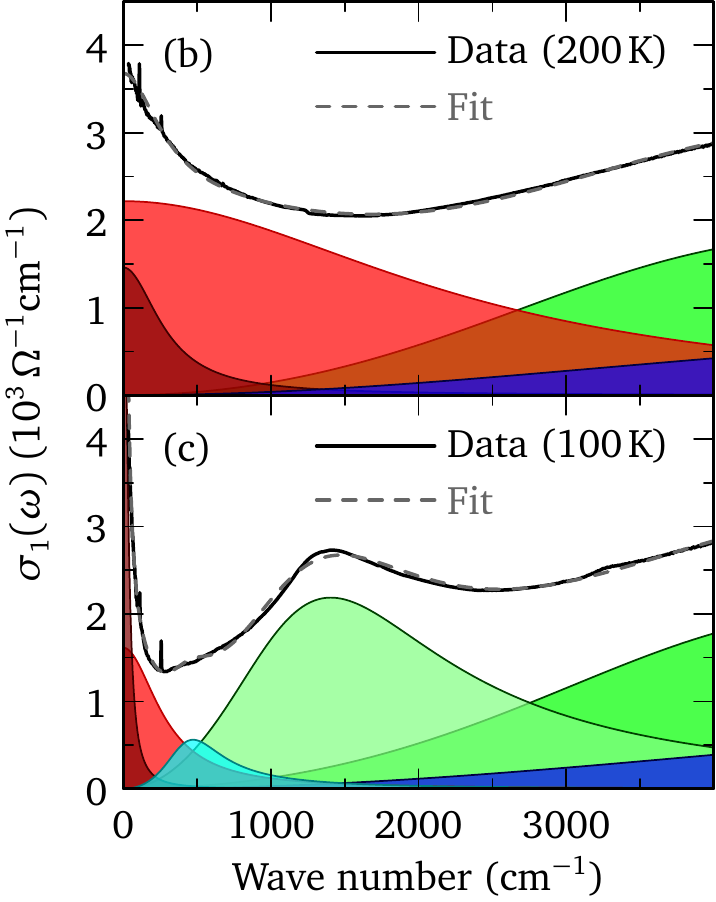} \hfil
\includegraphics[width=2.20in]{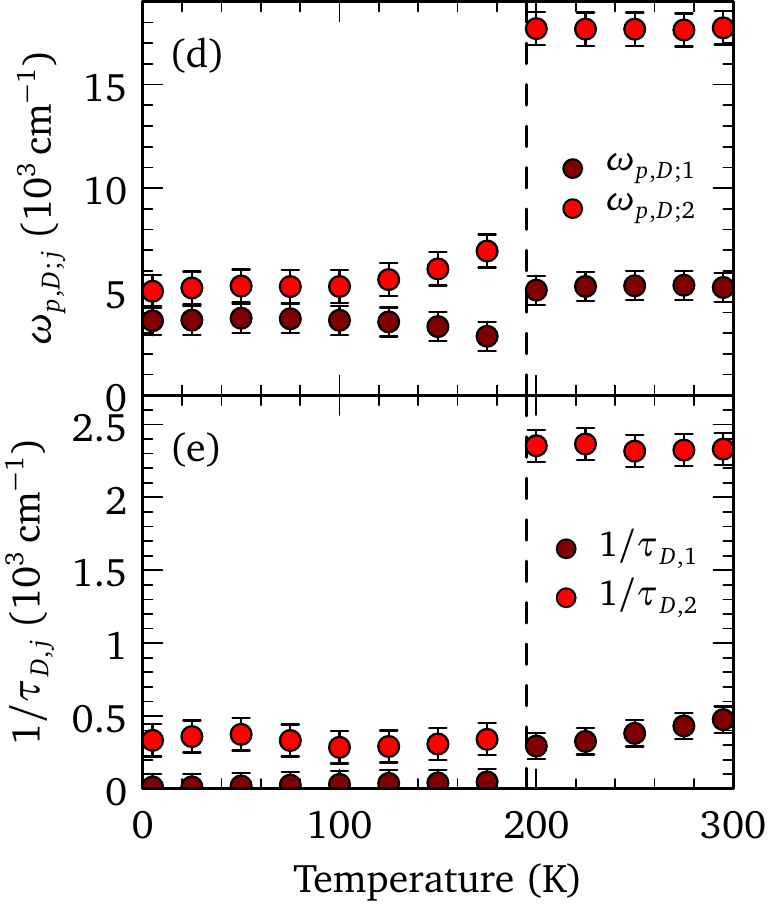}
\caption{(Color online) (a) The temperature dependence of the real part of the optical
conductivity for light polarized in the \emph{a-b} planes of SrFe$_2$As$_2$ above and
below $T_N=195$~K showing the partial gapping of the Fermi surface and the transfer of spectral weight.
The fitted individual contributions of the two-Drude model and Lorentz oscillators
compared to the real part of the optical conductivity at (b) 200~K and (c) 100~K.
The temperature dependence of the (d) plasma frequencies and (e) scattering
rates for the two-Drude model above and below $T_N$.
}
\label{fig:sr122}
\end{figure*}

%
%
The temperature dependence of the plasma frequencies and scattering rates for the two Drude
components are shown in Figs.~\ref{fig:ba122}(d) and \ref{fig:ba122}(e), respectively; above
$T_N$ the parameters are essentially temperature independent, which is not surprising given
the weak temperature dependence of the optical conductivity.  Below $T_N$,  the
plasma frequency for the broad Drude component decreases dramatically from $\omega_{p,D;2}
\simeq 12\,300 \rightarrow 3800$~cm$^{-1}$, while the plasma frequency for the narrow
component decreases only slightly,  $\omega_{p,D;1}\simeq 4200 \rightarrow  3400$~cm$^{-1}$.
The total carrier concentration is observed to decrease from $\omega_{p,D}\simeq 12\,900 \rightarrow
5100$~cm$^{-1}$.  This roughly 85\% decrease in the number of free carriers is in agreement with
previous estimates \cite{hu08,akrap09}.
The values for the plasma frequencies of the narrow and broad Drude components, and the
intensities of the two new oscillators obey the constraint in Eq.~(\ref{eq:wp}), with
$\omega_p\simeq 13\,200\pm 500$~cm$^{-1}$  ($T\ll{T_N})$.  A closer examination of the values
returned from the fits reveals that the missing spectral weight from the narrow Drude
component appears to be captured by $\omega_{01}$,
\begin{equation}
\omega_{p,D;1}^2(T\gtrsim T_N) \simeq \omega_{p,D;1}^2(T\ll{T_N})+\Omega_{01}^2
\end{equation}
and likewise the loss of spectral weight from the broad Drude component appears to be
captured by $\omega_{02}$,
\begin{equation}
\omega_{p,D;2}^2(T\gtrsim T_N) \simeq \omega_{p,D;2}^2(T\ll{T_N})+\Omega_{02}^2.
\end{equation}
%
%
Allowing that the Fermi surface reconstruction below $T_N$ results in the partial gapping
of these two pockets, it is reasonable to associate $\omega_{01}$ and $\omega_{02}$ with
gap-like features in the optical conductivity that gap-like features that appear to be
more or less isotropic in the \emph{a-b} planes \cite{nakajima11,yin11a}.
The average optical gap for the narrow Drude band is therefore estimated to be $\Delta_1 \simeq 44$~meV,
while for the broad Drude band it is $\Delta_2 \simeq 112$~meV.  These estimates are in good
agreement with the peaks observed in the Raman response \cite{chauviere11} and the optical
conductivity \cite{hu08,charnukha13}, as well as the values for $\Delta_1$ and $\Delta_2$
determined from them (in this work the average value for the gap is always associated with
the peak in the conductivity).

%
%
The scattering rate for the broad Drude term drops abruptly below $T_N$, $1/\tau_{D,2}
\simeq 1200 \rightarrow 190$~cm$^{-1}$, while the scattering rate for the narrow Drude
component drops by over an order of magnitude, $1/\tau_{D,1} \simeq 90 \rightarrow
3$~cm$^{-1}$.  As Fig.~\ref{fig:ba122}(c) demonstrates, it is rather difficult to
determine small values of $1/\tau_D$ from fits to only the real part of the Drude
optical conductivity,
\begin{equation}
  \sigma_{1,D}(\omega)=\frac{\sigma_0}{1+\omega^2\tau_D^2},
\end{equation}
which has the form of a Lorentzian centered at zero frequency with a full width at half
maximum of $1/\tau_D$ and $\sigma_0= 2\pi \omega_{p,D}^2 \tau_D/Z_0$.  This difficulty is
further illustrated in the fit to the data at 5~K in Fig.~\ref{fig:drude}(a), where most of
the spectral weight of the narrow Drude component lies below 100~cm$^{-1}$; if the fit
was restricted to only $\sigma_1(\omega)$, the narrow scattering rate would be nearly impossible
to determine with any degree of confidence.  However, in our analysis the real and imaginary
parts of the optical conductivity are fit simultaneously.  The imaginary part of the Drude
conductivity,
\begin{equation}
  \sigma_{2,D}(\omega)=\frac{\sigma_0\,\omega\tau_D}{1+\omega^2\tau_D^2},
\end{equation}
is considerably broader than the real part, as Fig.~\ref{fig:drude}(b) indicates,
allowing values of $1/\tau_D \lesssim 10$~cm$^{-1}$ to be fit reliably.  Thus, despite
the loss of free carriers below $T_N$, the decrease in the scattering rate for
the narrow Drude component is responsible for the increasingly metallic behavior at low
temperature \cite{hu08}.

%
%
\subsection{SrFe$_\mathbf{2}$As$_\mathbf{2}$}
The temperature dependence of the real part of the optical conductivity for SrFe$_2$As$_2$
($T_N\simeq 195$~K) for light polarized in the \emph{a-b} planes is shown in Fig.~\ref{fig:sr122}(a)
in the infrared region.  The overall temperature dependence is quite similar to that of
BaFe$_2$As$_2$.  Above $T_N$, the conductivity is metallic and displays little temperature
dependence, while below $T_N$ the dramatic narrowing of free carrier contribution and
decrease in the low-frequency conductivity leads to the redistribution of spectral weight
over a much larger energy scale with a prominent peak appearing at $\simeq 1400$~cm$^{-1}$
(the reflectance and the optical conductivity are shown over a broader energy range in Figs.~S4
and S5 in the Supplemental Material); however, this feature was observed at a much lower energy,
$\simeq 900$~cm$^{-1}$, in BaFe$_2$As$_2$.

%
%
The optical conductivity has been fit using the two-Drude model.  The result for the fit just
above $T_N$ at 200~K is shown in Fig.~\ref{fig:sr122}(b) where it is decomposed into the
individual contributions from the Drude and Lorentz components.  The free-carrier response
is reproduced using a narrow and a broad Drude component, with $\omega_{p,D;1}\simeq
5100$~cm$^{-1}$ and $1/\tau_{D,1}\simeq 300$~cm$^{-1}$, and $\omega_{p,D;2} \simeq
17\,700$~cm$^{-1}$ and $1/\tau_{D,2} \simeq 2360$~cm$^{-1}$, respectively, and a series
of oscillators at $\omega_1\simeq 5000$~cm$^{-1}$ and $\omega_2\simeq 11\,200$~cm$^{-1}$
[Table~\ref{tab:fits}(b)].  The plasma frequencies for the Drude components are both
larger than what was observed in BaFe$_2$As$_2$, with the combined value for
$\omega_{p}\simeq 18\,400\pm 600$~cm$^{-1}$. These results are consistent with other
optical studies of this material \cite{hu08,charnukha13}.

%
%
The fits to the complex conductivity for $T<T_N$ are again performed using the constraint
in Eq.~(\ref{eq:wp}).  The results of the fit at 100~K, shown for the real part of the
optical conductivity in Fig.~\ref{fig:sr122}(c), indicate that both Drude components
decrease in strength and narrow below $T_N$ at the same time that spectral weight is
transferred into two new Lorentz oscillators at $\omega_{01}\simeq 470$~cm$^{-1}$ and
$\omega_2\simeq 1450$~cm$^{-1}$ [Table~\ref{tab:fits}(b)].  The detailed temperature
dependence of the plasma frequencies and scattering rates are shown in Figs.~\ref{fig:sr122}(d)
and \ref{fig:sr122}(e), respectively.
As previously noted, for $T>T_N$, the plasma frequencies display little or no
temperature dependence; however, for $T<T_N$, $\omega_{p,D;1} \simeq 5100 \rightarrow
3600$~cm$^{-1}$ for the narrow Drude component, while a much larger decrease,
$\omega_{p,D;2} \simeq 17\,700 \rightarrow 5000$~cm$^{-1}$, is observed for the broad Drude
component.  The scattering rate for the narrow Drude component decreases significantly,
$1/\tau_{D,1} \simeq 300$~cm$^{-1}$ just above $T_N$ to $\simeq 13$~cm$^{-1}$ at low
temperature.  The decrease in the scattering rate for the broad Drude component,
$1/\tau_{D,2}\simeq 2350 \rightarrow 330$~cm$^{-1}$, while significant, is less dramatic.
The total carrier concentration is observed to decrease from $\omega_{p,D} \simeq
18\,400 \rightarrow 6200$~cm$^{-1}$; this $\approx 89$\% decrease in the number of free
carriers for $T\ll T_N$ is somewhat larger than what was observed in BaFe$_2$As$_2$ \cite{hu08}.
The values for the plasma frequencies of the narrow and broad Drude components, and the
intensities of the two new oscillators sum to $\omega_p \simeq 17\,320\pm 600$~cm$^{-1}$
($T\ll{T_N}$), indicating that the spectral weight from the narrow and broad Drude components
has been almost entirely transferred into the two new oscillators, $\omega_{01}$ and $\omega_{02}$.
%
%
The estimates for the optical gap energies are $\Delta_1 \simeq 58$~meV and
$\Delta_2 \simeq 180$~meV, respectively.  The value for the large gap is in good
agreement with Raman results \cite{yang14}; however, both of these values are
somewhat larger than previous optical estimates \cite{charnukha13}.  A possible
source of uncertainty is that the gap features in this material are much broader than
in BaFe$_2$As$_2$, making them more difficult to fit unless controls such as the
conservation of spectral weight described in Eq.(\ref{eq:wp}) are introduced.

%
%
\begin{figure*}[tbh]
%
%
\includegraphics[width=2.50in]{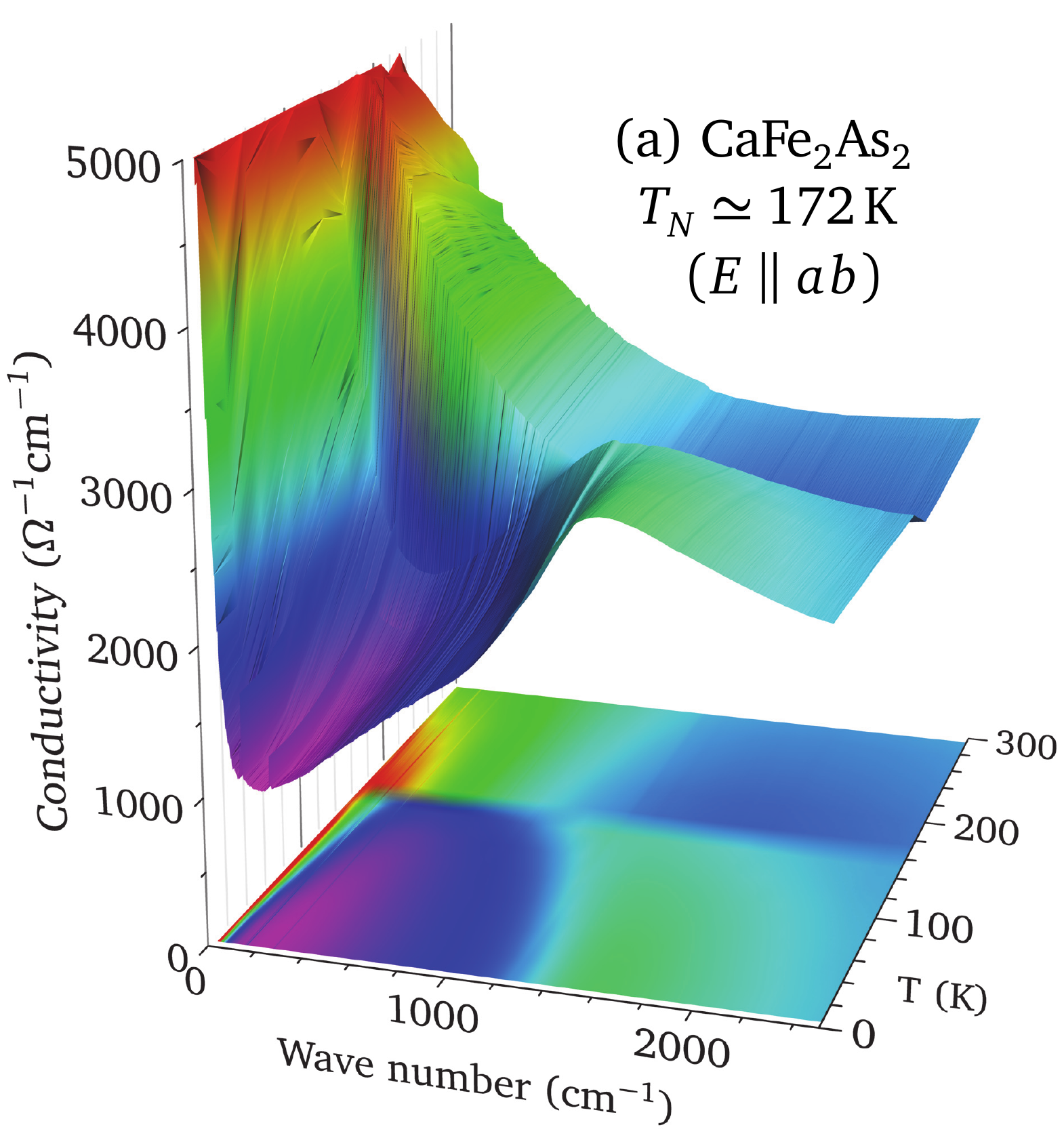} \hfil
\includegraphics[width=2.05in]{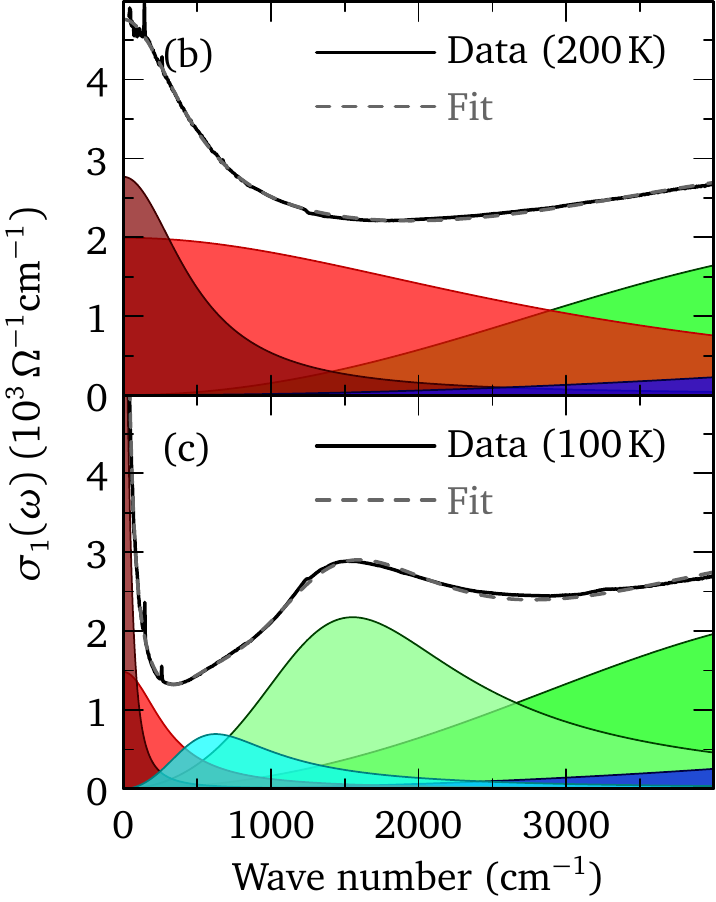} \hfil
\includegraphics[width=2.20in]{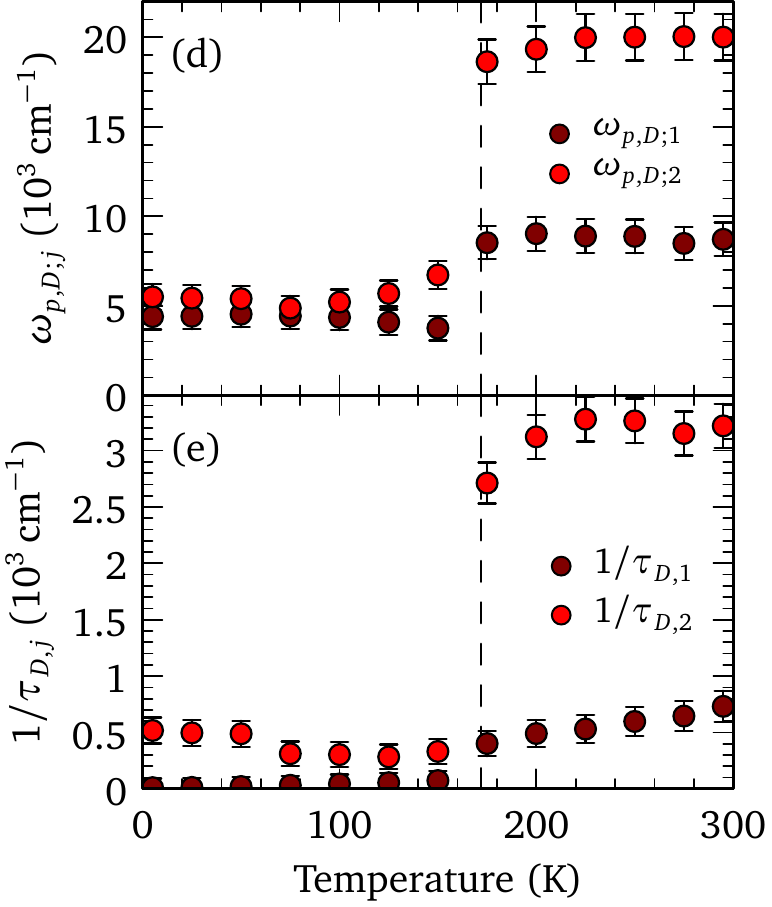}
\caption{(Color online) (a) The temperature dependence of the real part of the optical
conductivity for light polarized in the \emph{a-b} planes of CaFe$_2$As$_2$ above and
below $T_N$ showing the partial gapping of the Fermi surface and the transfer of spectral weight.
The fitted individual contributions of the two-Drude model and Lorentz oscillators
compared to the real part of the optical conductivity at (b) 200~K and (c) 100~K.
The temperature dependence of the (d) plasma frequencies and (e) scattering
rates for  the two-Drude model above and below $T_N$.
}
\label{fig:ca122}
\end{figure*}

%
%
\subsection{CaFe$_\mathbf{2}$As$_\mathbf{2}$}
The temperature dependence of the real part of the optical conductivity for
CaFe$_2$As$_2$ ($T_N \simeq 172$~K) with light polarized in the \emph{a-b} planes is
shown in Fig.~\ref{fig:ca122}(a) in the infrared region.  Unlike the previous
two materials, the conductivity displays a relatively large temperature dependence
above $T_N$ (the reflectance and the optical conductivity are shown over a broader
energy range in Figs.~S6 and S7 in the Supplemental Material); this is entirely due
to the strong temperature dependence of the narrow Drude component and $1/\tau_{D,1}$,
[Table~\ref{tab:fits}(c)].  Below $T_N$, there is once again the characteristic
narrowing of the free-carrier response coupled with the dramatic suppression of
the low-frequency conductivity and the transfer of spectral weight into the peak
that emerges at $\simeq 1720$~cm$^{-1}$.

%
%
The result of the fit to the complex conductivity above $T_N$ at 200~K using
the two-Drude model is compared to the real part in Fig.~\ref{fig:ca122}(b), where
it is decomposed into its individual contributions.  As in previous cases, the
free-carrier response is described quite well by a narrow and a broad Drude
component of $\omega_{p,D;1}\simeq 8900$~cm$^{-1}$ and $1/\tau_{D,1}\simeq 490$~cm$^{-1}$,
and $\omega_{p,D;2}\simeq 19\,300$~cm$^{-1}$ and  $1/\tau_{D,2}\simeq 3100$~cm$^{-1}$,
and a series of oscillators at $\omega_1\simeq 5700$~cm$^{-1}$ and $\omega_2\simeq
12\,800$~cm$^{-1}$ [Table~\ref{tab:fits}(c)].
The resulting value of $\omega_{p}\simeq 21\,200\pm 1200$~cm$^{-1}$ is larger than
was observed in either of the other two materials, in agreement with a previous
work \cite{charnukha13}.  Above $T_N$, the scattering rate for the broad Drude component
has a weak temperature dependence, while the narrow component has a strong temperature
dependence, decreasing from $1/\tau_{D,1} \simeq 730$~cm$^{-1}$ at room temperature to
$\simeq 400$~cm$^{-1}$ just above $T_N$ [Table~\ref{tab:fits}(c)].

%
%
Below $T_N$ the two-Drude model is fit to the complex conductivity using the
previously described technique.  The results of the fit at 100~K, shown for
the real part of the optical conductivity in Fig.~\ref{fig:ca122}(c), once
again show that both Drude components decrease in strength and narrow at the
same time that spectral weight is transferred into two new Lorentz oscillators
at $\omega_{01}\simeq 660$~cm$^{-1}$ and $\omega_{02}\simeq 1670$~cm$^{-1}$
[Table~\ref{tab:fits}(c)].  The detailed temperature dependence of the plasma
frequencies and scattering rates are shown in Figs.~\ref{fig:ca122}(d) and
\ref{fig:ca122}(e), respectively.
While the plasma frequencies display little or no temperature dependence for $T>T_N$,
below $T_N$ the narrow Drude component decreases somewhat, $\omega_{p,D;1} \simeq 8520
\rightarrow 4400$~cm$^{-1}$, while a more dramatic decrease, $\omega_{p,D;2} \simeq
18\,640 \rightarrow 5500$~cm$^{-1}$, is observed for the broad Drude component.
The scattering rate for the narrow Drude component decreases dramatically,
$1/\tau_{D,1} \simeq 400$~cm$^{-1}$ just above $T_N$ to $\simeq 11$~cm$^{-1}$ at low
temperature; the decrease in the scattering rate for the broad Drude component,
$1/\tau_{D,2}\simeq 2700 \rightarrow 520$~cm$^{-1}$, while not as dramatic, is
still significant.
The total carrier concentration is observed to decrease from $\omega_{p,D} \simeq
21\,200 \rightarrow 7040$~cm$^{-1}$; this $\approx 89$\% decrease in the total number
of free carriers for $T\ll T_N$ is similar to what was observed in SrFe$_2$As$_2$.

%
%
\begin{table*}[t]
\caption{The results for the constrained two-Drude fits to the complex conductivity in
$A$Fe$_2$As$_2$ for $A=$Ba ($T_N\simeq 138$~K), Sr ($T_N\simeq 195$~K), and Ca
($T_N\simeq 172$~K), for $T\gtrsim{T_N}$, and $T\ll{T_N}$.  The peaks of the two new
Lorentz oscillators that appear in the infrared region below $T_N$, $\omega_{01}$ and
$\omega_{02}$, are associated with the formation SDW-like gaps, $\Delta_1$ and
$\Delta_2$, respectively.  All of these features scale roughly with the Pauling electronegativity,
$\chi_{\rm P}$.  All units are in cm$^{-1}$, except for $\chi_{\rm P}$ and the last two columns.}
\begin{ruledtabular}
\begin{tabular}{cc c cccc c cccc c ccc ccc cc}
     &                & & \multicolumn{4}{c}{$T\gtrsim{T_N}$} & & \multicolumn{13}{c}{$T\ll{T_N}$} \\
 $A$ & $\chi_{\rm P}$ & & $\omega_{p,D;1}$ & $1/\tau_{D,1}$ & $\omega_{p,D;2}$ & $1/\tau_{D,2}$ &
     & $\omega_{p,D;1}$ & $1/\tau_{D,1}$ & $\omega_{p,D;2}$ & $1/\tau_{D,2}$ &
     & $\omega_{01}$ & $\gamma_{01}$ & $\Omega_{01}$ & $\omega_{02}$ & $\gamma_{02}$ & $\Omega_{02}$
     & $\Delta_1/{k_{\rm B}T_N}$ & $\Delta_2/{k_{\rm B}T_N}$ \\
\cline{1-2}  \cline{4-7} \cline{9-21}
%
%
Ba & 0.89 & & 4210 &  90 & 12300 & 1190 & & 3420 &  3 & 3830 & 190 & & 360 & 290 & 3790 &  900 & 1110 & 11530 & 3.8 & 9.4 \\
Sr & 0.95 & & 5070 & 290 & 17700 & 2350 & & 3600 & 13 & 5020 & 330 & & 470 & 470 & 4140 & 1450 & 1820 & 15650 & 3.5 & 10.7 \\
Ca & 1.00 & & 8520 & 400 & 18640 & 2710 & & 4400 & 11 & 5500 & 518 & & 660 & 820 & 5650 & 1670 & 1720 & 15310 & 5.5 & 14 \\
\end{tabular}
\end{ruledtabular}
%
\label{tab:gaps}
\end{table*}
%

%
%
The values for the plasma frequencies of the narrow and broad Drude components, and the
intensities of the two new oscillators sum to $\omega_p \simeq 16\,540$~cm$^{-1}$ ($T\ll{T_N}$),
indicating that a significant portion of the spectral weight has been shifted into the
two new oscillators, $\omega_{01}$ and $\omega_{02}$, leading to estimates for the
optical gap energies of $\simeq 82$~meV and $\simeq 207$~meV, respectively.
Both of these values are somewhat larger than previous optical estimates \cite{charnukha13},
and the value for the large gap is larger than the strong feature observed in the Raman
response \cite{zhang16}.  However, as in the case of SrFe$_2$As$_2$, the gap features
are rather broad, and in the absence of controls, are difficult to fit reliably.

\subsection{Common features}
Common to all these material is the result that above $T_N$ the optical conductivity
is reproduced by a strong, broad Drude response that shows little temperature
dependence, and a weaker, narrower Drude response where the scattering rate
displays a slight temperature dependence.  For the progression $A$Fe$_2$As$_2$, for
$A=\,$Ba, Sr and Ca, the plasma frequencies for both the narrow and broad Drude
components, the scattering rates, and the gap-like features are all increasing
(although we would remark that $1/\tau_{D,1}$ and $1/\tau_{D,2}$ are substantially
lower in BaFe$_2$As$_2$ than in either of the other two materials).
The trends in the electronic properties appear to follow the electronegativity
(electron affinity) of the alkali earth atoms (Table~\ref{tab:gaps}).  The Pauling
electronegativities of Ba, Sr, and Ca are $\chi_{\rm P}= 0.89$, 0.95 and 1.00,
respectively.  The increasing electron affinity also leads to a decrease in the
covalent radius of 1.98, 1.92, and 1.74~\AA , which is also connected to the decreasing
\emph{c}-axis lattice parameter of 13, 12.4 and 11.7~\AA\ for the Ba \cite{rotter08a},
Sr \cite{tegel08}, and Ca \cite{wu08} materials, respectively. Thus, while the
electronegativities of the alkali earth atoms are a useful guide for establishing
trends in these materials, we would caution that the electronic structure, especially
below $T_N$, is quite complicated, and that detailed structural properties should
also be taken into consideration.

%
%
In each material, below $T_N$, despite the appearance of new structure in the optical
conductivity that signals the reconstruction and partial gapping of the Fermi surface,
the two-Drude model continues to reproduce the free-carrier response quite well.
For the broad Drude component, both the plasma frequency and the scattering rate undergo
a significant reduction below $T_N$, while for the narrow Drude component the plasma
frequency decreases only slightly, while the scattering rate decreases by over an
order of magnitude.  More specifically, $1/\tau_{D,1}$ has a strong temperature
dependence below $T_N$, while $1/\tau_{D,2}$ undergoes an abrupt drop just below $T_N$,
below which it remains relatively constant (Table~\ref{tab:fits}).
The dramatic collapse of the scattering rate for the narrow Drude component for $T\ll{T_N}$
is reminiscent of what is observed in other materials where a Fermi surface
reconstruction leads to large portions of the Fermi surface being removed, with a
concomitant loss of free carriers; in some special cases this is referred to as a
``nodal metal'' \cite{ando01,lee05,homes13}, but more generally it describes any
semimetal with a very small Fermi surface.  In multiband materials, it is not uncommon
for one (or more) of the scattering rates to be quite small, typically on the order
of several cm$^{-1}$ \cite{homes15a,homes15b}.  The very small values of $1/\tau_{D,1}$ at
low temperatures is similar to what is seen in some Weyl and Dirac semimetals, where
scattering rates are only a few cm$^{-1}$ \cite{chen15,xu16,akrap16};
indeed, the observation \cite{richard10} and calculation \cite{yin11a} of Dirac cone-like
dispersion of the electronic bands of BaFe$_2$As$_2$ below $T_N$ suggests that this is a
natural comparison.

In all three materials, structure is observed in the optical conductivity that is
associated with the partial gapping of the Fermi surface appears below $T_N$, a small gap
($\Delta_1$) and a large gap ($\Delta_2$); these features may be associated transitions
between relatively flat bands located at high-symmetry points ($\Gamma$ and $M^\prime$)
\cite{yin11a}.  The values range from $\Delta_1 \simeq 44 - 82$~meV for the small gap to
$\Delta_2 \simeq 112 - 207$~meV for the large gap.  This yields values of
$\Delta_1/k_{\rm B}T_N \simeq 3.8 - 5.5$ and $\Delta_2/k_{\rm B}T_N \simeq 9.4 - 14$,
where $k_{\rm B}$ is Boltzmann's constant.  As previously noted,  both $\Delta_1$ and
$\Delta_2$ are increasing across this family of materials; however, the ratio of the gaps
shows little variation, with $\Delta_2/\Delta_1 \simeq 2.5-3$, suggesting that the gaps
scale with the electronic bandwidth.

%
%
\section{Conclusions}
The temperature dependence of the detailed optical properties of BaFe$_2$As$_2$, SrFe$_2$As$_2$,
and CaFe$_2$As$_2$ single crystals have been determined over a wide energy range above and
below $T_N \simeq 138$, 195 and 172~K, respectively, for light polarized in the \emph{a-b} planes.
%
%
The complex optical properties may be reliably fit using two Drude components
in combination with a series of Lorentz oscillators.  Above $T_N$ in all three materials, the
free-carrier response consists of a weak, narrow Drude term, and a much stronger, broader
Drude term, both of which display only a weak temperature dependence.  The plasma frequencies
of both the narrow and broad terms are observed to increase in the Ba, Sr, and Ca family
of materials.
%
%
Below $T_N$ the Fermi surface reconstruction produces dramatic changes in the
complex conductivity.  While the materials are increasingly metallic at low temperature,
there is a decrease in the low-frequency spectral weight from both the narrow and broad Drude
components and a commensurate transfer to the gap-like features ($\Delta_1$ and $\Delta_2$)
observed at higher energies.  The complex conductivity may only be reliably fit using the
two-Drude model if the constraint that the spectral weight is constant below roughly
5000~cm$^{-1}$ is introduced.  The loss of spectral weight from the narrow Drude component
is apparently transferred to peak in the optical conductivity associated with the
low-energy gap $\Delta_1$, while the loss of spectral weight from the broad Drude
component is apparently transferred the high-energy gap $\Delta_2$.

Below $T_N$, both the plasma frequency and the scattering rate in the broad Drude term
decrease substantially; the plasma frequency in the narrow Drude term experiences
a slight decrease, but scattering rate decreases by over an order of magnitude,
and in the case of BaFe$_2$As$_2$, is only a few cm$^{-1}$ for $T\ll T_N$.  Dirac
semimetals often display extremely small scattering rates, suggesting that the
extraordinarily low value for the scattering rate in these materials may be related to
the Dirac cone-like dispersion observed in the electronic bands below $T_N$.

%
%
\begin{acknowledgements}
We would like to acknowledge useful discussions with E. Bascones and Y. Gallais.
A.~A. acknowledges funding from the Ambizione grant of the Swiss National Science Foundation.
Work at the Ames Laboratory (S.~L.~B. and P.~C.~C.) was supported by the U.S. Department of Energy (DOE),
Office of Science, Basic Energy Sciences, Materials Sciences and Engineering Division.
The Ames Laboratory is operated for the U.S. Department of Energy by Iowa State University
under contract No. DE-AC02-07CH11358.  We would like to thanks Alex Thaler and Sheng Ran for
help in samples' synthesis.
Work at Brookhaven National Laboratory was supported by the Office of Science, U.S. Department
of Energy under Contract No. DE-SC0012704.
\end{acknowledgements}

%
%
%

\begin{thebibliography}{71}%
\makeatletter
\providecommand \@ifxundefined [1]{%
 \@ifx{#1\undefined}
}%
\providecommand \@ifnum [1]{%
 \ifnum #1\expandafter \@firstoftwo
 \else \expandafter \@secondoftwo
 \fi
}%
\providecommand \@ifx [1]{%
 \ifx #1\expandafter \@firstoftwo
 \else \expandafter \@secondoftwo
 \fi
}%
\providecommand \natexlab [1]{#1}%
\providecommand \enquote  [1]{``#1''}%
\providecommand \bibnamefont  [1]{#1}%
\providecommand \bibfnamefont [1]{#1}%
\providecommand \citenamefont [1]{#1}%
\providecommand \href@noop [0]{\@secondoftwo}%
\providecommand \href [0]{\begingroup \@sanitize@url \@href}%
\providecommand \@href[1]{\@@startlink{#1}\@@href}%
\providecommand \@@href[1]{\endgroup#1\@@endlink}%
\providecommand \@sanitize@url [0]{\catcode `\\12\catcode `\$12\catcode
  `\&12\catcode `\#12\catcode `\^12\catcode `\_12\catcode `\%12\relax}%
\providecommand \@@startlink[1]{}%
\providecommand \@@endlink[0]{}%
\providecommand \url  [0]{\begingroup\@sanitize@url \@url }%
\providecommand \@url [1]{\endgroup\@href {#1}{\urlprefix }}%
\providecommand \urlprefix  [0]{URL }%
\providecommand \Eprint [0]{\href }%
\providecommand \doibase [0]{http://dx.doi.org/}%
\providecommand \selectlanguage [0]{\@gobble}%
\providecommand \bibinfo  [0]{\@secondoftwo}%
\providecommand \bibfield  [0]{\@secondoftwo}%
\providecommand \translation [1]{[#1]}%
\providecommand \BibitemOpen [0]{}%
\providecommand \bibitemStop [0]{}%
\providecommand \bibitemNoStop [0]{.\EOS\space}%
\providecommand \EOS [0]{\spacefactor3000\relax}%
\providecommand \BibitemShut  [1]{\csname bibitem#1\endcsname}%
\let\auto@bib@innerbib\@empty
\bibitem [{\citenamefont {Johnston}(2010)}]{johnston10}%
  \BibitemOpen
  \bibfield  {author} {\bibinfo {author} {\bibfnamefont {David~C.}\
  \bibnamefont {Johnston}},\ }\bibfield  {title} {\enquote {\bibinfo {title}
  {The puzzle of high temperature superconductivity in layered iron pnictides
  and chalcogenides},}\ }\href {\doibase 10.1080/00018732.2010.513480}
  {\bibfield  {journal} {\bibinfo  {journal} {Adv. Phys.}\ }\textbf {\bibinfo
  {volume} {59}},\ \bibinfo {pages} {803--1061} (\bibinfo {year}
  {2010})}\BibitemShut {NoStop}%
\bibitem [{\citenamefont {Paglione}\ and\ \citenamefont
  {Greene}(2010)}]{paglione10}%
  \BibitemOpen
  \bibfield  {author} {\bibinfo {author} {\bibfnamefont {Johnpierre}\
  \bibnamefont {Paglione}}\ and\ \bibinfo {author} {\bibfnamefont {Richard~L.}\
  \bibnamefont {Greene}},\ }\bibfield  {title} {\enquote {\bibinfo {title}
  {High-temperature superconductivity in iron-based materials},}\ }\href
  {\doibase 10.1038/nphys1759} {\bibfield  {journal} {\bibinfo  {journal} {Nat.
  Phys.}\ }\textbf {\bibinfo {volume} {6}},\ \bibinfo {pages} {645--658}
  (\bibinfo {year} {2010})}\BibitemShut {NoStop}%
\bibitem [{\citenamefont {Canfield}\ and\ \citenamefont
  {Bud'ko}(2010)}]{canfield10}%
  \BibitemOpen
  \bibfield  {author} {\bibinfo {author} {\bibfnamefont {Paul~C.}\ \bibnamefont
  {Canfield}}\ and\ \bibinfo {author} {\bibfnamefont {Sergey~L.}\ \bibnamefont
  {Bud'ko}},\ }\bibfield  {title} {\enquote {\bibinfo {title} {{FeAs-Based
  Superconductivity: A Case Study of the Effects of Transition Metal Doping on
  BaFe$_2$As$_2$}},}\ }\href {\doibase
  10.1146/annurev-conmatphys-070909-104041} {\bibfield  {journal} {\bibinfo
  {journal} {Ann. Rev. Cond. Mat. Phys.}\ }\textbf {\bibinfo {volume} {1}},\
  \bibinfo {pages} {27--50} (\bibinfo {year} {2010})}\BibitemShut {NoStop}%
\bibitem [{\citenamefont {Si}\ \emph {et~al.}(2016)\citenamefont {Si},
  \citenamefont {Yu},\ and\ \citenamefont {Abrahams}}]{si16}%
  \BibitemOpen
  \bibfield  {author} {\bibinfo {author} {\bibfnamefont {Qimiao}\ \bibnamefont
  {Si}}, \bibinfo {author} {\bibfnamefont {Rong}\ \bibnamefont {Yu}}, \ and\
  \bibinfo {author} {\bibfnamefont {Elihu}\ \bibnamefont {Abrahams}},\
  }\bibfield  {title} {\enquote {\bibinfo {title} {High-temperature
  superconductivity in iron pnictides and chalcogenides},}\ }\href {\doibase
  10.1038/natrevmats.2016.17} {\bibfield  {journal} {\bibinfo  {journal} {Nat.
  Rev. Mater.}\ }\textbf {\bibinfo {volume} {1}},\ \bibinfo {pages} {16017}
  (\bibinfo {year} {2016})}\BibitemShut {NoStop}%
\bibitem [{\citenamefont {Ishikawa}\ \emph {et~al.}(2009)\citenamefont
  {Ishikawa}, \citenamefont {Eguchi}, \citenamefont {Kodama}, \citenamefont
  {Fujimaki}, \citenamefont {Einaga}, \citenamefont {Ohmura}, \citenamefont
  {Nakayama}, \citenamefont {Mitsuda},\ and\ \citenamefont
  {Yamada}}]{ishikawa09}%
  \BibitemOpen
  \bibfield  {author} {\bibinfo {author} {\bibfnamefont {Fumihiro}\
  \bibnamefont {Ishikawa}}, \bibinfo {author} {\bibfnamefont {Naoya}\
  \bibnamefont {Eguchi}}, \bibinfo {author} {\bibfnamefont {Michihiro}\
  \bibnamefont {Kodama}}, \bibinfo {author} {\bibfnamefont {Koji}\ \bibnamefont
  {Fujimaki}}, \bibinfo {author} {\bibfnamefont {Mari}\ \bibnamefont {Einaga}},
  \bibinfo {author} {\bibfnamefont {Ayako}\ \bibnamefont {Ohmura}}, \bibinfo
  {author} {\bibfnamefont {Atsuko}\ \bibnamefont {Nakayama}}, \bibinfo {author}
  {\bibfnamefont {Akihiro}\ \bibnamefont {Mitsuda}}, \ and\ \bibinfo {author}
  {\bibfnamefont {Yuh}\ \bibnamefont {Yamada}},\ }\bibfield  {title} {\enquote
  {\bibinfo {title} {Zero-resistance superconducting phase in
  {BaFe}$_{2}${As}$_{2}$ under high pressure},}\ }\href {\doibase
  10.1103/PhysRevB.79.172506} {\bibfield  {journal} {\bibinfo  {journal} {Phys.
  Rev. B}\ }\textbf {\bibinfo {volume} {79}},\ \bibinfo {pages} {172506}
  (\bibinfo {year} {2009})}\BibitemShut {NoStop}%
\bibitem [{\citenamefont {Alireza}\ \emph {et~al.}(2009)\citenamefont
  {Alireza}, \citenamefont {Ko}, \citenamefont {Gillett}, \citenamefont
  {Petrone}, \citenamefont {Cole}, \citenamefont {Sebastian},\ and\
  \citenamefont {Lonzarich}}]{alireza09}%
  \BibitemOpen
  \bibfield  {author} {\bibinfo {author} {\bibfnamefont {Patricia~L.}\
  \bibnamefont {Alireza}}, \bibinfo {author} {\bibfnamefont {Y.~T.~Chris}\
  \bibnamefont {Ko}}, \bibinfo {author} {\bibfnamefont {Jack}\ \bibnamefont
  {Gillett}}, \bibinfo {author} {\bibfnamefont {Chiara~M.}\ \bibnamefont
  {Petrone}}, \bibinfo {author} {\bibfnamefont {Jacqui~M.}\ \bibnamefont
  {Cole}}, \bibinfo {author} {\bibfnamefont {Suchitra~E.}\ \bibnamefont
  {Sebastian}}, \ and\ \bibinfo {author} {\bibfnamefont {Gilbert~G.}\
  \bibnamefont {Lonzarich}},\ }\bibfield  {title} {\enquote {\bibinfo {title}
  {{Superconductivity up to 29~K in SrFe$_2$As$_2$ and BaFe$_2$As$_2$ at high
  pressures}},}\ }\href {\doibase 10.1088/0953-8984/21/1/012208} {\bibfield
  {journal} {\bibinfo  {journal} {J. Phys: Cond. Matter}\ }\textbf {\bibinfo
  {volume} {21}},\ \bibinfo {pages} {012208} (\bibinfo {year}
  {2009})}\BibitemShut {NoStop}%
\bibitem [{\citenamefont {Colombier}\ \emph {et~al.}(2009)\citenamefont
  {Colombier}, \citenamefont {Bud'ko}, \citenamefont {Ni},\ and\ \citenamefont
  {Canfield}}]{colombier09}%
  \BibitemOpen
  \bibfield  {author} {\bibinfo {author} {\bibfnamefont {E.}~\bibnamefont
  {Colombier}}, \bibinfo {author} {\bibfnamefont {S.~L.}\ \bibnamefont
  {Bud'ko}}, \bibinfo {author} {\bibfnamefont {N.}~\bibnamefont {Ni}}, \ and\
  \bibinfo {author} {\bibfnamefont {P.~C.}\ \bibnamefont {Canfield}},\
  }\bibfield  {title} {\enquote {\bibinfo {title} {Complete pressure-dependent
  phase diagrams for {SrFe}$_{2}${As}$_{2}$ and {BaFe}$_{2}${As}$_{2}$},}\
  }\href {\doibase 10.1103/PhysRevB.79.224518} {\bibfield  {journal} {\bibinfo
  {journal} {Phys. Rev. B}\ }\textbf {\bibinfo {volume} {79}},\ \bibinfo
  {pages} {224518} (\bibinfo {year} {2009})}\BibitemShut {NoStop}%
\bibitem [{\citenamefont {Yamazaki}\ \emph {et~al.}(2010)\citenamefont
  {Yamazaki}, \citenamefont {Takeshita}, \citenamefont {Kobayashi},
  \citenamefont {Fukazawa}, \citenamefont {Kohori}, \citenamefont {Kihou},
  \citenamefont {Lee}, \citenamefont {Kito}, \citenamefont {Iyo},\ and\
  \citenamefont {Eisaki}}]{yamazaki10}%
  \BibitemOpen
  \bibfield  {author} {\bibinfo {author} {\bibfnamefont {Takehiro}\
  \bibnamefont {Yamazaki}}, \bibinfo {author} {\bibfnamefont {Nao}\
  \bibnamefont {Takeshita}}, \bibinfo {author} {\bibfnamefont {Ryosuke}\
  \bibnamefont {Kobayashi}}, \bibinfo {author} {\bibfnamefont {Hideto}\
  \bibnamefont {Fukazawa}}, \bibinfo {author} {\bibfnamefont {Yoh}\
  \bibnamefont {Kohori}}, \bibinfo {author} {\bibfnamefont {Kunihiro}\
  \bibnamefont {Kihou}}, \bibinfo {author} {\bibfnamefont {Chul-Ho}\
  \bibnamefont {Lee}}, \bibinfo {author} {\bibfnamefont {Hijiri}\ \bibnamefont
  {Kito}}, \bibinfo {author} {\bibfnamefont {Akira}\ \bibnamefont {Iyo}}, \
  and\ \bibinfo {author} {\bibfnamefont {Hiroshi}\ \bibnamefont {Eisaki}},\
  }\bibfield  {title} {\enquote {\bibinfo {title} {Appearance of
  pressure-induced superconductivity in {BaFe}$_{2}${As}$_{2}$ under
  hydrostatic conditions and its extremely high sensitivity to uniaxial
  stress},}\ }\href {\doibase 10.1103/PhysRevB.81.224511} {\bibfield  {journal}
  {\bibinfo  {journal} {Phys. Rev. B}\ }\textbf {\bibinfo {volume} {81}},\
  \bibinfo {pages} {224511} (\bibinfo {year} {2010})}\BibitemShut {NoStop}%
\bibitem [{\citenamefont {Sefat}\ \emph {et~al.}(2008)\citenamefont {Sefat},
  \citenamefont {Jin}, \citenamefont {McGuire}, \citenamefont {Sales},
  \citenamefont {Singh},\ and\ \citenamefont {Mandrus}}]{sefat08}%
  \BibitemOpen
  \bibfield  {author} {\bibinfo {author} {\bibfnamefont {Athena~S.}\
  \bibnamefont {Sefat}}, \bibinfo {author} {\bibfnamefont {Rongying}\
  \bibnamefont {Jin}}, \bibinfo {author} {\bibfnamefont {Michael~A.}\
  \bibnamefont {McGuire}}, \bibinfo {author} {\bibfnamefont {Brian~C.}\
  \bibnamefont {Sales}}, \bibinfo {author} {\bibfnamefont {David~J.}\
  \bibnamefont {Singh}}, \ and\ \bibinfo {author} {\bibfnamefont {David}\
  \bibnamefont {Mandrus}},\ }\bibfield  {title} {\enquote {\bibinfo {title}
  {{Superconductivity at 22~K in Co-Doped BaFe$_{2}$As$_{2}$ Crystals}},}\
  }\href {\doibase 10.1103/PhysRevLett.101.117004} {\bibfield  {journal}
  {\bibinfo  {journal} {Phys. Rev. Lett.}\ }\textbf {\bibinfo {volume} {101}},\
  \bibinfo {pages} {117004} (\bibinfo {year} {2008})}\BibitemShut {NoStop}%
\bibitem [{\citenamefont {Ni}\ \emph {et~al.}(2008{\natexlab{a}})\citenamefont
  {Ni}, \citenamefont {Tillman}, \citenamefont {Yan}, \citenamefont {Kracher},
  \citenamefont {Hannahs}, \citenamefont {Bud'ko},\ and\ \citenamefont
  {Canfield}}]{ni08a}%
  \BibitemOpen
  \bibfield  {author} {\bibinfo {author} {\bibfnamefont {N.}~\bibnamefont
  {Ni}}, \bibinfo {author} {\bibfnamefont {M.~E.}\ \bibnamefont {Tillman}},
  \bibinfo {author} {\bibfnamefont {J.-Q.}\ \bibnamefont {Yan}}, \bibinfo
  {author} {\bibfnamefont {A.}~\bibnamefont {Kracher}}, \bibinfo {author}
  {\bibfnamefont {S.~T.}\ \bibnamefont {Hannahs}}, \bibinfo {author}
  {\bibfnamefont {S.~L.}\ \bibnamefont {Bud'ko}}, \ and\ \bibinfo {author}
  {\bibfnamefont {P.~C.}\ \bibnamefont {Canfield}},\ }\bibfield  {title}
  {\enquote {\bibinfo {title} {Effects of {Co} substitution on thermodynamic
  and transport properties and anisotropic ${H}_{c2}$ in
  {Ba}({Fe}$_{1-x}${Co}$_{x}$)$_{2}${As}$_{2}$ single crystals},}\ }\href
  {\doibase 10.1103/PhysRevB.78.214515} {\bibfield  {journal} {\bibinfo
  {journal} {Phys. Rev. B}\ }\textbf {\bibinfo {volume} {78}},\ \bibinfo
  {pages} {214515} (\bibinfo {year} {2008}{\natexlab{a}})}\BibitemShut
  {NoStop}%
\bibitem [{\citenamefont {Chu}\ \emph {et~al.}(2009)\citenamefont {Chu},
  \citenamefont {Analytis}, \citenamefont {Kucharczyk},\ and\ \citenamefont
  {Fisher}}]{chu09}%
  \BibitemOpen
  \bibfield  {author} {\bibinfo {author} {\bibfnamefont {Jiun-Haw}\
  \bibnamefont {Chu}}, \bibinfo {author} {\bibfnamefont {James~G.}\
  \bibnamefont {Analytis}}, \bibinfo {author} {\bibfnamefont {Chris}\
  \bibnamefont {Kucharczyk}}, \ and\ \bibinfo {author} {\bibfnamefont {Ian~R.}\
  \bibnamefont {Fisher}},\ }\bibfield  {title} {\enquote {\bibinfo {title}
  {{Determination of the phase diagram of the electron-doped superconductor
  $\text{Ba}{({\text{Fe}}_{1\ensuremath{-}x}{\text{Co}}_{x})}_{2}{\text{As}}_{2}$}},}\
  }\href {\doibase 10.1103/PhysRevB.79.014506} {\bibfield  {journal} {\bibinfo
  {journal} {Phys. Rev. B}\ }\textbf {\bibinfo {volume} {79}},\ \bibinfo
  {pages} {014506} (\bibinfo {year} {2009})}\BibitemShut {NoStop}%
\bibitem [{\citenamefont {Canfield}\ \emph {et~al.}(2009)\citenamefont
  {Canfield}, \citenamefont {Bud'ko}, \citenamefont {Ni}, \citenamefont {Yan},\
  and\ \citenamefont {Kracher}}]{canfield09}%
  \BibitemOpen
  \bibfield  {author} {\bibinfo {author} {\bibfnamefont {P.~C.}\ \bibnamefont
  {Canfield}}, \bibinfo {author} {\bibfnamefont {S.~L.}\ \bibnamefont
  {Bud'ko}}, \bibinfo {author} {\bibfnamefont {Ni}~\bibnamefont {Ni}}, \bibinfo
  {author} {\bibfnamefont {J.~Q.}\ \bibnamefont {Yan}}, \ and\ \bibinfo
  {author} {\bibfnamefont {A.}~\bibnamefont {Kracher}},\ }\bibfield  {title}
  {\enquote {\bibinfo {title} {Decoupling of the superconducting and
  magnetic/structural phase transitions in electron-doped
  {BaFe}$_{2}${As}$_{2}$},}\ }\href {\doibase 10.1103/PhysRevB.80.060501}
  {\bibfield  {journal} {\bibinfo  {journal} {Phys. Rev. B}\ }\textbf {\bibinfo
  {volume} {80}},\ \bibinfo {pages} {060501} (\bibinfo {year}
  {2009})}\BibitemShut {NoStop}%
\bibitem [{\citenamefont {Rotter}\ \emph
  {et~al.}(2008{\natexlab{a}})\citenamefont {Rotter}, \citenamefont {Tegel},\
  and\ \citenamefont {Johrendt}}]{rotter08b}%
  \BibitemOpen
  \bibfield  {author} {\bibinfo {author} {\bibfnamefont {Marianne}\
  \bibnamefont {Rotter}}, \bibinfo {author} {\bibfnamefont {Marcus}\
  \bibnamefont {Tegel}}, \ and\ \bibinfo {author} {\bibfnamefont {Dirk}\
  \bibnamefont {Johrendt}},\ }\bibfield  {title} {\enquote {\bibinfo {title}
  {{Superconductivity at 38~K in the Iron Arsenide
  (Ba$_{1-x}$K$_x$)Fe$_2$As$_2$}},}\ }\href {\doibase
  10.1103/PhysRevLett.101.107006} {\bibfield  {journal} {\bibinfo  {journal}
  {Phys. Rev. Lett.}\ }\textbf {\bibinfo {volume} {101}},\ \bibinfo {pages}
  {107006} (\bibinfo {year} {2008}{\natexlab{a}})}\BibitemShut {NoStop}%
\bibitem [{\citenamefont {Torikachvili}\ \emph {et~al.}(2008)\citenamefont
  {Torikachvili}, \citenamefont {Bud'ko}, \citenamefont {Ni},\ and\
  \citenamefont {Canfield}}]{torikachvili08}%
  \BibitemOpen
  \bibfield  {author} {\bibinfo {author} {\bibfnamefont {M.~S.}\ \bibnamefont
  {Torikachvili}}, \bibinfo {author} {\bibfnamefont {S.~L.}\ \bibnamefont
  {Bud'ko}}, \bibinfo {author} {\bibfnamefont {N.}~\bibnamefont {Ni}}, \ and\
  \bibinfo {author} {\bibfnamefont {P.~C.}\ \bibnamefont {Canfield}},\
  }\bibfield  {title} {\enquote {\bibinfo {title} {{Effect of pressure on the
  structural phase transition and superconductivity in
  $({\text{Ba}}_{1\ensuremath{-}x}{\text{K}}_{x}){\text{Fe}}_{2}{\text{As}}_{2}$
  ($x=0$ and 0.45) and ${\text{SrFe}}_{2}{\text{As}}_{2}$ single crystals}},}\
  }\href {\doibase 10.1103/PhysRevB.78.104527} {\bibfield  {journal} {\bibinfo
  {journal} {Phys. Rev. B}\ }\textbf {\bibinfo {volume} {78}},\ \bibinfo
  {pages} {104527} (\bibinfo {year} {2008})}\BibitemShut {NoStop}%
\bibitem [{\citenamefont {{Chen, H.}}\ \emph {et~al.}(2009)\citenamefont
  {{Chen, H.}}, \citenamefont {{Ren, Y.}}, \citenamefont {{Qiu, Y.}},
  \citenamefont {{Bao, Wei}}, \citenamefont {{Liu, R. H.}}, \citenamefont {{Wu,
  G.}}, \citenamefont {{Wu, T.}}, \citenamefont {{Xie, Y. L.}}, \citenamefont
  {{Wang, X. F.}}, \citenamefont {{Huang, Q.}},\ and\ \citenamefont {{Chen, X.
  H.}}}]{chenh09}%
  \BibitemOpen
  \bibfield  {author} {\bibinfo {author} {\bibnamefont {{Chen, H.}}}, \bibinfo
  {author} {\bibnamefont {{Ren, Y.}}}, \bibinfo {author} {\bibnamefont {{Qiu,
  Y.}}}, \bibinfo {author} {\bibnamefont {{Bao, Wei}}}, \bibinfo {author}
  {\bibnamefont {{Liu, R. H.}}}, \bibinfo {author} {\bibnamefont {{Wu, G.}}},
  \bibinfo {author} {\bibnamefont {{Wu, T.}}}, \bibinfo {author} {\bibnamefont
  {{Xie, Y. L.}}}, \bibinfo {author} {\bibnamefont {{Wang, X. F.}}}, \bibinfo
  {author} {\bibnamefont {{Huang, Q.}}}, \ and\ \bibinfo {author} {\bibnamefont
  {{Chen, X. H.}}},\ }\bibfield  {title} {\enquote {\bibinfo {title}
  {{Coexistence of the spin-density wave and superconductivity in
  Ba$_{1-x}$K$_x$Fe$_2$As$_2$}},}\ }\href {\doibase 10.1209/0295-5075/85/17006}
  {\bibfield  {journal} {\bibinfo  {journal} {EPL}\ }\textbf {\bibinfo {volume}
  {85}},\ \bibinfo {pages} {17006} (\bibinfo {year} {2009})}\BibitemShut
  {NoStop}%
\bibitem [{\citenamefont {Jiang}\ \emph {et~al.}(2009)\citenamefont {Jiang},
  \citenamefont {Xing}, \citenamefont {Xuan}, \citenamefont {Wang},
  \citenamefont {Ren}, \citenamefont {Feng}, \citenamefont {Dai}, \citenamefont
  {Xu},\ and\ \citenamefont {Cao}}]{jiang09}%
  \BibitemOpen
  \bibfield  {author} {\bibinfo {author} {\bibfnamefont {Shuai}\ \bibnamefont
  {Jiang}}, \bibinfo {author} {\bibfnamefont {Hui}\ \bibnamefont {Xing}},
  \bibinfo {author} {\bibfnamefont {Guofang}\ \bibnamefont {Xuan}}, \bibinfo
  {author} {\bibfnamefont {Cao}\ \bibnamefont {Wang}}, \bibinfo {author}
  {\bibfnamefont {Zhi}\ \bibnamefont {Ren}}, \bibinfo {author} {\bibfnamefont
  {Chunmu}\ \bibnamefont {Feng}}, \bibinfo {author} {\bibfnamefont {Jianhui}\
  \bibnamefont {Dai}}, \bibinfo {author} {\bibfnamefont {Zhu'an}\ \bibnamefont
  {Xu}}, \ and\ \bibinfo {author} {\bibfnamefont {Guanghan}\ \bibnamefont
  {Cao}},\ }\bibfield  {title} {\enquote {\bibinfo {title} {Superconductivity
  up to 30~{K} in the vicinity of the quantum critical point in
  {BaFe}$_2$({As}$_{1-x}${P}$_x$)$_2$},}\ }\href {\doibase
  10.1088/0953-8984/21/38/382203} {\bibfield  {journal} {\bibinfo  {journal}
  {J. Phys.: Condens. Matter}\ }\textbf {\bibinfo {volume} {21}},\ \bibinfo
  {pages} {382203} (\bibinfo {year} {2009})}\BibitemShut {NoStop}%
\bibitem [{\citenamefont {Rullier-Albenque}\ \emph {et~al.}(2010)\citenamefont
  {Rullier-Albenque}, \citenamefont {Colson}, \citenamefont {Forget},
  \citenamefont {Thu\'ery},\ and\ \citenamefont {Poissonnet}}]{rullier10}%
  \BibitemOpen
  \bibfield  {author} {\bibinfo {author} {\bibfnamefont {F.}~\bibnamefont
  {Rullier-Albenque}}, \bibinfo {author} {\bibfnamefont {D.}~\bibnamefont
  {Colson}}, \bibinfo {author} {\bibfnamefont {A.}~\bibnamefont {Forget}},
  \bibinfo {author} {\bibfnamefont {P.}~\bibnamefont {Thu\'ery}}, \ and\
  \bibinfo {author} {\bibfnamefont {S.}~\bibnamefont {Poissonnet}},\ }\bibfield
   {title} {\enquote {\bibinfo {title} {{Hole and electron contributions to the
  transport properties of {Ba}({Fe}$_{1-x}${Ru}$_{x}$)$_{2}${As}$_{2}$ single
  crystals}},}\ }\href {\doibase 10.1103/PhysRevB.81.224503} {\bibfield
  {journal} {\bibinfo  {journal} {Phys. Rev. B}\ }\textbf {\bibinfo {volume}
  {81}},\ \bibinfo {pages} {224503} (\bibinfo {year} {2010})}\BibitemShut
  {NoStop}%
\bibitem [{\citenamefont {Thaler}\ \emph {et~al.}(2010)\citenamefont {Thaler},
  \citenamefont {Ni}, \citenamefont {Kracher}, \citenamefont {Yan},
  \citenamefont {Bud'ko},\ and\ \citenamefont {Canfield}}]{thaler10}%
  \BibitemOpen
  \bibfield  {author} {\bibinfo {author} {\bibfnamefont {A.}~\bibnamefont
  {Thaler}}, \bibinfo {author} {\bibfnamefont {N.}~\bibnamefont {Ni}}, \bibinfo
  {author} {\bibfnamefont {A.}~\bibnamefont {Kracher}}, \bibinfo {author}
  {\bibfnamefont {J.~Q.}\ \bibnamefont {Yan}}, \bibinfo {author} {\bibfnamefont
  {S.~L.}\ \bibnamefont {Bud'ko}}, \ and\ \bibinfo {author} {\bibfnamefont
  {P.~C.}\ \bibnamefont {Canfield}},\ }\bibfield  {title} {\enquote {\bibinfo
  {title} {{Physical and magnetic properties of
  $\text{Ba}{({\text{Fe}}_{1\ensuremath{-}x}{\text{Ru}}_{x})}_{2}{\text{As}}_{2}$
  single crystals}},}\ }\href {\doibase 10.1103/PhysRevB.82.014534} {\bibfield
  {journal} {\bibinfo  {journal} {Phys. Rev. B}\ }\textbf {\bibinfo {volume}
  {82}},\ \bibinfo {pages} {014534} (\bibinfo {year} {2010})}\BibitemShut
  {NoStop}%
\bibitem [{\citenamefont {Nakai}\ \emph {et~al.}(2010)\citenamefont {Nakai},
  \citenamefont {Iye}, \citenamefont {Kitagawa}, \citenamefont {Ishida},
  \citenamefont {Ikeda}, \citenamefont {Kasahara}, \citenamefont {Shishido},
  \citenamefont {Shibauchi}, \citenamefont {Matsuda},\ and\ \citenamefont
  {Terashima}}]{nakai10}%
  \BibitemOpen
  \bibfield  {author} {\bibinfo {author} {\bibfnamefont {Y.}~\bibnamefont
  {Nakai}}, \bibinfo {author} {\bibfnamefont {T.}~\bibnamefont {Iye}}, \bibinfo
  {author} {\bibfnamefont {S.}~\bibnamefont {Kitagawa}}, \bibinfo {author}
  {\bibfnamefont {K.}~\bibnamefont {Ishida}}, \bibinfo {author} {\bibfnamefont
  {H.}~\bibnamefont {Ikeda}}, \bibinfo {author} {\bibfnamefont
  {S.}~\bibnamefont {Kasahara}}, \bibinfo {author} {\bibfnamefont
  {H.}~\bibnamefont {Shishido}}, \bibinfo {author} {\bibfnamefont
  {T.}~\bibnamefont {Shibauchi}}, \bibinfo {author} {\bibfnamefont
  {Y.}~\bibnamefont {Matsuda}}, \ and\ \bibinfo {author} {\bibfnamefont
  {T.}~\bibnamefont {Terashima}},\ }\bibfield  {title} {\enquote {\bibinfo
  {title} {{Unconventional Superconductivity and Antiferromagnetic Quantum
  Critical Behavior in the Isovalent-Doped
  {BaFe}$_{2}$({As}$_{1-x}${P}$_{x}$)$_{2}$}},}\ }\href {\doibase
  10.1103/PhysRevLett.105.107003} {\bibfield  {journal} {\bibinfo  {journal}
  {Phys. Rev. Lett.}\ }\textbf {\bibinfo {volume} {105}},\ \bibinfo {pages}
  {107003} (\bibinfo {year} {2010})}\BibitemShut {NoStop}%
\bibitem [{\citenamefont {Rotter}\ \emph
  {et~al.}(2008{\natexlab{b}})\citenamefont {Rotter}, \citenamefont {Tegel},
  \citenamefont {Johrendt}, \citenamefont {Schellenberg}, \citenamefont
  {Hermes},\ and\ \citenamefont {P\"ottgen}}]{rotter08a}%
  \BibitemOpen
  \bibfield  {author} {\bibinfo {author} {\bibfnamefont {Marianne}\
  \bibnamefont {Rotter}}, \bibinfo {author} {\bibfnamefont {Marcus}\
  \bibnamefont {Tegel}}, \bibinfo {author} {\bibfnamefont {Dirk}\ \bibnamefont
  {Johrendt}}, \bibinfo {author} {\bibfnamefont {Inga}\ \bibnamefont
  {Schellenberg}}, \bibinfo {author} {\bibfnamefont {Wilfried}\ \bibnamefont
  {Hermes}}, \ and\ \bibinfo {author} {\bibfnamefont {Rainer}\ \bibnamefont
  {P\"ottgen}},\ }\bibfield  {title} {\enquote {\bibinfo {title}
  {{Spin-density-wave anomaly at 140~K in the ternary iron arsenide
  {BaFe}$_2${As}$_2$}},}\ }\href {\doibase 10.1103/PhysRevB.78.020503}
  {\bibfield  {journal} {\bibinfo  {journal} {Phys. Rev. B}\ }\textbf {\bibinfo
  {volume} {78}},\ \bibinfo {pages} {020503(R)} (\bibinfo {year}
  {2008}{\natexlab{b}})}\BibitemShut {NoStop}%
\bibitem [{\citenamefont {Wang}\ \emph {et~al.}(2009)\citenamefont {Wang},
  \citenamefont {Wu}, \citenamefont {Wu}, \citenamefont {Chen}, \citenamefont
  {Xie}, \citenamefont {Ying}, \citenamefont {Yan}, \citenamefont {Liu},\ and\
  \citenamefont {Chen}}]{wang09}%
  \BibitemOpen
  \bibfield  {author} {\bibinfo {author} {\bibfnamefont {X.~F.}\ \bibnamefont
  {Wang}}, \bibinfo {author} {\bibfnamefont {T.}~\bibnamefont {Wu}}, \bibinfo
  {author} {\bibfnamefont {G.}~\bibnamefont {Wu}}, \bibinfo {author}
  {\bibfnamefont {H.}~\bibnamefont {Chen}}, \bibinfo {author} {\bibfnamefont
  {Y.~L.}\ \bibnamefont {Xie}}, \bibinfo {author} {\bibfnamefont {J.~J.}\
  \bibnamefont {Ying}}, \bibinfo {author} {\bibfnamefont {Y.~J.}\ \bibnamefont
  {Yan}}, \bibinfo {author} {\bibfnamefont {R.~H.}\ \bibnamefont {Liu}}, \ and\
  \bibinfo {author} {\bibfnamefont {X.~H.}\ \bibnamefont {Chen}},\ }\bibfield
  {title} {\enquote {\bibinfo {title} {{Anisotropy in the Electrical
  Resistivity and Susceptibility of Superconducting BaFe$_2$As$_2$ Single
  Crystals}},}\ }\href {\doibase 10.1103/PhysRevLett.102.117005} {\bibfield
  {journal} {\bibinfo  {journal} {Phys. Rev. Lett.}\ }\textbf {\bibinfo
  {volume} {102}},\ \bibinfo {pages} {117005} (\bibinfo {year}
  {2009})}\BibitemShut {NoStop}%
\bibitem [{\citenamefont {Fisher}\ \emph {et~al.}(2011)\citenamefont {Fisher},
  \citenamefont {Degiorgi},\ and\ \citenamefont {Shen}}]{fisher11}%
  \BibitemOpen
  \bibfield  {author} {\bibinfo {author} {\bibfnamefont {I.~R.}\ \bibnamefont
  {Fisher}}, \bibinfo {author} {\bibfnamefont {L.}~\bibnamefont {Degiorgi}}, \
  and\ \bibinfo {author} {\bibfnamefont {Z.~X.}\ \bibnamefont {Shen}},\
  }\bibfield  {title} {\enquote {\bibinfo {title} {{In-plane electronic
  anisotropy of underdoped `122' Fe-arsenide superconductors revealed by
  measurements of detwinned single crystals}},}\ }\href {\doibase
  10.1088/0034-4885/74/12/124506} {\bibfield  {journal} {\bibinfo  {journal}
  {Rep. Prog. Phys.}\ }\textbf {\bibinfo {volume} {74}},\ \bibinfo {pages}
  {124506} (\bibinfo {year} {2011})}\BibitemShut {NoStop}%
\bibitem [{\citenamefont {Chu}\ \emph {et~al.}(2010)\citenamefont {Chu},
  \citenamefont {Analytis}, \citenamefont {Greve}, \citenamefont {McMahon},
  \citenamefont {Islam}, \citenamefont {Yamamoto},\ and\ \citenamefont
  {Fisher}}]{chu10}%
  \BibitemOpen
  \bibfield  {author} {\bibinfo {author} {\bibfnamefont {Jiun-Haw}\
  \bibnamefont {Chu}}, \bibinfo {author} {\bibfnamefont {James~G.}\
  \bibnamefont {Analytis}}, \bibinfo {author} {\bibfnamefont {Kristiaan~De}\
  \bibnamefont {Greve}}, \bibinfo {author} {\bibfnamefont {Peter~L.}\
  \bibnamefont {McMahon}}, \bibinfo {author} {\bibfnamefont {Zahirul}\
  \bibnamefont {Islam}}, \bibinfo {author} {\bibfnamefont {Yoshihisa}\
  \bibnamefont {Yamamoto}}, \ and\ \bibinfo {author} {\bibfnamefont {Ian~R.}\
  \bibnamefont {Fisher}},\ }\bibfield  {title} {\enquote {\bibinfo {title}
  {{In-Plane Resistivity Anisotropy in an Underdoped Iron Arsenide
  Superconductor}},}\ }\href {\doibase 10.1126/science.1190482} {\bibfield
  {journal} {\bibinfo  {journal} {Science}\ }\textbf {\bibinfo {volume}
  {329}},\ \bibinfo {pages} {824--826} (\bibinfo {year} {2010})}\BibitemShut
  {NoStop}%
\bibitem [{\citenamefont {Ishida}\ \emph {et~al.}(2013)\citenamefont {Ishida},
  \citenamefont {Nakajima}, \citenamefont {Liang}, \citenamefont {Kihou},
  \citenamefont {Lee}, \citenamefont {Iyo}, \citenamefont {Eisaki},
  \citenamefont {Kakeshita}, \citenamefont {Tomioka}, \citenamefont {Ito},\
  and\ \citenamefont {Uchida}}]{ishida13}%
  \BibitemOpen
  \bibfield  {author} {\bibinfo {author} {\bibfnamefont {S.}~\bibnamefont
  {Ishida}}, \bibinfo {author} {\bibfnamefont {M.}~\bibnamefont {Nakajima}},
  \bibinfo {author} {\bibfnamefont {T.}~\bibnamefont {Liang}}, \bibinfo
  {author} {\bibfnamefont {K.}~\bibnamefont {Kihou}}, \bibinfo {author}
  {\bibfnamefont {C.~H.}\ \bibnamefont {Lee}}, \bibinfo {author} {\bibfnamefont
  {A.}~\bibnamefont {Iyo}}, \bibinfo {author} {\bibfnamefont {H.}~\bibnamefont
  {Eisaki}}, \bibinfo {author} {\bibfnamefont {T.}~\bibnamefont {Kakeshita}},
  \bibinfo {author} {\bibfnamefont {Y.}~\bibnamefont {Tomioka}}, \bibinfo
  {author} {\bibfnamefont {T.}~\bibnamefont {Ito}}, \ and\ \bibinfo {author}
  {\bibfnamefont {S.}~\bibnamefont {Uchida}},\ }\bibfield  {title} {\enquote
  {\bibinfo {title} {{Anisotropy of the In-Plane Resistivity of Underdoped
  {Ba}({Fe}$_{1-x}${Co}$_{x}$)$_{2}${As}$_{2}$ Superconductors Induced by
  Impurity Scattering in the Antiferromagnetic Orthorhombic Phase}},}\ }\href
  {\doibase 10.1103/PhysRevLett.110.207001} {\bibfield  {journal} {\bibinfo
  {journal} {Phys. Rev. Lett.}\ }\textbf {\bibinfo {volume} {110}},\ \bibinfo
  {pages} {207001} (\bibinfo {year} {2013})}\BibitemShut {NoStop}%
\bibitem [{\citenamefont {Tegel}\ \emph {et~al.}(2008)\citenamefont {Tegel},
  \citenamefont {Rotter}, \citenamefont {Wei$\beta$}, \citenamefont
  {Schappacher}, \citenamefont {P\"{o}ttgen},\ and\ \citenamefont
  {Johrendt}}]{tegel08}%
  \BibitemOpen
  \bibfield  {author} {\bibinfo {author} {\bibfnamefont {Marcus}\ \bibnamefont
  {Tegel}}, \bibinfo {author} {\bibfnamefont {Marianne}\ \bibnamefont
  {Rotter}}, \bibinfo {author} {\bibfnamefont {Veronika}\ \bibnamefont
  {Wei$\beta$}}, \bibinfo {author} {\bibfnamefont {Falko~M}\ \bibnamefont
  {Schappacher}}, \bibinfo {author} {\bibfnamefont {Rainer}\ \bibnamefont
  {P\"{o}ttgen}}, \ and\ \bibinfo {author} {\bibfnamefont {Dirk}\ \bibnamefont
  {Johrendt}},\ }\bibfield  {title} {\enquote {\bibinfo {title} {Structural and
  magnetic phase transitions in the ternary iron arsenides {SrFe}$_2${As}$_2$
  and {EuFe}$_2${As}$_2$},}\ }\href {\doibase 10.1088/0953-8984/20/45/452201}
  {\bibfield  {journal} {\bibinfo  {journal} {J. Phys.: Condens. Matter}\
  }\textbf {\bibinfo {volume} {20}},\ \bibinfo {pages} {452201} (\bibinfo
  {year} {2008})}\BibitemShut {NoStop}%
\bibitem [{\citenamefont {Zhao}\ \emph {et~al.}(2008)\citenamefont {Zhao},
  \citenamefont {Ratcliff}, \citenamefont {Lynn}, \citenamefont {Chen},
  \citenamefont {Luo}, \citenamefont {Wang}, \citenamefont {Hu},\ and\
  \citenamefont {Dai}}]{zhao08}%
  \BibitemOpen
  \bibfield  {author} {\bibinfo {author} {\bibfnamefont {Jun}\ \bibnamefont
  {Zhao}}, \bibinfo {author} {\bibfnamefont {W.}~\bibnamefont {Ratcliff}},
  \bibinfo {author} {\bibfnamefont {J.~W.}\ \bibnamefont {Lynn}}, \bibinfo
  {author} {\bibfnamefont {G.~F.}\ \bibnamefont {Chen}}, \bibinfo {author}
  {\bibfnamefont {J.~L.}\ \bibnamefont {Luo}}, \bibinfo {author} {\bibfnamefont
  {N.~L.}\ \bibnamefont {Wang}}, \bibinfo {author} {\bibfnamefont {Jiangping}\
  \bibnamefont {Hu}}, \ and\ \bibinfo {author} {\bibfnamefont {Pengcheng}\
  \bibnamefont {Dai}},\ }\bibfield  {title} {\enquote {\bibinfo {title} {Spin
  and lattice structures of single-crystalline {SrFe}$_{2}${As}$_{2}$},}\
  }\href {\doibase 10.1103/PhysRevB.78.140504} {\bibfield  {journal} {\bibinfo
  {journal} {Phys. Rev. B}\ }\textbf {\bibinfo {volume} {78}},\ \bibinfo
  {pages} {140504} (\bibinfo {year} {2008})}\BibitemShut {NoStop}%
\bibitem [{\citenamefont {Goldman}\ \emph {et~al.}(2008)\citenamefont
  {Goldman}, \citenamefont {Argyriou}, \citenamefont {Ouladdiaf}, \citenamefont
  {Chatterji}, \citenamefont {Kreyssig}, \citenamefont {Nandi}, \citenamefont
  {Ni}, \citenamefont {Bud'ko}, \citenamefont {Canfield},\ and\ \citenamefont
  {McQueeney}}]{goldman08}%
  \BibitemOpen
  \bibfield  {author} {\bibinfo {author} {\bibfnamefont {A.~I.}\ \bibnamefont
  {Goldman}}, \bibinfo {author} {\bibfnamefont {D.~N.}\ \bibnamefont
  {Argyriou}}, \bibinfo {author} {\bibfnamefont {B.}~\bibnamefont {Ouladdiaf}},
  \bibinfo {author} {\bibfnamefont {T.}~\bibnamefont {Chatterji}}, \bibinfo
  {author} {\bibfnamefont {A.}~\bibnamefont {Kreyssig}}, \bibinfo {author}
  {\bibfnamefont {S.}~\bibnamefont {Nandi}}, \bibinfo {author} {\bibfnamefont
  {N.}~\bibnamefont {Ni}}, \bibinfo {author} {\bibfnamefont {S.~L.}\
  \bibnamefont {Bud'ko}}, \bibinfo {author} {\bibfnamefont {P.~C.}\
  \bibnamefont {Canfield}}, \ and\ \bibinfo {author} {\bibfnamefont {R.~J.}\
  \bibnamefont {McQueeney}},\ }\bibfield  {title} {\enquote {\bibinfo {title}
  {{Lattice and magnetic instabilities in {CaFe}$_{2}${As}$_{2}$: A
  single-crystal neutron diffraction study}},}\ }\href {\doibase
  10.1103/PhysRevB.78.100506} {\bibfield  {journal} {\bibinfo  {journal} {Phys.
  Rev. B}\ }\textbf {\bibinfo {volume} {78}},\ \bibinfo {pages} {100506}
  (\bibinfo {year} {2008})}\BibitemShut {NoStop}%
\bibitem [{\citenamefont {Tanatar}\ \emph {et~al.}(2009)\citenamefont
  {Tanatar}, \citenamefont {Ni}, \citenamefont {Samolyuk}, \citenamefont
  {Bud'ko}, \citenamefont {Canfield},\ and\ \citenamefont
  {Prozorov}}]{tanatar09}%
  \BibitemOpen
  \bibfield  {author} {\bibinfo {author} {\bibfnamefont {M.~A.}\ \bibnamefont
  {Tanatar}}, \bibinfo {author} {\bibfnamefont {N.}~\bibnamefont {Ni}},
  \bibinfo {author} {\bibfnamefont {G.~D.}\ \bibnamefont {Samolyuk}}, \bibinfo
  {author} {\bibfnamefont {S.~L.}\ \bibnamefont {Bud'ko}}, \bibinfo {author}
  {\bibfnamefont {P.~C.}\ \bibnamefont {Canfield}}, \ and\ \bibinfo {author}
  {\bibfnamefont {R.}~\bibnamefont {Prozorov}},\ }\bibfield  {title} {\enquote
  {\bibinfo {title} {{Resistivity anisotropy of $A${Fe}$_2${As}$_2$ ($A=\,$Ca,
  Sr, Ba): Direct versus Montgomery technique measurements}},}\ }\href
  {\doibase 10.1103/PhysRevB.79.134528} {\bibfield  {journal} {\bibinfo
  {journal} {Phys. Rev. B}\ }\textbf {\bibinfo {volume} {79}},\ \bibinfo
  {pages} {134528} (\bibinfo {year} {2009})}\BibitemShut {NoStop}%
\bibitem [{\citenamefont {Tanatar}\ \emph {et~al.}(2010)\citenamefont
  {Tanatar}, \citenamefont {Blomberg}, \citenamefont {Kreyssig}, \citenamefont
  {Kim}, \citenamefont {Ni}, \citenamefont {Thaler}, \citenamefont {Bud'ko},
  \citenamefont {Canfield}, \citenamefont {Goldman}, \citenamefont {Mazin},\
  and\ \citenamefont {Prozorov}}]{tanatar10}%
  \BibitemOpen
  \bibfield  {author} {\bibinfo {author} {\bibfnamefont {M.~A.}\ \bibnamefont
  {Tanatar}}, \bibinfo {author} {\bibfnamefont {E.~C.}\ \bibnamefont
  {Blomberg}}, \bibinfo {author} {\bibfnamefont {A.}~\bibnamefont {Kreyssig}},
  \bibinfo {author} {\bibfnamefont {M.~G.}\ \bibnamefont {Kim}}, \bibinfo
  {author} {\bibfnamefont {N.}~\bibnamefont {Ni}}, \bibinfo {author}
  {\bibfnamefont {A.}~\bibnamefont {Thaler}}, \bibinfo {author} {\bibfnamefont
  {S.~L.}\ \bibnamefont {Bud'ko}}, \bibinfo {author} {\bibfnamefont {P.~C.}\
  \bibnamefont {Canfield}}, \bibinfo {author} {\bibfnamefont {A.~I.}\
  \bibnamefont {Goldman}}, \bibinfo {author} {\bibfnamefont {I.~I.}\
  \bibnamefont {Mazin}}, \ and\ \bibinfo {author} {\bibfnamefont
  {R.}~\bibnamefont {Prozorov}},\ }\bibfield  {title} {\enquote {\bibinfo
  {title} {{Uniaxial-strain mechanical detwinning of {CaFe}$_{2}${As}$_{2}$ and
  {BaFe}$_{2}${As}$_{2}$ crystals: Optical and transport study}},}\ }\href
  {\doibase 10.1103/PhysRevB.81.184508} {\bibfield  {journal} {\bibinfo
  {journal} {Phys. Rev. B}\ }\textbf {\bibinfo {volume} {81}},\ \bibinfo
  {pages} {184508} (\bibinfo {year} {2010})}\BibitemShut {NoStop}%
\bibitem [{\citenamefont {Blomberg}\ \emph {et~al.}(2011)\citenamefont
  {Blomberg}, \citenamefont {Tanatar}, \citenamefont {Kreyssig}, \citenamefont
  {Ni}, \citenamefont {Thaler}, \citenamefont {Hu}, \citenamefont {Bud'ko},
  \citenamefont {Canfield}, \citenamefont {Goldman},\ and\ \citenamefont
  {Prozorov}}]{blomberg11}%
  \BibitemOpen
  \bibfield  {author} {\bibinfo {author} {\bibfnamefont {E.~C.}\ \bibnamefont
  {Blomberg}}, \bibinfo {author} {\bibfnamefont {M.~A.}\ \bibnamefont
  {Tanatar}}, \bibinfo {author} {\bibfnamefont {A.}~\bibnamefont {Kreyssig}},
  \bibinfo {author} {\bibfnamefont {N.}~\bibnamefont {Ni}}, \bibinfo {author}
  {\bibfnamefont {A.}~\bibnamefont {Thaler}}, \bibinfo {author} {\bibfnamefont
  {Rongwei}\ \bibnamefont {Hu}}, \bibinfo {author} {\bibfnamefont {S.~L.}\
  \bibnamefont {Bud'ko}}, \bibinfo {author} {\bibfnamefont {P.~C.}\
  \bibnamefont {Canfield}}, \bibinfo {author} {\bibfnamefont {A.~I.}\
  \bibnamefont {Goldman}}, \ and\ \bibinfo {author} {\bibfnamefont
  {R.}~\bibnamefont {Prozorov}},\ }\bibfield  {title} {\enquote {\bibinfo
  {title} {In-plane anisotropy of electrical resistivity in strain-detwinned
  {SrFe}$_{2}${As}$_{2}$},}\ }\href {\doibase 10.1103/PhysRevB.83.134505}
  {\bibfield  {journal} {\bibinfo  {journal} {Phys. Rev. B}\ }\textbf {\bibinfo
  {volume} {83}},\ \bibinfo {pages} {134505} (\bibinfo {year}
  {2011})}\BibitemShut {NoStop}%
\bibitem [{\citenamefont {Hu}\ \emph {et~al.}(2008)\citenamefont {Hu},
  \citenamefont {Dong}, \citenamefont {Li}, \citenamefont {Li}, \citenamefont
  {Zheng}, \citenamefont {Chen}, \citenamefont {Luo},\ and\ \citenamefont
  {Wang}}]{hu08}%
  \BibitemOpen
  \bibfield  {author} {\bibinfo {author} {\bibfnamefont {W.~Z.}\ \bibnamefont
  {Hu}}, \bibinfo {author} {\bibfnamefont {J.}~\bibnamefont {Dong}}, \bibinfo
  {author} {\bibfnamefont {G.}~\bibnamefont {Li}}, \bibinfo {author}
  {\bibfnamefont {Z.}~\bibnamefont {Li}}, \bibinfo {author} {\bibfnamefont
  {P.}~\bibnamefont {Zheng}}, \bibinfo {author} {\bibfnamefont {G.~F.}\
  \bibnamefont {Chen}}, \bibinfo {author} {\bibfnamefont {J.~L.}\ \bibnamefont
  {Luo}}, \ and\ \bibinfo {author} {\bibfnamefont {N.~L.}\ \bibnamefont
  {Wang}},\ }\bibfield  {title} {\enquote {\bibinfo {title} {{Origin of the
  Spin Density Wave Instability in $A${Fe}$_2${As}$_2$ ($A=\,$Ba, Sr) as
  Revealed by Optical Spectroscopy}},}\ }\href {\doibase
  10.1103/PhysRevLett.101.257005} {\bibfield  {journal} {\bibinfo  {journal}
  {Phys. Rev. Lett.}\ }\textbf {\bibinfo {volume} {101}},\ \bibinfo {pages}
  {257005} (\bibinfo {year} {2008})}\BibitemShut {NoStop}%
\bibitem [{\citenamefont {Akrap}\ \emph {et~al.}(2009)\citenamefont {Akrap},
  \citenamefont {Tu}, \citenamefont {Li}, \citenamefont {Cao}, \citenamefont
  {Xu},\ and\ \citenamefont {Homes}}]{akrap09}%
  \BibitemOpen
  \bibfield  {author} {\bibinfo {author} {\bibfnamefont {A.}~\bibnamefont
  {Akrap}}, \bibinfo {author} {\bibfnamefont {J.~J.}\ \bibnamefont {Tu}},
  \bibinfo {author} {\bibfnamefont {L.~J.}\ \bibnamefont {Li}}, \bibinfo
  {author} {\bibfnamefont {G.~H.}\ \bibnamefont {Cao}}, \bibinfo {author}
  {\bibfnamefont {Z.~A.}\ \bibnamefont {Xu}}, \ and\ \bibinfo {author}
  {\bibfnamefont {C.~C.}\ \bibnamefont {Homes}},\ }\bibfield  {title} {\enquote
  {\bibinfo {title} {{Infrared phonon anomaly in {BaFe}$_{2}${As}$_{2}$}},}\
  }\href {\doibase 10.1103/PhysRevB.80.180502} {\bibfield  {journal} {\bibinfo
  {journal} {Phys. Rev. B}\ }\textbf {\bibinfo {volume} {80}},\ \bibinfo
  {pages} {180502} (\bibinfo {year} {2009})}\BibitemShut {NoStop}%
\bibitem [{\citenamefont {Pfuner}\ \emph {et~al.}(2009)\citenamefont {Pfuner},
  \citenamefont {Analytis}, \citenamefont {Chu}, \citenamefont {Fisher},\ and\
  \citenamefont {Degiorgi}}]{pfuner09}%
  \BibitemOpen
  \bibfield  {author} {\bibinfo {author} {\bibfnamefont {F.}~\bibnamefont
  {Pfuner}}, \bibinfo {author} {\bibfnamefont {J.~G.}\ \bibnamefont
  {Analytis}}, \bibinfo {author} {\bibfnamefont {J.-H.}\ \bibnamefont {Chu}},
  \bibinfo {author} {\bibfnamefont {I.~R.}\ \bibnamefont {Fisher}}, \ and\
  \bibinfo {author} {\bibfnamefont {L.}~\bibnamefont {Degiorgi}},\ }\bibfield
  {title} {\enquote {\bibinfo {title} {Charge dynamics of the spin-density-wave
  state in {BaFe}$_2${As}$_2$},}\ }\href {\doibase 10.1140/epjb/e2009-00062-2}
  {\bibfield  {journal} {\bibinfo  {journal} {Eur. Phys. J. B}\ }\textbf
  {\bibinfo {volume} {67}},\ \bibinfo {pages} {513--517} (\bibinfo {year}
  {2009})}\BibitemShut {NoStop}%
\bibitem [{\citenamefont {Chen}\ \emph {et~al.}(2010)\citenamefont {Chen},
  \citenamefont {Dong}, \citenamefont {Ruan}, \citenamefont {Hu}, \citenamefont
  {Cheng}, \citenamefont {Hu}, \citenamefont {Zheng}, \citenamefont {Fang},
  \citenamefont {Dai},\ and\ \citenamefont {Wang}}]{chen10}%
  \BibitemOpen
  \bibfield  {author} {\bibinfo {author} {\bibfnamefont {Z.~G.}\ \bibnamefont
  {Chen}}, \bibinfo {author} {\bibfnamefont {T.}~\bibnamefont {Dong}}, \bibinfo
  {author} {\bibfnamefont {R.~H.}\ \bibnamefont {Ruan}}, \bibinfo {author}
  {\bibfnamefont {B.~F.}\ \bibnamefont {Hu}}, \bibinfo {author} {\bibfnamefont
  {B.}~\bibnamefont {Cheng}}, \bibinfo {author} {\bibfnamefont {W.~Z.}\
  \bibnamefont {Hu}}, \bibinfo {author} {\bibfnamefont {P.}~\bibnamefont
  {Zheng}}, \bibinfo {author} {\bibfnamefont {Z.}~\bibnamefont {Fang}},
  \bibinfo {author} {\bibfnamefont {X.}~\bibnamefont {Dai}}, \ and\ \bibinfo
  {author} {\bibfnamefont {N.~L.}\ \bibnamefont {Wang}},\ }\bibfield  {title}
  {\enquote {\bibinfo {title} {{Measurement of the $c$-Axis Optical Reflectance
  of $A${Fe}$_{2}${As}$_{2}$ ($A=\,${Ba}, {Sr}) Single Crystals: Evidence of
  Different Mechanisms for the Formation of Two Energy Gaps}},}\ }\href
  {\doibase 10.1103/PhysRevLett.105.097003} {\bibfield  {journal} {\bibinfo
  {journal} {Phys. Rev. Lett.}\ }\textbf {\bibinfo {volume} {105}},\ \bibinfo
  {pages} {097003} (\bibinfo {year} {2010})}\BibitemShut {NoStop}%
\bibitem [{\citenamefont {Schafgans}\ \emph {et~al.}(2011)\citenamefont
  {Schafgans}, \citenamefont {Pursley}, \citenamefont {LaForge}, \citenamefont
  {Sefat}, \citenamefont {Mandrus},\ and\ \citenamefont {Basov}}]{schafgans11}%
  \BibitemOpen
  \bibfield  {author} {\bibinfo {author} {\bibfnamefont {A.~A.}\ \bibnamefont
  {Schafgans}}, \bibinfo {author} {\bibfnamefont {B.~C.}\ \bibnamefont
  {Pursley}}, \bibinfo {author} {\bibfnamefont {A.~D.}\ \bibnamefont
  {LaForge}}, \bibinfo {author} {\bibfnamefont {A.~S.}\ \bibnamefont {Sefat}},
  \bibinfo {author} {\bibfnamefont {D.}~\bibnamefont {Mandrus}}, \ and\
  \bibinfo {author} {\bibfnamefont {D.~N.}\ \bibnamefont {Basov}},\ }\bibfield
  {title} {\enquote {\bibinfo {title} {Phonon splitting and anomalous
  enhancement of infrared-active modes in {BaFe}$_{2}${As}$_{2}$},}\ }\href
  {\doibase 10.1103/PhysRevB.84.052501} {\bibfield  {journal} {\bibinfo
  {journal} {Phys. Rev. B}\ }\textbf {\bibinfo {volume} {84}},\ \bibinfo
  {pages} {052501} (\bibinfo {year} {2011})}\BibitemShut {NoStop}%
\bibitem [{\citenamefont {Moon}\ \emph {et~al.}(2012)\citenamefont {Moon},
  \citenamefont {Schafgans}, \citenamefont {Kasahara}, \citenamefont
  {Shibauchi}, \citenamefont {Terashima}, \citenamefont {Matsuda},
  \citenamefont {Tanatar}, \citenamefont {Prozorov}, \citenamefont {Thaler},
  \citenamefont {Canfield}, \citenamefont {Sefat}, \citenamefont {Mandrus},\
  and\ \citenamefont {Basov}}]{moon12}%
  \BibitemOpen
  \bibfield  {author} {\bibinfo {author} {\bibfnamefont {S.~J.}\ \bibnamefont
  {Moon}}, \bibinfo {author} {\bibfnamefont {A.~A.}\ \bibnamefont {Schafgans}},
  \bibinfo {author} {\bibfnamefont {S.}~\bibnamefont {Kasahara}}, \bibinfo
  {author} {\bibfnamefont {T.}~\bibnamefont {Shibauchi}}, \bibinfo {author}
  {\bibfnamefont {T.}~\bibnamefont {Terashima}}, \bibinfo {author}
  {\bibfnamefont {Y.}~\bibnamefont {Matsuda}}, \bibinfo {author} {\bibfnamefont
  {M.~A.}\ \bibnamefont {Tanatar}}, \bibinfo {author} {\bibfnamefont
  {R.}~\bibnamefont {Prozorov}}, \bibinfo {author} {\bibfnamefont
  {A.}~\bibnamefont {Thaler}}, \bibinfo {author} {\bibfnamefont {P.~C.}\
  \bibnamefont {Canfield}}, \bibinfo {author} {\bibfnamefont {A.~S.}\
  \bibnamefont {Sefat}}, \bibinfo {author} {\bibfnamefont {D.}~\bibnamefont
  {Mandrus}}, \ and\ \bibinfo {author} {\bibfnamefont {D.~N.}\ \bibnamefont
  {Basov}},\ }\bibfield  {title} {\enquote {\bibinfo {title} {{Infrared
  Measurement of the Pseudogap of P-Doped and Co-Doped High-Temperature
  {BaFe}$_{2}${As}$_{2}$ Superconductors}},}\ }\href {\doibase
  10.1103/PhysRevLett.109.027006} {\bibfield  {journal} {\bibinfo  {journal}
  {Phys. Rev. Lett.}\ }\textbf {\bibinfo {volume} {109}},\ \bibinfo {pages}
  {027006} (\bibinfo {year} {2012})}\BibitemShut {NoStop}%
\bibitem [{\citenamefont {Nakajima}\ \emph {et~al.}(2011)\citenamefont
  {Nakajima}, \citenamefont {Liang}, \citenamefont {Ishida}, \citenamefont
  {Tomioka}, \citenamefont {Kihou}, \citenamefont {Lee}, \citenamefont {Iyo},
  \citenamefont {Eisaki}, \citenamefont {Kakeshita}, \citenamefont {Ito},\ and\
  \citenamefont {Uchida}}]{nakajima11}%
  \BibitemOpen
  \bibfield  {author} {\bibinfo {author} {\bibfnamefont {M.}~\bibnamefont
  {Nakajima}}, \bibinfo {author} {\bibfnamefont {T.}~\bibnamefont {Liang}},
  \bibinfo {author} {\bibfnamefont {S.}~\bibnamefont {Ishida}}, \bibinfo
  {author} {\bibfnamefont {Y.}~\bibnamefont {Tomioka}}, \bibinfo {author}
  {\bibfnamefont {K.}~\bibnamefont {Kihou}}, \bibinfo {author} {\bibfnamefont
  {C.~H.}\ \bibnamefont {Lee}}, \bibinfo {author} {\bibfnamefont
  {A.}~\bibnamefont {Iyo}}, \bibinfo {author} {\bibfnamefont {H.}~\bibnamefont
  {Eisaki}}, \bibinfo {author} {\bibfnamefont {T.}~\bibnamefont {Kakeshita}},
  \bibinfo {author} {\bibfnamefont {T.}~\bibnamefont {Ito}}, \ and\ \bibinfo
  {author} {\bibfnamefont {S.}~\bibnamefont {Uchida}},\ }\bibfield  {title}
  {\enquote {\bibinfo {title} {Unprecedented anisotropic metallic state in
  undoped iron arsenide {BaFe}$_2${As}$_2$ revealed by optical spectroscopy},}\
  }\href {\doibase 10.1073/pnas.1100102108} {\bibfield  {journal} {\bibinfo
  {journal} {PNAS}\ }\textbf {\bibinfo {volume} {108}},\ \bibinfo {pages}
  {12238--–12242} (\bibinfo {year} {2011})}\BibitemShut {NoStop}%
\bibitem [{\citenamefont {Dusza}\ \emph {et~al.}(2011)\citenamefont {Dusza},
  \citenamefont {Lucarelli}, \citenamefont {Pfuner}, \citenamefont {Chu},
  \citenamefont {Fisher},\ and\ \citenamefont {Degiorgi}}]{dusza11}%
  \BibitemOpen
  \bibfield  {author} {\bibinfo {author} {\bibfnamefont {A.}~\bibnamefont
  {Dusza}}, \bibinfo {author} {\bibfnamefont {A.}~\bibnamefont {Lucarelli}},
  \bibinfo {author} {\bibfnamefont {F.}~\bibnamefont {Pfuner}}, \bibinfo
  {author} {\bibfnamefont {J.-H.}\ \bibnamefont {Chu}}, \bibinfo {author}
  {\bibfnamefont {I.~R.}\ \bibnamefont {Fisher}}, \ and\ \bibinfo {author}
  {\bibfnamefont {L.}~\bibnamefont {Degiorgi}},\ }\bibfield  {title} {\enquote
  {\bibinfo {title} {{Anisotropic charge dynamics in detwinned
  Ba(Fe$_{1-x}$Co$_x$)$_2$As$_2$}},}\ }\href {\doibase
  10.1209/0295-5075/93/37002} {\bibfield  {journal} {\bibinfo  {journal} {EPL}\
  }\textbf {\bibinfo {volume} {93}},\ \bibinfo {pages} {37002} (\bibinfo {year}
  {2011})}\BibitemShut {NoStop}%
\bibitem [{\citenamefont {Charnukha}\ \emph {et~al.}(2013)\citenamefont
  {Charnukha}, \citenamefont {Pr\"opper}, \citenamefont {Larkin}, \citenamefont
  {Sun}, \citenamefont {Li}, \citenamefont {Lin}, \citenamefont {Wolf},
  \citenamefont {Keimer},\ and\ \citenamefont {Boris}}]{charnukha13}%
  \BibitemOpen
  \bibfield  {author} {\bibinfo {author} {\bibfnamefont {A.}~\bibnamefont
  {Charnukha}}, \bibinfo {author} {\bibfnamefont {D.}~\bibnamefont
  {Pr\"opper}}, \bibinfo {author} {\bibfnamefont {T.~I.}\ \bibnamefont
  {Larkin}}, \bibinfo {author} {\bibfnamefont {D.~L.}\ \bibnamefont {Sun}},
  \bibinfo {author} {\bibfnamefont {Z.~W.}\ \bibnamefont {Li}}, \bibinfo
  {author} {\bibfnamefont {C.~T.}\ \bibnamefont {Lin}}, \bibinfo {author}
  {\bibfnamefont {T.}~\bibnamefont {Wolf}}, \bibinfo {author} {\bibfnamefont
  {B.}~\bibnamefont {Keimer}}, \ and\ \bibinfo {author} {\bibfnamefont {A.~V.}\
  \bibnamefont {Boris}},\ }\bibfield  {title} {\enquote {\bibinfo {title}
  {{Spin-density-wave-induced anomalies in the optical conductivity of
  $A${Fe}$_{2}${As}$_{2}$, ($A=\text{Ca}$, Sr, Ba) single-crystalline iron
  pnictides}},}\ }\href {\doibase 10.1103/PhysRevB.88.184511} {\bibfield
  {journal} {\bibinfo  {journal} {Phys. Rev. B}\ }\textbf {\bibinfo {volume}
  {88}},\ \bibinfo {pages} {184511} (\bibinfo {year} {2013})}\BibitemShut
  {NoStop}%
\bibitem [{\citenamefont {Nakajima}\ \emph {et~al.}(2014)\citenamefont
  {Nakajima}, \citenamefont {Ishida}, \citenamefont {Tanaka}, \citenamefont
  {Kihou}, \citenamefont {Tomioka}, \citenamefont {Saito}, \citenamefont {Lee},
  \citenamefont {Fukazawa}, \citenamefont {Kohori}, \citenamefont {Kakeshita},
  \citenamefont {Iyo}, \citenamefont {Ito}, \citenamefont {Eisaki},\ and\
  \citenamefont {Uchida}}]{nakajima14}%
  \BibitemOpen
  \bibfield  {author} {\bibinfo {author} {\bibfnamefont {M.}~\bibnamefont
  {Nakajima}}, \bibinfo {author} {\bibfnamefont {S.}~\bibnamefont {Ishida}},
  \bibinfo {author} {\bibfnamefont {T.}~\bibnamefont {Tanaka}}, \bibinfo
  {author} {\bibfnamefont {K.}~\bibnamefont {Kihou}}, \bibinfo {author}
  {\bibfnamefont {Y.}~\bibnamefont {Tomioka}}, \bibinfo {author} {\bibfnamefont
  {T.}~\bibnamefont {Saito}}, \bibinfo {author} {\bibfnamefont {C.~H.}\
  \bibnamefont {Lee}}, \bibinfo {author} {\bibfnamefont {H.}~\bibnamefont
  {Fukazawa}}, \bibinfo {author} {\bibfnamefont {Y.}~\bibnamefont {Kohori}},
  \bibinfo {author} {\bibfnamefont {T.}~\bibnamefont {Kakeshita}}, \bibinfo
  {author} {\bibfnamefont {A.}~\bibnamefont {Iyo}}, \bibinfo {author}
  {\bibfnamefont {T.}~\bibnamefont {Ito}}, \bibinfo {author} {\bibfnamefont
  {H.}~\bibnamefont {Eisaki}}, \ and\ \bibinfo {author} {\bibfnamefont
  {S.}~\bibnamefont {Uchida}},\ }\bibfield  {title} {\enquote {\bibinfo {title}
  {Normal-state charge dynamics in doped {BaFe}$_2${As}$_2$: Roles of doping
  and necessary ingredients for superconductivity},}\ }\href {\doibase
  10.1038/srep05873} {\bibfield  {journal} {\bibinfo  {journal} {Sci. Rep.}\
  }\textbf {\bibinfo {volume} {4}},\ \bibinfo {pages} {5873} (\bibinfo {year}
  {2014})}\BibitemShut {NoStop}%
\bibitem [{\citenamefont {Wang}\ \emph {et~al.}(2014)\citenamefont {Wang},
  \citenamefont {Wang}, \citenamefont {Dong}, \citenamefont {Chen},\ and\
  \citenamefont {Wang}}]{wang14}%
  \BibitemOpen
  \bibfield  {author} {\bibinfo {author} {\bibfnamefont {X.~B.}\ \bibnamefont
  {Wang}}, \bibinfo {author} {\bibfnamefont {H.~P.}\ \bibnamefont {Wang}},
  \bibinfo {author} {\bibfnamefont {T.}~\bibnamefont {Dong}}, \bibinfo {author}
  {\bibfnamefont {R.~Y.}\ \bibnamefont {Chen}}, \ and\ \bibinfo {author}
  {\bibfnamefont {N.~L.}\ \bibnamefont {Wang}},\ }\bibfield  {title} {\enquote
  {\bibinfo {title} {Optical spectroscopy study of the collapsed tetragonal
  phase of {CaFe}$_{2}(${As}$_{0.935}${P}$_{0.065})_{2}$ single crystals},}\
  }\href {\doibase 10.1103/PhysRevB.90.144513} {\bibfield  {journal} {\bibinfo
  {journal} {Phys. Rev. B}\ }\textbf {\bibinfo {volume} {90}},\ \bibinfo
  {pages} {144513} (\bibinfo {year} {2014})}\BibitemShut {NoStop}%
\bibitem [{\citenamefont {Singh}(2008)}]{singh08}%
  \BibitemOpen
  \bibfield  {author} {\bibinfo {author} {\bibfnamefont {D.~J.}\ \bibnamefont
  {Singh}},\ }\bibfield  {title} {\enquote {\bibinfo {title} {{Electronic
  structure and doping in BaFe$_2$As$_2$ and LiFeAs: Density functional
  calculations}},}\ }\href {\doibase 10.1103/PhysRevB.78.094511} {\bibfield
  {journal} {\bibinfo  {journal} {Phys. Rev. B}\ }\textbf {\bibinfo {volume}
  {78}},\ \bibinfo {pages} {094511} (\bibinfo {year} {2008})}\BibitemShut
  {NoStop}%
\bibitem [{\citenamefont {Fink}\ \emph {et~al.}(2009)\citenamefont {Fink},
  \citenamefont {Thirupathaiah}, \citenamefont {Ovsyannikov}, \citenamefont
  {D{\"u}rr}, \citenamefont {Follath}, \citenamefont {Huang}, \citenamefont
  {de~Jong}, \citenamefont {Golden}, \citenamefont {Zhang}, \citenamefont
  {Jeschke}, \citenamefont {Valent\'{i}}, \citenamefont {Felser}, \citenamefont
  {Dastjani~Farahani}, \citenamefont {Rotter},\ and\ \citenamefont
  {Johrendt}}]{fink09}%
  \BibitemOpen
  \bibfield  {author} {\bibinfo {author} {\bibfnamefont {J.}~\bibnamefont
  {Fink}}, \bibinfo {author} {\bibfnamefont {S.}~\bibnamefont {Thirupathaiah}},
  \bibinfo {author} {\bibfnamefont {R.}~\bibnamefont {Ovsyannikov}}, \bibinfo
  {author} {\bibfnamefont {H.~A.}\ \bibnamefont {D{\"u}rr}}, \bibinfo {author}
  {\bibfnamefont {R.}~\bibnamefont {Follath}}, \bibinfo {author} {\bibfnamefont
  {Y.}~\bibnamefont {Huang}}, \bibinfo {author} {\bibfnamefont
  {S.}~\bibnamefont {de~Jong}}, \bibinfo {author} {\bibfnamefont {M.~S.}\
  \bibnamefont {Golden}}, \bibinfo {author} {\bibfnamefont {Yu-Zhong}\
  \bibnamefont {Zhang}}, \bibinfo {author} {\bibfnamefont {H.~O.}\ \bibnamefont
  {Jeschke}}, \bibinfo {author} {\bibfnamefont {R.}~\bibnamefont
  {Valent\'{i}}}, \bibinfo {author} {\bibfnamefont {C.}~\bibnamefont {Felser}},
  \bibinfo {author} {\bibfnamefont {S.}~\bibnamefont {Dastjani~Farahani}},
  \bibinfo {author} {\bibfnamefont {M.}~\bibnamefont {Rotter}}, \ and\ \bibinfo
  {author} {\bibfnamefont {D.}~\bibnamefont {Johrendt}},\ }\bibfield  {title}
  {\enquote {\bibinfo {title} {Electronic structure studies of
  {BaFe}$_2${As}$_2$ by angle-resolved photoemission spectroscopy},}\ }\href
  {\doibase 10.1103/PhysRevB.79.155118} {\bibfield  {journal} {\bibinfo
  {journal} {Phys. Rev. B}\ }\textbf {\bibinfo {volume} {79}},\ \bibinfo
  {pages} {155118} (\bibinfo {year} {2009})}\BibitemShut {NoStop}%
\bibitem [{\citenamefont {Wu}\ \emph {et~al.}(2010)\citenamefont {Wu},
  \citenamefont {Bari{\v{s}}i{\'{c}}}, \citenamefont {Kallina}, \citenamefont
  {Faridian}, \citenamefont {Gorshunov}, \citenamefont {Drichko}, \citenamefont
  {Li}, \citenamefont {Lin}, \citenamefont {Cao}, \citenamefont {Xu},
  \citenamefont {Wang},\ and\ \citenamefont {Dressel}}]{wu10a}%
  \BibitemOpen
  \bibfield  {author} {\bibinfo {author} {\bibfnamefont {D.}~\bibnamefont
  {Wu}}, \bibinfo {author} {\bibfnamefont {N.}~\bibnamefont
  {Bari{\v{s}}i{\'{c}}}}, \bibinfo {author} {\bibfnamefont {P.}~\bibnamefont
  {Kallina}}, \bibinfo {author} {\bibfnamefont {A.}~\bibnamefont {Faridian}},
  \bibinfo {author} {\bibfnamefont {B.}~\bibnamefont {Gorshunov}}, \bibinfo
  {author} {\bibfnamefont {N.}~\bibnamefont {Drichko}}, \bibinfo {author}
  {\bibfnamefont {L.~J.}\ \bibnamefont {Li}}, \bibinfo {author} {\bibfnamefont
  {X.}~\bibnamefont {Lin}}, \bibinfo {author} {\bibfnamefont {G.~H.}\
  \bibnamefont {Cao}}, \bibinfo {author} {\bibfnamefont {Z.~A.}\ \bibnamefont
  {Xu}}, \bibinfo {author} {\bibfnamefont {N.~L.}\ \bibnamefont {Wang}}, \ and\
  \bibinfo {author} {\bibfnamefont {M.}~\bibnamefont {Dressel}},\ }\bibfield
  {title} {\enquote {\bibinfo {title} {Optical investigations of the normal and
  superconducting states reveal two electronic subsystems in iron pnictides},}\
  }\href {\doibase 10.1103/PhysRevB.81.100512} {\bibfield  {journal} {\bibinfo
  {journal} {Phys. Rev. B}\ }\textbf {\bibinfo {volume} {81}},\ \bibinfo
  {pages} {100512(R)} (\bibinfo {year} {2010})}\BibitemShut {NoStop}%
\bibitem [{\citenamefont {Yi}\ \emph {et~al.}(2009)\citenamefont {Yi},
  \citenamefont {Lu}, \citenamefont {Analytis}, \citenamefont {Chu},
  \citenamefont {Mo}, \citenamefont {He}, \citenamefont {Hashimoto},
  \citenamefont {Moore}, \citenamefont {Mazin}, \citenamefont {Singh},
  \citenamefont {Hussain}, \citenamefont {Fisher},\ and\ \citenamefont
  {Shen}}]{yi09}%
  \BibitemOpen
  \bibfield  {author} {\bibinfo {author} {\bibfnamefont {M.}~\bibnamefont
  {Yi}}, \bibinfo {author} {\bibfnamefont {D.~H.}\ \bibnamefont {Lu}}, \bibinfo
  {author} {\bibfnamefont {J.~G.}\ \bibnamefont {Analytis}}, \bibinfo {author}
  {\bibfnamefont {J.-H.}\ \bibnamefont {Chu}}, \bibinfo {author} {\bibfnamefont
  {S.-K.}\ \bibnamefont {Mo}}, \bibinfo {author} {\bibfnamefont {R.-H.}\
  \bibnamefont {He}}, \bibinfo {author} {\bibfnamefont {M.}~\bibnamefont
  {Hashimoto}}, \bibinfo {author} {\bibfnamefont {R.~G.}\ \bibnamefont
  {Moore}}, \bibinfo {author} {\bibfnamefont {I.~I.}\ \bibnamefont {Mazin}},
  \bibinfo {author} {\bibfnamefont {D.~J.}\ \bibnamefont {Singh}}, \bibinfo
  {author} {\bibfnamefont {Z.}~\bibnamefont {Hussain}}, \bibinfo {author}
  {\bibfnamefont {I.~R.}\ \bibnamefont {Fisher}}, \ and\ \bibinfo {author}
  {\bibfnamefont {Z.-X.}\ \bibnamefont {Shen}},\ }\bibfield  {title} {\enquote
  {\bibinfo {title} {{Unconventional electronic reconstruction in undoped
  (Ba,Sr)Fe$_2$As$_2$ across the spin density wave transition}},}\ }\href
  {\doibase 10.1103/PhysRevB.80.174510} {\bibfield  {journal} {\bibinfo
  {journal} {Phys. Rev. B}\ }\textbf {\bibinfo {volume} {80}},\ \bibinfo
  {pages} {174510} (\bibinfo {year} {2009})}\BibitemShut {NoStop}%
\bibitem [{\citenamefont {Yang}\ \emph {et~al.}(2009)\citenamefont {Yang},
  \citenamefont {Zhang}, \citenamefont {Ou}, \citenamefont {Zhao},
  \citenamefont {Shen}, \citenamefont {Zhou}, \citenamefont {Wei},
  \citenamefont {Chen}, \citenamefont {Xu}, \citenamefont {He}, \citenamefont
  {Chen}, \citenamefont {Wang}, \citenamefont {Wang}, \citenamefont {Wu},
  \citenamefont {Wu}, \citenamefont {Chen}, \citenamefont {Arita},
  \citenamefont {Shimada}, \citenamefont {Taniguchi}, \citenamefont {Lu},
  \citenamefont {Xiang},\ and\ \citenamefont {Feng}}]{yang09b}%
  \BibitemOpen
  \bibfield  {author} {\bibinfo {author} {\bibfnamefont {L.~X.}\ \bibnamefont
  {Yang}}, \bibinfo {author} {\bibfnamefont {Y.}~\bibnamefont {Zhang}},
  \bibinfo {author} {\bibfnamefont {H.~W.}\ \bibnamefont {Ou}}, \bibinfo
  {author} {\bibfnamefont {J.~F.}\ \bibnamefont {Zhao}}, \bibinfo {author}
  {\bibfnamefont {D.~W.}\ \bibnamefont {Shen}}, \bibinfo {author}
  {\bibfnamefont {B.}~\bibnamefont {Zhou}}, \bibinfo {author} {\bibfnamefont
  {J.}~\bibnamefont {Wei}}, \bibinfo {author} {\bibfnamefont {F.}~\bibnamefont
  {Chen}}, \bibinfo {author} {\bibfnamefont {M.}~\bibnamefont {Xu}}, \bibinfo
  {author} {\bibfnamefont {C.}~\bibnamefont {He}}, \bibinfo {author}
  {\bibfnamefont {Y.}~\bibnamefont {Chen}}, \bibinfo {author} {\bibfnamefont
  {Z.~D.}\ \bibnamefont {Wang}}, \bibinfo {author} {\bibfnamefont {X.~F.}\
  \bibnamefont {Wang}}, \bibinfo {author} {\bibfnamefont {T.}~\bibnamefont
  {Wu}}, \bibinfo {author} {\bibfnamefont {G.}~\bibnamefont {Wu}}, \bibinfo
  {author} {\bibfnamefont {X.~H.}\ \bibnamefont {Chen}}, \bibinfo {author}
  {\bibfnamefont {M.}~\bibnamefont {Arita}}, \bibinfo {author} {\bibfnamefont
  {K.}~\bibnamefont {Shimada}}, \bibinfo {author} {\bibfnamefont
  {M.}~\bibnamefont {Taniguchi}}, \bibinfo {author} {\bibfnamefont {Z.~Y.}\
  \bibnamefont {Lu}}, \bibinfo {author} {\bibfnamefont {T.}~\bibnamefont
  {Xiang}}, \ and\ \bibinfo {author} {\bibfnamefont {D.~L.}\ \bibnamefont
  {Feng}},\ }\bibfield  {title} {\enquote {\bibinfo {title} {{Electronic
  Structure and Unusual Exchange Splitting in the Spin-Density-Wave State of
  the {BaFe}$_{2}${As}$_{2}$ Parent Compound of Iron-Based Superconductors}},}\
  }\href {\doibase 10.1103/PhysRevLett.102.107002} {\bibfield  {journal}
  {\bibinfo  {journal} {Phys. Rev. Lett.}\ }\textbf {\bibinfo {volume} {102}},\
  \bibinfo {pages} {107002} (\bibinfo {year} {2009})}\BibitemShut {NoStop}%
\bibitem [{\citenamefont {Richard}\ \emph {et~al.}(2010)\citenamefont
  {Richard}, \citenamefont {Nakayama}, \citenamefont {Sato}, \citenamefont
  {Neupane}, \citenamefont {Xu}, \citenamefont {Bowen}, \citenamefont {Chen},
  \citenamefont {Luo}, \citenamefont {Wang}, \citenamefont {Dai}, \citenamefont
  {Fang}, \citenamefont {Ding},\ and\ \citenamefont {Takahashi}}]{richard10}%
  \BibitemOpen
  \bibfield  {author} {\bibinfo {author} {\bibfnamefont {P.}~\bibnamefont
  {Richard}}, \bibinfo {author} {\bibfnamefont {K.}~\bibnamefont {Nakayama}},
  \bibinfo {author} {\bibfnamefont {T.}~\bibnamefont {Sato}}, \bibinfo {author}
  {\bibfnamefont {M.}~\bibnamefont {Neupane}}, \bibinfo {author} {\bibfnamefont
  {Y.-M.}\ \bibnamefont {Xu}}, \bibinfo {author} {\bibfnamefont {J.~H.}\
  \bibnamefont {Bowen}}, \bibinfo {author} {\bibfnamefont {G.~F.}\ \bibnamefont
  {Chen}}, \bibinfo {author} {\bibfnamefont {J.~L.}\ \bibnamefont {Luo}},
  \bibinfo {author} {\bibfnamefont {N.~L.}\ \bibnamefont {Wang}}, \bibinfo
  {author} {\bibfnamefont {X.}~\bibnamefont {Dai}}, \bibinfo {author}
  {\bibfnamefont {Z.}~\bibnamefont {Fang}}, \bibinfo {author} {\bibfnamefont
  {H.}~\bibnamefont {Ding}}, \ and\ \bibinfo {author} {\bibfnamefont
  {T.}~\bibnamefont {Takahashi}},\ }\bibfield  {title} {\enquote {\bibinfo
  {title} {{Observation of Dirac Cone Electronic Dispersion in
  {BaFe}$_{2}${As}$_{2}$}},}\ }\href {\doibase 10.1103/PhysRevLett.104.137001}
  {\bibfield  {journal} {\bibinfo  {journal} {Phys. Rev. Lett.}\ }\textbf
  {\bibinfo {volume} {104}},\ \bibinfo {pages} {137001} (\bibinfo {year}
  {2010})}\BibitemShut {NoStop}%
\bibitem [{\citenamefont {Shimojima}\ \emph {et~al.}(2010)\citenamefont
  {Shimojima}, \citenamefont {Ishizaka}, \citenamefont {Ishida}, \citenamefont
  {Katayama}, \citenamefont {Ohgushi}, \citenamefont {Kiss}, \citenamefont
  {Okawa}, \citenamefont {Togashi}, \citenamefont {Wang}, \citenamefont {Chen},
  \citenamefont {Watanabe}, \citenamefont {Kadota}, \citenamefont {Oguchi},
  \citenamefont {Chainani},\ and\ \citenamefont {Shin}}]{shimojima10}%
  \BibitemOpen
  \bibfield  {author} {\bibinfo {author} {\bibfnamefont {T.}~\bibnamefont
  {Shimojima}}, \bibinfo {author} {\bibfnamefont {K.}~\bibnamefont {Ishizaka}},
  \bibinfo {author} {\bibfnamefont {Y.}~\bibnamefont {Ishida}}, \bibinfo
  {author} {\bibfnamefont {N.}~\bibnamefont {Katayama}}, \bibinfo {author}
  {\bibfnamefont {K.}~\bibnamefont {Ohgushi}}, \bibinfo {author} {\bibfnamefont
  {T.}~\bibnamefont {Kiss}}, \bibinfo {author} {\bibfnamefont {M.}~\bibnamefont
  {Okawa}}, \bibinfo {author} {\bibfnamefont {T.}~\bibnamefont {Togashi}},
  \bibinfo {author} {\bibfnamefont {X.-Y.}\ \bibnamefont {Wang}}, \bibinfo
  {author} {\bibfnamefont {C.-T.}\ \bibnamefont {Chen}}, \bibinfo {author}
  {\bibfnamefont {S.}~\bibnamefont {Watanabe}}, \bibinfo {author}
  {\bibfnamefont {R.}~\bibnamefont {Kadota}}, \bibinfo {author} {\bibfnamefont
  {T.}~\bibnamefont {Oguchi}}, \bibinfo {author} {\bibfnamefont
  {A.}~\bibnamefont {Chainani}}, \ and\ \bibinfo {author} {\bibfnamefont
  {S.}~\bibnamefont {Shin}},\ }\bibfield  {title} {\enquote {\bibinfo {title}
  {Orbital-dependent modifications of electronic structure across the
  magnetostructural transition in {BaFe}$_2${As}$_2$},}\ }\href {\doibase
  10.1103/PhysRevLett.104.057002} {\bibfield  {journal} {\bibinfo  {journal}
  {Phys. Rev. Lett.}\ }\textbf {\bibinfo {volume} {104}},\ \bibinfo {pages}
  {057002} (\bibinfo {year} {2010})}\BibitemShut {NoStop}%
\bibitem [{\citenamefont {Fuglsang~Jensen}\ \emph {et~al.}(2011)\citenamefont
  {Fuglsang~Jensen}, \citenamefont {Brouet}, \citenamefont {Papalazarou},
  \citenamefont {Nicolaou}, \citenamefont {Taleb-Ibrahimi}, \citenamefont
  {Le~F\`evre}, \citenamefont {Bertran}, \citenamefont {Forget},\ and\
  \citenamefont {Colson}}]{jensen11}%
  \BibitemOpen
  \bibfield  {author} {\bibinfo {author} {\bibfnamefont {M.}~\bibnamefont
  {Fuglsang~Jensen}}, \bibinfo {author} {\bibfnamefont {V.}~\bibnamefont
  {Brouet}}, \bibinfo {author} {\bibfnamefont {E.}~\bibnamefont {Papalazarou}},
  \bibinfo {author} {\bibfnamefont {A.}~\bibnamefont {Nicolaou}}, \bibinfo
  {author} {\bibfnamefont {A.}~\bibnamefont {Taleb-Ibrahimi}}, \bibinfo
  {author} {\bibfnamefont {P.}~\bibnamefont {Le~F\`evre}}, \bibinfo {author}
  {\bibfnamefont {F.}~\bibnamefont {Bertran}}, \bibinfo {author} {\bibfnamefont
  {A.}~\bibnamefont {Forget}}, \ and\ \bibinfo {author} {\bibfnamefont
  {D.}~\bibnamefont {Colson}},\ }\bibfield  {title} {\enquote {\bibinfo {title}
  {Angle-resolved photoemission study of the role of nesting and orbital
  orderings in the antiferromagnetic phase of {BaFe}$_{2}${As}$_{2}$},}\ }\href
  {\doibase 10.1103/PhysRevB.84.014509} {\bibfield  {journal} {\bibinfo
  {journal} {Phys. Rev. B}\ }\textbf {\bibinfo {volume} {84}},\ \bibinfo
  {pages} {014509} (\bibinfo {year} {2011})}\BibitemShut {NoStop}%
\bibitem [{\citenamefont {Dhaka}\ \emph {et~al.}(2014)\citenamefont {Dhaka},
  \citenamefont {Jiang}, \citenamefont {Ran}, \citenamefont {Bud'ko},
  \citenamefont {Canfield}, \citenamefont {Harmon}, \citenamefont {Kaminski},
  \citenamefont {Tomi\ifmmode~\acute{c}\else \'{c}\fi{}}, \citenamefont
  {Valent\'{\i}},\ and\ \citenamefont {Lee}}]{dhaka14}%
  \BibitemOpen
  \bibfield  {author} {\bibinfo {author} {\bibfnamefont {R.~S.}\ \bibnamefont
  {Dhaka}}, \bibinfo {author} {\bibfnamefont {Rui}\ \bibnamefont {Jiang}},
  \bibinfo {author} {\bibfnamefont {S.}~\bibnamefont {Ran}}, \bibinfo {author}
  {\bibfnamefont {S.~L.}\ \bibnamefont {Bud'ko}}, \bibinfo {author}
  {\bibfnamefont {P.~C.}\ \bibnamefont {Canfield}}, \bibinfo {author}
  {\bibfnamefont {B.~N.}\ \bibnamefont {Harmon}}, \bibinfo {author}
  {\bibfnamefont {Adam}\ \bibnamefont {Kaminski}}, \bibinfo {author}
  {\bibfnamefont {Milan}\ \bibnamefont {Tomi\ifmmode~\acute{c}\else
  \'{c}\fi{}}}, \bibinfo {author} {\bibfnamefont {Roser}\ \bibnamefont
  {Valent\'{\i}}}, \ and\ \bibinfo {author} {\bibfnamefont {Yongbin}\
  \bibnamefont {Lee}},\ }\bibfield  {title} {\enquote {\bibinfo {title}
  {Dramatic changes in the electronic structure upon transition to the
  collapsed tetragonal phase in {CaFe}$_{2}${As}$_{2}$},}\ }\href {\doibase
  10.1103/PhysRevB.89.020511} {\bibfield  {journal} {\bibinfo  {journal} {Phys.
  Rev. B}\ }\textbf {\bibinfo {volume} {89}},\ \bibinfo {pages} {020511}
  (\bibinfo {year} {2014})}\BibitemShut {NoStop}%
\bibitem [{\citenamefont {Gofryk}\ \emph {et~al.}(2014)\citenamefont {Gofryk},
  \citenamefont {Saparov}, \citenamefont {Durakiewicz}, \citenamefont
  {Chikina}, \citenamefont {Danzenb\"acher}, \citenamefont {Vyalikh},
  \citenamefont {Graf},\ and\ \citenamefont {Sefat}}]{gofryk14}%
  \BibitemOpen
  \bibfield  {author} {\bibinfo {author} {\bibfnamefont {K.}~\bibnamefont
  {Gofryk}}, \bibinfo {author} {\bibfnamefont {B.}~\bibnamefont {Saparov}},
  \bibinfo {author} {\bibfnamefont {T.}~\bibnamefont {Durakiewicz}}, \bibinfo
  {author} {\bibfnamefont {A.}~\bibnamefont {Chikina}}, \bibinfo {author}
  {\bibfnamefont {S.}~\bibnamefont {Danzenb\"acher}}, \bibinfo {author}
  {\bibfnamefont {D.~V.}\ \bibnamefont {Vyalikh}}, \bibinfo {author}
  {\bibfnamefont {M.~J.}\ \bibnamefont {Graf}}, \ and\ \bibinfo {author}
  {\bibfnamefont {A.~S.}\ \bibnamefont {Sefat}},\ }\bibfield  {title} {\enquote
  {\bibinfo {title} {Fermi-surface reconstruction and complex phase equilibria
  in {CaFe}$_{2}${As}$_{2}$},}\ }\href {\doibase
  10.1103/PhysRevLett.112.186401} {\bibfield  {journal} {\bibinfo  {journal}
  {Phys. Rev. Lett.}\ }\textbf {\bibinfo {volume} {112}},\ \bibinfo {pages}
  {186401} (\bibinfo {year} {2014})}\BibitemShut {NoStop}%
\bibitem [{\citenamefont {Yan}\ \emph {et~al.}(2008)\citenamefont {Yan},
  \citenamefont {Kreyssig}, \citenamefont {Nandi}, \citenamefont {Ni},
  \citenamefont {Bud'ko}, \citenamefont {Kracher}, \citenamefont {McQueeney},
  \citenamefont {McCallum}, \citenamefont {Lograsso}, \citenamefont {Goldman},\
  and\ \citenamefont {Canfield}}]{yan08}%
  \BibitemOpen
  \bibfield  {author} {\bibinfo {author} {\bibfnamefont {J.-Q.}\ \bibnamefont
  {Yan}}, \bibinfo {author} {\bibfnamefont {A.}~\bibnamefont {Kreyssig}},
  \bibinfo {author} {\bibfnamefont {S.}~\bibnamefont {Nandi}}, \bibinfo
  {author} {\bibfnamefont {N.}~\bibnamefont {Ni}}, \bibinfo {author}
  {\bibfnamefont {S.~L.}\ \bibnamefont {Bud'ko}}, \bibinfo {author}
  {\bibfnamefont {A.}~\bibnamefont {Kracher}}, \bibinfo {author} {\bibfnamefont
  {R.~J.}\ \bibnamefont {McQueeney}}, \bibinfo {author} {\bibfnamefont {R.~W.}\
  \bibnamefont {McCallum}}, \bibinfo {author} {\bibfnamefont {T.~A.}\
  \bibnamefont {Lograsso}}, \bibinfo {author} {\bibfnamefont {A.~I.}\
  \bibnamefont {Goldman}}, \ and\ \bibinfo {author} {\bibfnamefont {P.~C.}\
  \bibnamefont {Canfield}},\ }\bibfield  {title} {\enquote {\bibinfo {title}
  {Structural transition and anisotropic properties of single-crystalline
  {SrFe}$_{2}${As}$_{2}$},}\ }\href {\doibase 10.1103/PhysRevB.78.024516}
  {\bibfield  {journal} {\bibinfo  {journal} {Phys. Rev. B}\ }\textbf {\bibinfo
  {volume} {78}},\ \bibinfo {pages} {024516} (\bibinfo {year}
  {2008})}\BibitemShut {NoStop}%
\bibitem [{\citenamefont {Ni}\ \emph {et~al.}(2008{\natexlab{b}})\citenamefont
  {Ni}, \citenamefont {Nandi}, \citenamefont {Kreyssig}, \citenamefont
  {Goldman}, \citenamefont {Mun}, \citenamefont {Bud'ko},\ and\ \citenamefont
  {Canfield}}]{ni08b}%
  \BibitemOpen
  \bibfield  {author} {\bibinfo {author} {\bibfnamefont {N.}~\bibnamefont
  {Ni}}, \bibinfo {author} {\bibfnamefont {S.}~\bibnamefont {Nandi}}, \bibinfo
  {author} {\bibfnamefont {A.}~\bibnamefont {Kreyssig}}, \bibinfo {author}
  {\bibfnamefont {A.~I.}\ \bibnamefont {Goldman}}, \bibinfo {author}
  {\bibfnamefont {E.~D.}\ \bibnamefont {Mun}}, \bibinfo {author} {\bibfnamefont
  {S.~L.}\ \bibnamefont {Bud'ko}}, \ and\ \bibinfo {author} {\bibfnamefont
  {P.~C.}\ \bibnamefont {Canfield}},\ }\bibfield  {title} {\enquote {\bibinfo
  {title} {First-order structural phase transition in
  {CaFe}$_{2}${As}$_{2}$},}\ }\href {\doibase 10.1103/PhysRevB.78.014523}
  {\bibfield  {journal} {\bibinfo  {journal} {Phys. Rev. B}\ }\textbf {\bibinfo
  {volume} {78}},\ \bibinfo {pages} {014523} (\bibinfo {year}
  {2008}{\natexlab{b}})}\BibitemShut {NoStop}%
\bibitem [{\citenamefont {Homes}\ \emph {et~al.}(1993)\citenamefont {Homes},
  \citenamefont {Reedyk}, \citenamefont {Crandles},\ and\ \citenamefont
  {Timusk}}]{homes93}%
  \BibitemOpen
  \bibfield  {author} {\bibinfo {author} {\bibfnamefont {C.~C.}\ \bibnamefont
  {Homes}}, \bibinfo {author} {\bibfnamefont {M.}~\bibnamefont {Reedyk}},
  \bibinfo {author} {\bibfnamefont {D.~A.}\ \bibnamefont {Crandles}}, \ and\
  \bibinfo {author} {\bibfnamefont {T.}~\bibnamefont {Timusk}},\ }\bibfield
  {title} {\enquote {\bibinfo {title} {Technique for measuring the reflectance
  of irregular, submillimeter-sized samples},}\ }\href {\doibase
  10.1364/AO.32.002976} {\bibfield  {journal} {\bibinfo  {journal} {Appl.
  Opt.}\ }\textbf {\bibinfo {volume} {32}},\ \bibinfo {pages} {2976--2983}
  (\bibinfo {year} {1993})}\BibitemShut {NoStop}%
\bibitem [{\citenamefont {Dressel}\ and\ \citenamefont
  {Gr{\"u}ner}(2001)}]{dressel-book}%
  \BibitemOpen
  \bibfield  {author} {\bibinfo {author} {\bibfnamefont {M.}~\bibnamefont
  {Dressel}}\ and\ \bibinfo {author} {\bibfnamefont {G.}~\bibnamefont
  {Gr{\"u}ner}},\ }\href@noop {} {\emph {\bibinfo {title} {Electrodynamics of
  Solids}}}\ (\bibinfo  {publisher} {Cambridge University Press},\ \bibinfo
  {address} {Cambridge},\ \bibinfo {year} {2001})\BibitemShut {NoStop}%
\bibitem [{\citenamefont {Wooten}(1972)}]{wooten}%
  \BibitemOpen
  \bibfield  {author} {\bibinfo {author} {\bibfnamefont {F.}~\bibnamefont
  {Wooten}},\ }\href@noop {} {\emph {\bibinfo {title} {Optical Properties of
  Solids}}}\ (\bibinfo  {publisher} {Academic Press},\ \bibinfo {address} {New
  York},\ \bibinfo {year} {1972})\ pp.\ \bibinfo {pages} {244--250}\BibitemShut
  {NoStop}%
\bibitem [{\citenamefont {Valenzuela}\ \emph {et~al.}(2013)\citenamefont
  {Valenzuela}, \citenamefont {Calder\'on}, \citenamefont {Le\'on},\ and\
  \citenamefont {Bascones}}]{valenzuela13}%
  \BibitemOpen
  \bibfield  {author} {\bibinfo {author} {\bibfnamefont {B.}~\bibnamefont
  {Valenzuela}}, \bibinfo {author} {\bibfnamefont {M.~J.}\ \bibnamefont
  {Calder\'on}}, \bibinfo {author} {\bibfnamefont {G.}~\bibnamefont {Le\'on}},
  \ and\ \bibinfo {author} {\bibfnamefont {E.}~\bibnamefont {Bascones}},\
  }\bibfield  {title} {\enquote {\bibinfo {title} {{Optical conductivity and
  Raman scattering of iron superconductors}},}\ }\href {\doibase
  10.1103/PhysRevB.87.075136} {\bibfield  {journal} {\bibinfo  {journal} {Phys.
  Rev. B}\ }\textbf {\bibinfo {volume} {87}},\ \bibinfo {pages} {075136}
  (\bibinfo {year} {2013})}\BibitemShut {NoStop}%
\bibitem [{\citenamefont {Calder\'on}\ \emph {et~al.}(2014)\citenamefont
  {Calder\'on}, \citenamefont {Medici}, \citenamefont {Valenzuela},\ and\
  \citenamefont {Bascones}}]{calderon14}%
  \BibitemOpen
  \bibfield  {author} {\bibinfo {author} {\bibfnamefont {M.~J.}\ \bibnamefont
  {Calder\'on}}, \bibinfo {author} {\bibfnamefont {L.~de'}\ \bibnamefont
  {Medici}}, \bibinfo {author} {\bibfnamefont {B.}~\bibnamefont {Valenzuela}},
  \ and\ \bibinfo {author} {\bibfnamefont {E.}~\bibnamefont {Bascones}},\
  }\bibfield  {title} {\enquote {\bibinfo {title} {{Correlation, doping, and
  interband effects on the optical conductivity of iron superconductors}},}\
  }\href {\doibase 10.1103/PhysRevB.90.115128} {\bibfield  {journal} {\bibinfo
  {journal} {Phys. Rev. B}\ }\textbf {\bibinfo {volume} {90}},\ \bibinfo
  {pages} {115128} (\bibinfo {year} {2014})}\BibitemShut {NoStop}%
\bibitem [{\citenamefont {Yin}\ \emph {et~al.}(2011)\citenamefont {Yin},
  \citenamefont {Haule},\ and\ \citenamefont {Kotliar}}]{yin11a}%
  \BibitemOpen
  \bibfield  {author} {\bibinfo {author} {\bibfnamefont {Z.~P.}\ \bibnamefont
  {Yin}}, \bibinfo {author} {\bibfnamefont {K.}~\bibnamefont {Haule}}, \ and\
  \bibinfo {author} {\bibfnamefont {G.}~\bibnamefont {Kotliar}},\ }\bibfield
  {title} {\enquote {\bibinfo {title} {Magnetism and charge dynamics in iron
  pnictides},}\ }\href {\doibase 10.1038/nphys1923} {\bibfield  {journal}
  {\bibinfo  {journal} {Nat. Phys.}\ }\textbf {\bibinfo {volume} {7}},\
  \bibinfo {pages} {294--297} (\bibinfo {year} {2011})}\BibitemShut {NoStop}%
\bibitem [{\citenamefont {Chauvi\`ere}\ \emph {et~al.}(2011)\citenamefont
  {Chauvi\`ere}, \citenamefont {Gallais}, \citenamefont {Cazayous},
  \citenamefont {M\'easson}, \citenamefont {Sacuto}, \citenamefont {Colson},\
  and\ \citenamefont {Forget}}]{chauviere11}%
  \BibitemOpen
  \bibfield  {author} {\bibinfo {author} {\bibfnamefont {L.}~\bibnamefont
  {Chauvi\`ere}}, \bibinfo {author} {\bibfnamefont {Y.}~\bibnamefont
  {Gallais}}, \bibinfo {author} {\bibfnamefont {M.}~\bibnamefont {Cazayous}},
  \bibinfo {author} {\bibfnamefont {M.~A.}\ \bibnamefont {M\'easson}}, \bibinfo
  {author} {\bibfnamefont {A.}~\bibnamefont {Sacuto}}, \bibinfo {author}
  {\bibfnamefont {D.}~\bibnamefont {Colson}}, \ and\ \bibinfo {author}
  {\bibfnamefont {A.}~\bibnamefont {Forget}},\ }\bibfield  {title} {\enquote
  {\bibinfo {title} {{Raman scattering study of spin-density-wave order and
  electron-phonon coupling in Ba(Fe$_{1-x}$Co$_x$)$_{2}$As$_{2}$}},}\ }\href
  {\doibase 10.1103/PhysRevB.84.104508} {\bibfield  {journal} {\bibinfo
  {journal} {Phys. Rev. B}\ }\textbf {\bibinfo {volume} {84}},\ \bibinfo
  {pages} {104508} (\bibinfo {year} {2011})}\BibitemShut {NoStop}%
\bibitem [{\citenamefont {Yang}\ \emph {et~al.}(2014)\citenamefont {Yang},
  \citenamefont {Gallais}, \citenamefont {Rullier-Albenque}, \citenamefont
  {M\'easson}, \citenamefont {Cazayous}, \citenamefont {Sacuto}, \citenamefont
  {Shi}, \citenamefont {Colson},\ and\ \citenamefont {Forget}}]{yang14}%
  \BibitemOpen
  \bibfield  {author} {\bibinfo {author} {\bibfnamefont {Y.-X.}\ \bibnamefont
  {Yang}}, \bibinfo {author} {\bibfnamefont {Y.}~\bibnamefont {Gallais}},
  \bibinfo {author} {\bibfnamefont {F.}~\bibnamefont {Rullier-Albenque}},
  \bibinfo {author} {\bibfnamefont {M.-A.}\ \bibnamefont {M\'easson}}, \bibinfo
  {author} {\bibfnamefont {M.}~\bibnamefont {Cazayous}}, \bibinfo {author}
  {\bibfnamefont {A.}~\bibnamefont {Sacuto}}, \bibinfo {author} {\bibfnamefont
  {J.}~\bibnamefont {Shi}}, \bibinfo {author} {\bibfnamefont {D.}~\bibnamefont
  {Colson}}, \ and\ \bibinfo {author} {\bibfnamefont {A.}~\bibnamefont
  {Forget}},\ }\bibfield  {title} {\enquote {\bibinfo {title}
  {{Temperature-induced change in the Fermi surface topology in the spin
  density wave phase of Sr(Fe$_{1-x}$Co$_x$)$_2$As$_2$}},}\ }\href {\doibase
  10.1103/PhysRevB.89.125130} {\bibfield  {journal} {\bibinfo  {journal} {Phys.
  Rev. B}\ }\textbf {\bibinfo {volume} {89}},\ \bibinfo {pages} {125130}
  (\bibinfo {year} {2014})}\BibitemShut {NoStop}%
\bibitem [{\citenamefont {Zhang}\ \emph {et~al.}(2016)\citenamefont {Zhang},
  \citenamefont {Yin}, \citenamefont {Ignatov}, \citenamefont {Bukowski},
  \citenamefont {Karpinski}, \citenamefont {Sefat}, \citenamefont {Ding},
  \citenamefont {Richard},\ and\ \citenamefont {Blumberg}}]{zhang16}%
  \BibitemOpen
  \bibfield  {author} {\bibinfo {author} {\bibfnamefont {W.-L.}\ \bibnamefont
  {Zhang}}, \bibinfo {author} {\bibfnamefont {Z.~P.}\ \bibnamefont {Yin}},
  \bibinfo {author} {\bibfnamefont {A.}~\bibnamefont {Ignatov}}, \bibinfo
  {author} {\bibfnamefont {Z.}~\bibnamefont {Bukowski}}, \bibinfo {author}
  {\bibfnamefont {Janusz}\ \bibnamefont {Karpinski}}, \bibinfo {author}
  {\bibfnamefont {Athena~S.}\ \bibnamefont {Sefat}}, \bibinfo {author}
  {\bibfnamefont {H.}~\bibnamefont {Ding}}, \bibinfo {author} {\bibfnamefont
  {P.}~\bibnamefont {Richard}}, \ and\ \bibinfo {author} {\bibfnamefont
  {G.}~\bibnamefont {Blumberg}},\ }\bibfield  {title} {\enquote {\bibinfo
  {title} {{Raman scattering study of spin-density-wave-induced anisotropic
  electronic properties in $A${Fe}$_2${As}$_2$ ($A=\mathrm{Ca}$, Eu)}},}\
  }\href {\doibase 10.1103/PhysRevB.93.205106} {\bibfield  {journal} {\bibinfo
  {journal} {Phys. Rev. B}\ }\textbf {\bibinfo {volume} {93}},\ \bibinfo
  {pages} {205106} (\bibinfo {year} {2016})}\BibitemShut {NoStop}%
\bibitem [{\citenamefont {Wu}\ \emph {et~al.}(2008)\citenamefont {Wu},
  \citenamefont {Chen}, \citenamefont {Wu}, \citenamefont {Xie}, \citenamefont
  {Yan}, \citenamefont {Liu}, \citenamefont {Wang}, \citenamefont {Ying},\ and\
  \citenamefont {Chen}}]{wu08}%
  \BibitemOpen
  \bibfield  {author} {\bibinfo {author} {\bibfnamefont {G.}~\bibnamefont
  {Wu}}, \bibinfo {author} {\bibfnamefont {H.}~\bibnamefont {Chen}}, \bibinfo
  {author} {\bibfnamefont {T.}~\bibnamefont {Wu}}, \bibinfo {author}
  {\bibfnamefont {Y.~L.}\ \bibnamefont {Xie}}, \bibinfo {author} {\bibfnamefont
  {Y.~J.}\ \bibnamefont {Yan}}, \bibinfo {author} {\bibfnamefont {R.~H.}\
  \bibnamefont {Liu}}, \bibinfo {author} {\bibfnamefont {X.~F.}\ \bibnamefont
  {Wang}}, \bibinfo {author} {\bibfnamefont {J.~J.}\ \bibnamefont {Ying}}, \
  and\ \bibinfo {author} {\bibfnamefont {X.~H.}\ \bibnamefont {Chen}},\
  }\bibfield  {title} {\enquote {\bibinfo {title} {{Different resistivity
  response to spin-density wave and superconductivity at 20~K in
  Ca$_{1-x}$Na$_x$Fe$_2$As$_2$}},}\ }\href {\doibase
  10.1088/0953-8984/20/42/422201} {\bibfield  {journal} {\bibinfo  {journal}
  {J. Phys.: Condens. Matter}\ }\textbf {\bibinfo {volume} {20}},\ \bibinfo
  {pages} {422201} (\bibinfo {year} {2008})}\BibitemShut {NoStop}%
\bibitem [{\citenamefont {Ando}\ \emph {et~al.}(2001)\citenamefont {Ando},
  \citenamefont {Lavrov}, \citenamefont {Komiya}, \citenamefont {Segawa},\ and\
  \citenamefont {Sun}}]{ando01}%
  \BibitemOpen
  \bibfield  {author} {\bibinfo {author} {\bibfnamefont {Yoichi}\ \bibnamefont
  {Ando}}, \bibinfo {author} {\bibfnamefont {A.~N.}\ \bibnamefont {Lavrov}},
  \bibinfo {author} {\bibfnamefont {Seiki}\ \bibnamefont {Komiya}}, \bibinfo
  {author} {\bibfnamefont {Kouji}\ \bibnamefont {Segawa}}, \ and\ \bibinfo
  {author} {\bibfnamefont {X.~F.}\ \bibnamefont {Sun}},\ }\bibfield  {title}
  {\enquote {\bibinfo {title} {{Mobility of the Doped Holes and the
  Antiferromagnetic Correlations in Underdoped High- ${T}_{c}$ Cuprates}},}\
  }\href {\doibase 10.1103/PhysRevLett.87.017001} {\bibfield  {journal}
  {\bibinfo  {journal} {Phys. Rev. Lett.}\ }\textbf {\bibinfo {volume} {87}},\
  \bibinfo {pages} {017001} (\bibinfo {year} {2001})}\BibitemShut {NoStop}%
\bibitem [{\citenamefont {Lee}\ \emph {et~al.}(2005)\citenamefont {Lee},
  \citenamefont {Segawa}, \citenamefont {Li}, \citenamefont {Padilla},
  \citenamefont {Dumm}, \citenamefont {Dordevic}, \citenamefont {Homes},
  \citenamefont {Ando},\ and\ \citenamefont {Basov}}]{lee05}%
  \BibitemOpen
  \bibfield  {author} {\bibinfo {author} {\bibfnamefont {Y.~S.}\ \bibnamefont
  {Lee}}, \bibinfo {author} {\bibfnamefont {Kouji}\ \bibnamefont {Segawa}},
  \bibinfo {author} {\bibfnamefont {Z.~Q.}\ \bibnamefont {Li}}, \bibinfo
  {author} {\bibfnamefont {W.~J.}\ \bibnamefont {Padilla}}, \bibinfo {author}
  {\bibfnamefont {M.}~\bibnamefont {Dumm}}, \bibinfo {author} {\bibfnamefont
  {S.~V.}\ \bibnamefont {Dordevic}}, \bibinfo {author} {\bibfnamefont {C.~C.}\
  \bibnamefont {Homes}}, \bibinfo {author} {\bibfnamefont {Yoichi}\
  \bibnamefont {Ando}}, \ and\ \bibinfo {author} {\bibfnamefont {D.~N.}\
  \bibnamefont {Basov}},\ }\bibfield  {title} {\enquote {\bibinfo {title}
  {{Electrodynamics of the nodal metal state in weakly doped high-${T}_{c}$
  cuprates}},}\ }\href {\doibase 10.1103/PhysRevB.72.054529} {\bibfield
  {journal} {\bibinfo  {journal} {Phys. Rev. B}\ }\textbf {\bibinfo {volume}
  {72}},\ \bibinfo {pages} {054529} (\bibinfo {year} {2005})}\BibitemShut
  {NoStop}%
\bibitem [{\citenamefont {Homes}\ \emph {et~al.}(2013)\citenamefont {Homes},
  \citenamefont {Tu}, \citenamefont {Li}, \citenamefont {Gu},\ and\
  \citenamefont {Akrap}}]{homes13}%
  \BibitemOpen
  \bibfield  {author} {\bibinfo {author} {\bibfnamefont {C.~C.}\ \bibnamefont
  {Homes}}, \bibinfo {author} {\bibfnamefont {J.~J.}\ \bibnamefont {Tu}},
  \bibinfo {author} {\bibfnamefont {J.}~\bibnamefont {Li}}, \bibinfo {author}
  {\bibfnamefont {G.~D.}\ \bibnamefont {Gu}}, \ and\ \bibinfo {author}
  {\bibfnamefont {A.}~\bibnamefont {Akrap}},\ }\bibfield  {title} {\enquote
  {\bibinfo {title} {Optical conductivity of nodal metals},}\ }\href {\doibase
  10.1038/srep03446} {\bibfield  {journal} {\bibinfo  {journal} {Sci. Rep.}\
  }\textbf {\bibinfo {volume} {3}},\ \bibinfo {pages} {3446} (\bibinfo {year}
  {2013})}\BibitemShut {NoStop}%
\bibitem [{\citenamefont {Homes}\ \emph
  {et~al.}(2015{\natexlab{a}})\citenamefont {Homes}, \citenamefont {Dai},
  \citenamefont {Wen}, \citenamefont {Xu},\ and\ \citenamefont
  {Gu}}]{homes15a}%
  \BibitemOpen
  \bibfield  {author} {\bibinfo {author} {\bibfnamefont {C.~C.}\ \bibnamefont
  {Homes}}, \bibinfo {author} {\bibfnamefont {Y.~M.}\ \bibnamefont {Dai}},
  \bibinfo {author} {\bibfnamefont {J.~S.}\ \bibnamefont {Wen}}, \bibinfo
  {author} {\bibfnamefont {Z.~J.}\ \bibnamefont {Xu}}, \ and\ \bibinfo {author}
  {\bibfnamefont {G.~D.}\ \bibnamefont {Gu}},\ }\bibfield  {title} {\enquote
  {\bibinfo {title} {{FeTe}$_{0.55}${Se}$_{0.45}$: {A} multiband superconductor
  in the clean and dirty limit},}\ }\href {\doibase 10.1103/PhysRevB.91.144503}
  {\bibfield  {journal} {\bibinfo  {journal} {Phys. Rev. B}\ }\textbf {\bibinfo
  {volume} {91}},\ \bibinfo {pages} {144503} (\bibinfo {year}
  {2015}{\natexlab{a}})}\BibitemShut {NoStop}%
\bibitem [{\citenamefont {Homes}\ \emph
  {et~al.}(2015{\natexlab{b}})\citenamefont {Homes}, \citenamefont {Ali},\ and\
  \citenamefont {Cava}}]{homes15b}%
  \BibitemOpen
  \bibfield  {author} {\bibinfo {author} {\bibfnamefont {C.~C.}\ \bibnamefont
  {Homes}}, \bibinfo {author} {\bibfnamefont {M.~N.}\ \bibnamefont {Ali}}, \
  and\ \bibinfo {author} {\bibfnamefont {R.~J.}\ \bibnamefont {Cava}},\
  }\bibfield  {title} {\enquote {\bibinfo {title} {{Optical properties of the
  perfectly compensated semimetal {WTe}$_{2}$}},}\ }\href {\doibase
  10.1103/PhysRevB.92.161109} {\bibfield  {journal} {\bibinfo  {journal} {Phys.
  Rev. B}\ }\textbf {\bibinfo {volume} {92}},\ \bibinfo {pages} {161109(R)}
  (\bibinfo {year} {2015}{\natexlab{b}})}\BibitemShut {NoStop}%
\bibitem [{\citenamefont {Chen}\ \emph {et~al.}(2015)\citenamefont {Chen},
  \citenamefont {Zhang}, \citenamefont {Schneeloch}, \citenamefont {Zhang},
  \citenamefont {Li}, \citenamefont {Gu},\ and\ \citenamefont {Wang}}]{chen15}%
  \BibitemOpen
  \bibfield  {author} {\bibinfo {author} {\bibfnamefont {R.~Y.}\ \bibnamefont
  {Chen}}, \bibinfo {author} {\bibfnamefont {S.~J.}\ \bibnamefont {Zhang}},
  \bibinfo {author} {\bibfnamefont {J.~A.}\ \bibnamefont {Schneeloch}},
  \bibinfo {author} {\bibfnamefont {C.}~\bibnamefont {Zhang}}, \bibinfo
  {author} {\bibfnamefont {Q.}~\bibnamefont {Li}}, \bibinfo {author}
  {\bibfnamefont {G.~D.}\ \bibnamefont {Gu}}, \ and\ \bibinfo {author}
  {\bibfnamefont {N.~L.}\ \bibnamefont {Wang}},\ }\bibfield  {title} {\enquote
  {\bibinfo {title} {{Optical spectroscopy study of the three-dimensional Dirac
  semimetal {ZrTe}$_{5}$}},}\ }\href {\doibase 10.1103/PhysRevB.92.075107}
  {\bibfield  {journal} {\bibinfo  {journal} {Phys. Rev. B}\ }\textbf {\bibinfo
  {volume} {92}},\ \bibinfo {pages} {075107} (\bibinfo {year}
  {2015})}\BibitemShut {NoStop}%
\bibitem [{\citenamefont {Xu}\ \emph {et~al.}(2016)\citenamefont {Xu},
  \citenamefont {Dai}, \citenamefont {Zhao}, \citenamefont {Wang},
  \citenamefont {Yang}, \citenamefont {Zhang}, \citenamefont {Liu},
  \citenamefont {Xiao}, \citenamefont {Chen}, \citenamefont {Taylor},
  \citenamefont {Yarotski}, \citenamefont {Prasankumar},\ and\ \citenamefont
  {Qiu}}]{xu16}%
  \BibitemOpen
  \bibfield  {author} {\bibinfo {author} {\bibfnamefont {B.}~\bibnamefont
  {Xu}}, \bibinfo {author} {\bibfnamefont {Y.~M.}\ \bibnamefont {Dai}},
  \bibinfo {author} {\bibfnamefont {L.~X.}\ \bibnamefont {Zhao}}, \bibinfo
  {author} {\bibfnamefont {K.}~\bibnamefont {Wang}}, \bibinfo {author}
  {\bibfnamefont {R.}~\bibnamefont {Yang}}, \bibinfo {author} {\bibfnamefont
  {W.}~\bibnamefont {Zhang}}, \bibinfo {author} {\bibfnamefont {J.~Y.}\
  \bibnamefont {Liu}}, \bibinfo {author} {\bibfnamefont {H.}~\bibnamefont
  {Xiao}}, \bibinfo {author} {\bibfnamefont {G.~F.}\ \bibnamefont {Chen}},
  \bibinfo {author} {\bibfnamefont {A.~J.}\ \bibnamefont {Taylor}}, \bibinfo
  {author} {\bibfnamefont {D.~A.}\ \bibnamefont {Yarotski}}, \bibinfo {author}
  {\bibfnamefont {R.~P.}\ \bibnamefont {Prasankumar}}, \ and\ \bibinfo {author}
  {\bibfnamefont {X.~G.}\ \bibnamefont {Qiu}},\ }\bibfield  {title} {\enquote
  {\bibinfo {title} {{Optical spectroscopy of the Weyl semimetal TaAs}},}\
  }\href {\doibase 10.1103/PhysRevB.93.121110} {\bibfield  {journal} {\bibinfo
  {journal} {Phys. Rev. B}\ }\textbf {\bibinfo {volume} {93}},\ \bibinfo
  {pages} {121110} (\bibinfo {year} {2016})}\BibitemShut {NoStop}%
\bibitem [{\citenamefont {Akrap}\ \emph {et~al.}(2016)\citenamefont {Akrap},
  \citenamefont {Hakl}, \citenamefont {Tchoumakov}, \citenamefont {Crassee},
  \citenamefont {Kuba}, \citenamefont {Goerbig}, \citenamefont {Homes},
  \citenamefont {Caha}, \citenamefont {Nov\'ak}, \citenamefont {Teppe},
  \citenamefont {Desrat}, \citenamefont {Koohpayeh}, \citenamefont {Wu},
  \citenamefont {Armitage}, \citenamefont {Nateprov}, \citenamefont
  {Arushanov}, \citenamefont {Gibson}, \citenamefont {Cava}, \citenamefont
  {van~der Marel}, \citenamefont {Piot}, \citenamefont {Faugeras},
  \citenamefont {Martinez}, \citenamefont {Potemski},\ and\ \citenamefont
  {Orlita}}]{akrap16}%
  \BibitemOpen
  \bibfield  {author} {\bibinfo {author} {\bibfnamefont {A.}~\bibnamefont
  {Akrap}}, \bibinfo {author} {\bibfnamefont {M.}~\bibnamefont {Hakl}},
  \bibinfo {author} {\bibfnamefont {S.}~\bibnamefont {Tchoumakov}}, \bibinfo
  {author} {\bibfnamefont {I.}~\bibnamefont {Crassee}}, \bibinfo {author}
  {\bibfnamefont {J.}~\bibnamefont {Kuba}}, \bibinfo {author} {\bibfnamefont
  {M.~O.}\ \bibnamefont {Goerbig}}, \bibinfo {author} {\bibfnamefont {C.~C.}\
  \bibnamefont {Homes}}, \bibinfo {author} {\bibfnamefont {O.}~\bibnamefont
  {Caha}}, \bibinfo {author} {\bibfnamefont {J.}~\bibnamefont {Nov\'ak}},
  \bibinfo {author} {\bibfnamefont {F.}~\bibnamefont {Teppe}}, \bibinfo
  {author} {\bibfnamefont {W.}~\bibnamefont {Desrat}}, \bibinfo {author}
  {\bibfnamefont {S.}~\bibnamefont {Koohpayeh}}, \bibinfo {author}
  {\bibfnamefont {L.}~\bibnamefont {Wu}}, \bibinfo {author} {\bibfnamefont
  {N.~P.}\ \bibnamefont {Armitage}}, \bibinfo {author} {\bibfnamefont
  {A.}~\bibnamefont {Nateprov}}, \bibinfo {author} {\bibfnamefont
  {E.}~\bibnamefont {Arushanov}}, \bibinfo {author} {\bibfnamefont {Q.~D.}\
  \bibnamefont {Gibson}}, \bibinfo {author} {\bibfnamefont {R.~J.}\
  \bibnamefont {Cava}}, \bibinfo {author} {\bibfnamefont {D.}~\bibnamefont
  {van~der Marel}}, \bibinfo {author} {\bibfnamefont {B.~A.}\ \bibnamefont
  {Piot}}, \bibinfo {author} {\bibfnamefont {C.}~\bibnamefont {Faugeras}},
  \bibinfo {author} {\bibfnamefont {G.}~\bibnamefont {Martinez}}, \bibinfo
  {author} {\bibfnamefont {M.}~\bibnamefont {Potemski}}, \ and\ \bibinfo
  {author} {\bibfnamefont {M.}~\bibnamefont {Orlita}},\ }\bibfield  {title}
  {\enquote {\bibinfo {title} {{Magneto-Optical Signature of Massless Kane
  Electrons in ${\mathrm{Cd}}_{3}{\mathrm{As}}_{2}$}},}\ }\href {\doibase
  10.1103/PhysRevLett.117.136401} {\bibfield  {journal} {\bibinfo  {journal}
  {Phys. Rev. Lett.}\ }\textbf {\bibinfo {volume} {117}},\ \bibinfo {pages}
  {136401} (\bibinfo {year} {2016})}\BibitemShut {NoStop}%
\end{thebibliography}
%
%

\newpage \ \\

\newpage
\vspace*{-2.0cm}
\hspace*{-2.5cm} {
  \centering
  \includegraphics[width=1.2\textwidth]{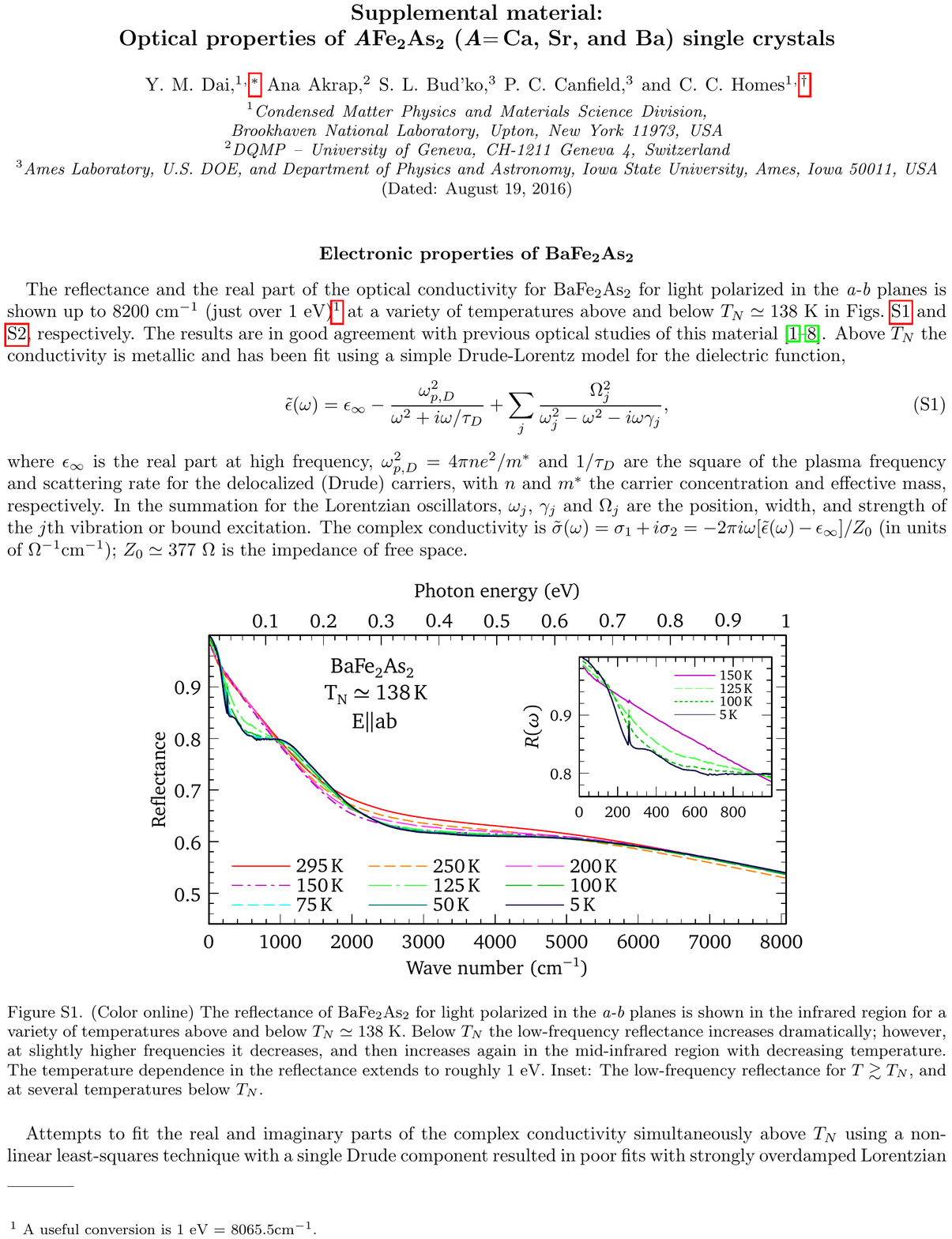} \\
  \ \\
}

\newpage
\vspace*{-2.0cm}
\hspace*{-2.5cm} {
  \centering
  \includegraphics[width=1.2\textwidth]{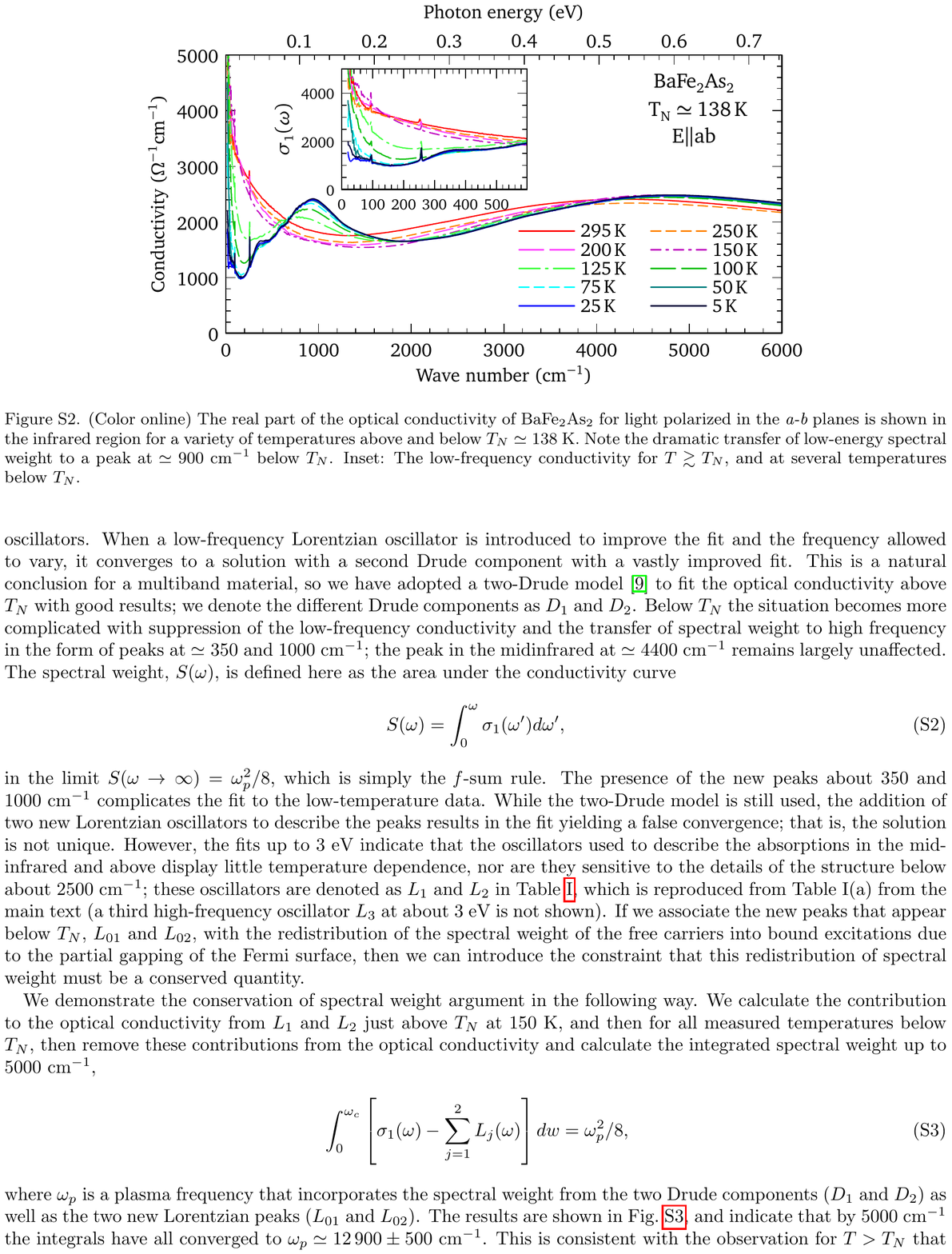} \\
  \ \\
}

\newpage
\vspace*{-2.0cm}
\hspace*{-2.5cm} {
  \centering
  \includegraphics[width=1.2\textwidth]{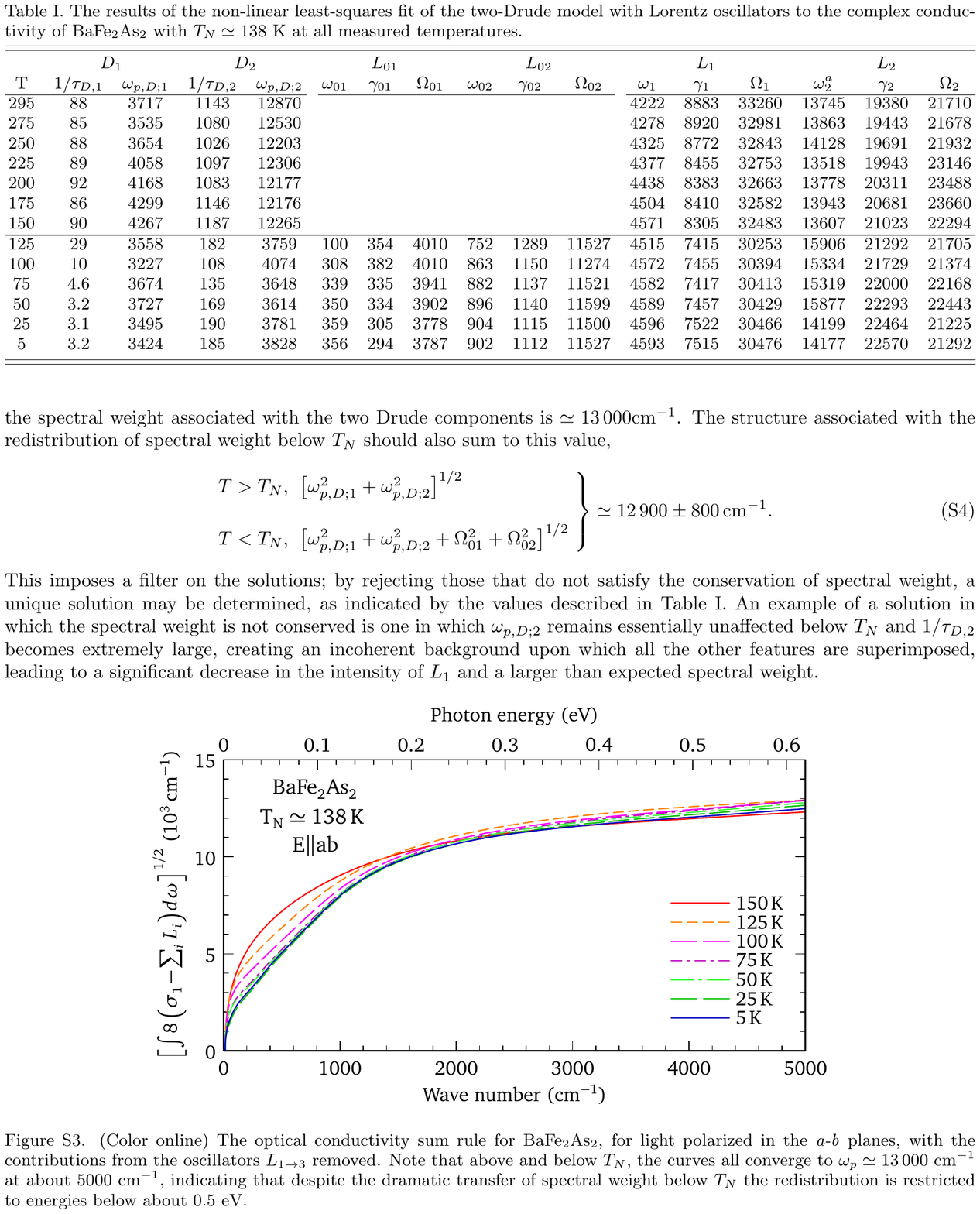} \\
  \ \\
}

\newpage
\vspace*{-2.0cm}
\hspace*{-2.5cm} {
  \centering
  \includegraphics[width=1.2\textwidth]{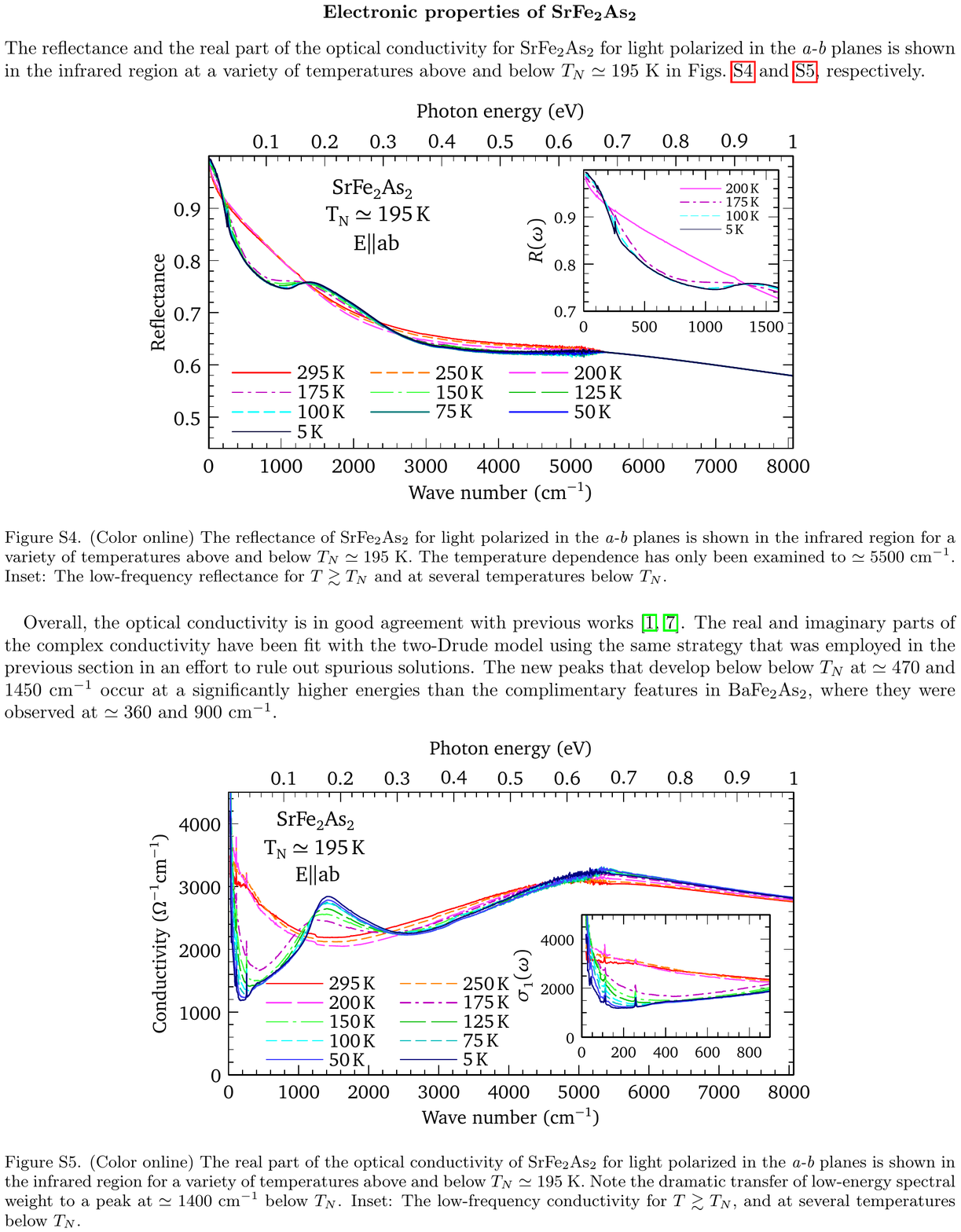} \\
  \ \\
}

\newpage
\vspace*{-2.0cm}
\hspace*{-2.5cm} {
  \centering
  \includegraphics[width=1.2\textwidth]{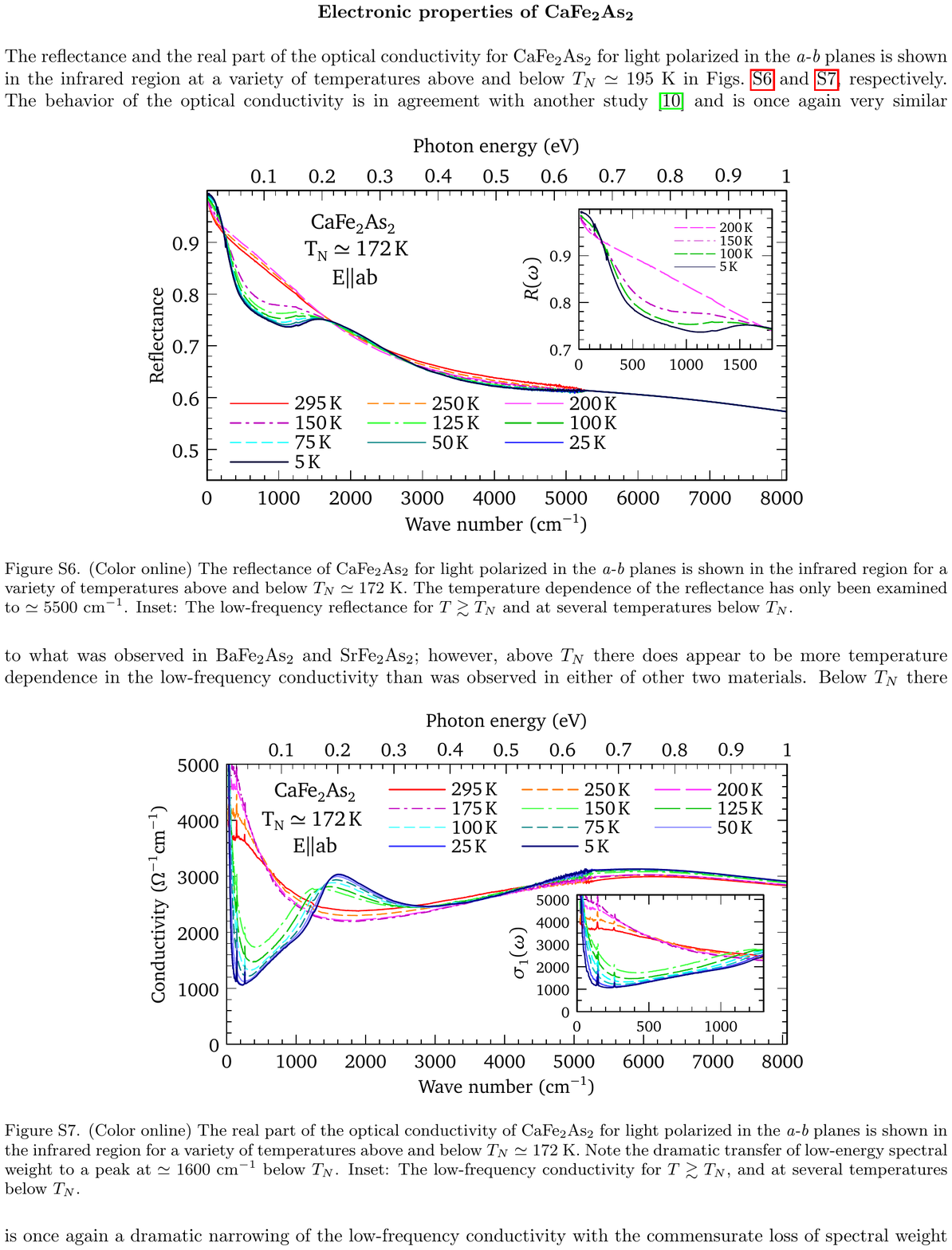} \\
  \ \\
}

\newpage
\vspace*{-2.0cm}
\hspace*{-2.5cm} {
  \centering
  \includegraphics[width=1.2\textwidth]{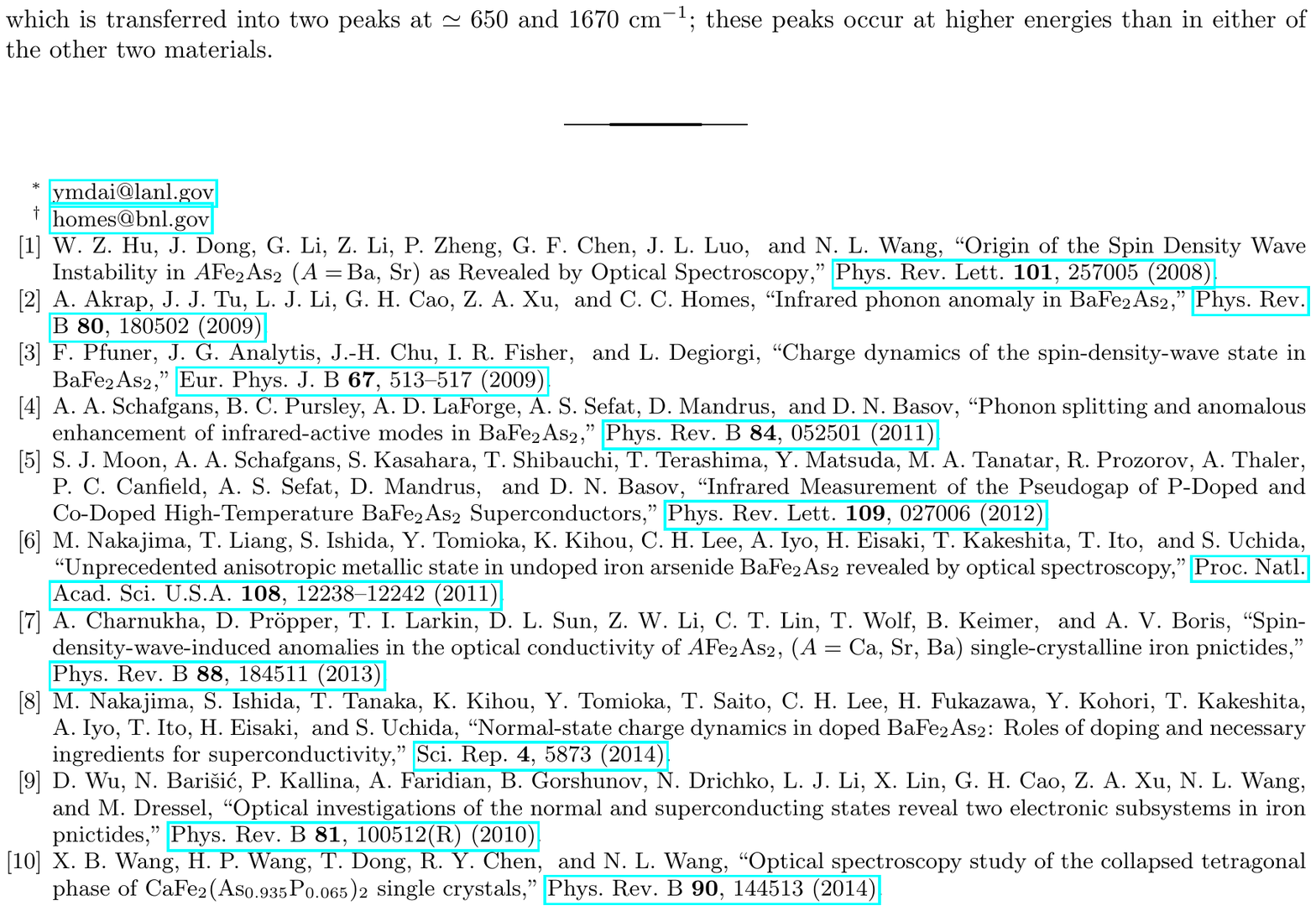} \\
  \ \\
}

\end{document}